\documentclass[
aps,
prd,
reprint,
amsmath,
amssymb,
superscriptaddress
]{revtex4-2}

\usepackage{amsmath,amssymb}
\usepackage{graphicx}
\usepackage[
colorlinks=true,
linkcolor=blue,
citecolor=blue,
urlcolor=blue
]{hyperref}

\begin{document}


\title{Selected open-charm decays of the narrow \(P_c\) states: Quark-interchange predictions and pion-exchange sensitivity}


\author{Vandan Patel}
\email[]{vandankp12998@gmail.com}
\affiliation{Department of Physics, Sardar Vallabhbhai National Institute of Technology, Surat, Gujarat-395007, India}

\author{Ajay Kumar Rai}
\email[]{raiajayk@gmail.com}
\affiliation{Department of Physics, Sardar Vallabhbhai National Institute of Technology, Surat, Gujarat-395007, India}


\date{\today}

\begin{abstract}
	The proximity of the narrow $P_c(4312)$, $P_c(4440)$, and $P_c(4457)$
	structures to the $\Sigma_c\bar D^{(*)}$ thresholds motivates their
	interpretation as hadronic molecules. We treat the $P_c(4312)$ as an
	$S$-wave $\Sigma_c\bar D$ state with $J^P=1/2^-$ and examine both
	$J^P=1/2^-$ and $3/2^-$ $\Sigma_c\bar D^*$ assignments for each heavier
	peak. The selected $S$-wave open-charm transitions to
	$\Lambda_c\bar D^{(*)}$ are calculated at Born order in the
	quark-interchange model, including the prior and post rearrangement diagrams
	and their color, flavor, spin, and spatial matrix elements. Across the input
	variations considered, the quark-interchange widths robustly favor
	$\Lambda_c\bar D^*$ over $\Lambda_c\bar D$ for every $J^P=1/2^-$
	assignment. At either heavier peak mass, the $\Lambda_c\bar D^*$ width for
	$J^P=3/2^-$ is suppressed by approximately one order of magnitude relative
	to $J^P=1/2^-$, providing a potential decay-based discriminator of the spin
	assignments. The large channel hierarchies arise from spatial overlaps and
	interference between spin-independent and hyperfine contributions and are
	not universal heavy-quark-spin-symmetry ratios. We also explore a retarded
	one-pion-exchange transition at first Born order. It affects
	$P_c(4312)\to\Lambda_c\bar D^*$ most strongly, while the coherent widths
	remain sensitive to the cutoff, energy-transfer prescription, short-distance
	subtraction, and unresolved relative matching sign. The quark-interchange
	hierarchies are therefore the central predictions, whereas the pion-exchange
	results quantify matching sensitivity. Measurements of the
	$\Lambda_c\bar D^*$ modes can directly test these patterns.
\end{abstract}


\maketitle

\section{Introduction}
\label{sec:introduction}

The spectroscopy of hadrons beyond conventional quark--antiquark and
three-quark configurations provides an important probe of nonperturbative
quantum chromodynamics. The expanding exotic-hadron spectrum includes both
tetraquark and pentaquark candidates, for which compact-multiquark and
hadronic-molecular descriptions are among the principal interpretations
~\cite{Chen2016Exotics,Lebed2017Exotics}. Phenomenological studies range
from Regge-trajectory analyses of tetraquark spectra across the light,
strange, charm, and bottom flavor sectors
~\cite{Patel2025LightTetra,Patel2025CharmTetra,Patel2025BottomTetra}
to constituent-model investigations of beauty-containing tetraquark
spectroscopy and decay dynamics~\cite{Lodha2026BeautyTetra} and of singly
charmed pentaquark spectra and meson--baryon decay patterns
~\cite{Rathod2026Pentaquark}. Since different structural hypotheses can
produce similar mass spectra, decay observables are especially valuable for
discriminating among them.

In the hidden-charm baryon sector, the LHCb
Collaboration first observed structures in the $J/\psi p$ invariant-mass
distribution of $\Lambda_b^0\to J/\psi p K^-$ decays~\cite{LHCb2015}. A
later analysis with a larger data sample revealed the narrow $P_c(4312)^+$
state and resolved the former $P_c(4450)^+$ enhancement into the nearby
$P_c(4440)^+$ and $P_c(4457)^+$ peaks~\cite{LHCb2019}. Their proximity to
the $\Sigma_c\bar D$ and $\Sigma_c\bar D^*$ thresholds makes a
hadronic-molecular interpretation particularly attractive, although compact
multiquark configurations and kinematic mechanisms have not been excluded;
see Refs.~\cite{Guo2018,Chen2023Review} for reviews.

In the simplest $S$-wave molecular picture, the $P_c(4312)$ is predominantly
an isospin-$1/2$ $\Sigma_c\bar D$ state with $J^P=1/2^-$, whereas the
$P_c(4440)$ and $P_c(4457)$ are associated, in an experimentally unresolved
order, with the $J^P=1/2^-$ and $3/2^-$ $\Sigma_c\bar D^*$ states. Such a
multiplet arises naturally in descriptions based on heavy-quark spin
symmetry (HQSS) and coupled-channel dynamics
~\cite{Liu2019HQSS,Xiao2019HQSS,Du2021}. The peak masses alone have not fixed
the spin ordering, and theoretical assignments remain model dependent.
This motivates the central question of the present work: can selected
open-charm decay hierarchies distinguish the two molecular spin assignments?

Decay observables contain information not supplied by the masses. HQSS gives
characteristic relations among hidden-charm $J/\psi N$ and $\eta_cN$ modes
and, under molecular assumptions, among open-charm
$\Lambda_c\bar D^{(*)}$ modes~\cite{Voloshin2019,Sakai2019}.
Effective-Lagrangian calculations have estimated a broader set of allowed
strong decays~\cite{LinZou2019}. At the constituent level, Wang
\textit{et al.}\ applied the quark-interchange (QI) model to
$P_c\to J/\psi p,\eta_cp$ and showed that hidden-charm decay ratios can probe
the molecular composition~\cite{Wang2020QI}. Because the QI transition is
generated by constituent interactions constrained by ordinary hadron
spectroscopy rather than channel-specific $P_c$--hadron couplings, it
provides a microscopic framework for studying open-charm rearrangement
decays as well. Coupled-channel and compositeness analyses have also treated the
$\Lambda_c\bar D^{(*)}$ channels explicitly and extracted their couplings,
decay strengths, or production rates
~\cite{GuoOller2019,Pan2023Production}.

Several approaches already indicate that $\Lambda_c\bar D^*$ is an
especially informative final state. Constituent-quark,
Bethe--Salpeter, and complex-scaling calculations find a pronounced
$\Lambda_c\bar D^*$ preference for the $P_c(4312)$, although their absolute
widths and dynamical treatments differ
~\cite{Dong2020,Li2023BS,Lin2023CSM}. Group-theoretical analyses likewise
identify spin-dependent restrictions in hidden- and open-charm
channels~\cite{Phumphan2021}. For the two heavier peaks, however, no common
spin assignment has emerged. Complex-scaling and pion-exchange
triangle-diagram studies favor
$J[P_c(4440)]=1/2$ and $J[P_c(4457)]=3/2$ under their respective
assumptions~\cite{Lin2023CSM,Yang2025}, whereas coupled-channel, Bethe--Salpeter, effective-Lagrangian, and
Born--Oppenheimer EFT analyses have supported the reverse ordering
~\cite{Du2021,Li2025BS,Brambilla2025BOEFT,Deng2026}.
Phenomenological constraints based on the disparity between the
$\Lambda_c\bar D$ and $\Lambda_c\bar D^*$ modes also favor
$J[P_c(4440)]=3/2$, while questioning whether $P_c(4457)$ belongs to the
same simple molecular multiplet~\cite{BurnsSwanson2022}.

Long-distance pion dynamics must also be interpreted with care. Iterated
tensor one-pion exchange (OPE) in nonperturbative multichannel calculations
requires suitable short-distance operators for regulator independence, and
on-shell $\Lambda_c\bar D^{(*)}\pi$ intermediate states generate three-body
cuts in a unitary description~\cite{Du2021,Lin2023CSM}. Dedicated effective-Lagrangian calculations also find sizable
$\bar D^{(*)}\Lambda_c\pi$ three-body widths and identify these modes as
promising search channels~\cite{LiuPanGeng2024}. Moreover, the
short-range part of OPE cannot be separated from contact or quark-exchange
dynamics in a model-independent way; its treatment can affect spin-dependent
conclusions~\cite{Xu2025QuarkExchange}. A Born-level QI--OPE combination
must therefore be presented as a controlled sensitivity study rather than a
complete coupled-channel construction.

The relevant open-charm final states are experimentally accessible. LHCb has
observed $\Lambda_b^0\to\Lambda_c^+\bar D^{(*)0}K^-$ decays and identified
them as useful normalization channels for future pentaquark searches in the
same final states~\cite{LHCb2024OpenCharm}. A complementary LHCb search for
prompt pentaquark production in charm-hadron final states found no
significant signal and established upper limits on the corresponding
yields~\cite{LHCb2024Prompt}. Theoretical studies also identify
$\Lambda_b^0\to\Lambda_c^+\bar D^{*0}K^-$ as a promising channel for future
searches~\cite{BurnsSwanson2022,Pan2023Production}. Measurements of the
resonant $\Lambda_c\bar D^{(*)}$ strengths can consequently test both
channel hierarchies and the spin dependence of the two heavier structures.

In this work, we calculate the selected transitions
$\Sigma_c\bar D^{(*)}\to\Lambda_c\bar D^{(*)}$ for the $P_c(4312)$,
$P_c(4440)$, and $P_c(4457)$, treating the initial states as pure $S$-wave
molecules. Our central calculation evaluates the $S$-wave QI amplitudes at
Born order, including all retained prior and post rearrangement diagrams and
their color, flavor, spin, and spatial matrix elements. We test the resulting
partial widths against variations of the molecular scale, threshold
prescription, and constituent inputs. The $\Lambda_c\bar D^*$ mode is
favored for every tested $J^P=1/2^-$ assignment and is strongly suppressed
for $J^P=3/2^-$ at a fixed heavier peak mass. We then supplement QI with a
direct retarded OPE kernel at first Born order. A pion-pole prescription and
the full regulated kernel are used to diagnose long-distance and
short-distance-matching sensitivity, including the unresolved global
QI--OPE sign. The QI patterns are our central predictions; the coherent
QI+OPE widths are alternative matching scenarios. The calculation is
restricted to the selected channels and to final $S$ waves. In particular,
$P_c(J^P=3/2^-)\to\Lambda_c\bar D$ begins in a $D$ wave and is not evaluated.

The paper is organized as follows. Section~\ref{sec:framework} defines the
molecular assignments, kinematics, and amplitude conventions.
Sections~\ref{sec:qi} and~\ref{sec:ope} construct the QI and OPE transition
mechanisms. Numerical inputs and validation checks are given in
Sec.~\ref{sec:validation}, followed by the results and discussion in
Sec.~\ref{sec:results}. Section~\ref{sec:summary} summarizes the conclusions
and outlook.

\section{Molecular assignments, kinematics, and conventions}
\label{sec:framework}

We work in natural units, $\hbar=c=1$. This section fixes the channel states,
partial-wave truncation, molecular wave functions, and amplitude normalization
used throughout the calculation. We first specify the spin--isospin basis and
the allowed final partial waves, then define the initial radial wave function,
and finally connect the transition kernels constructed in
Secs.~\ref{sec:qi} and~\ref{sec:ope} to the partial widths.

The three narrow structures reported by LHCb~\cite{LHCb2019} are treated as
near-threshold hadronic molecules with total isospin $I=\tfrac12$. The
configurations considered are
\begin{align}
	P_c(4312)^+ &: \quad
	\bigl|\Sigma_c\bar D;\,I=\tfrac12,\,
	J^P=\tfrac12^-\bigr\rangle,
	\label{eq:pc4312_assignment}\\
	P_c(4440)^+,\;P_c(4457)^+ &: \quad
	\bigl|\Sigma_c\bar D^*;\,I=\tfrac12\bigr\rangle,
	\nonumber\\[-0.4ex]
	&\hspace{1.5em}J^P=\tfrac12^-
	\quad\text{or}\quad J^P=\tfrac32^- .
	\label{eq:pc44_assignment}
\end{align}
Their proximity to the corresponding $\Sigma_c\bar D^{(*)}$ thresholds
motivates these assignments. For ground-state constituents in a relative
$S$ wave, $\Sigma_c\bar D$ has $J^P=\tfrac12^-$, whereas
$\Sigma_c\bar D^*$ allows $J^P=\tfrac12^-$ and $\tfrac32^-$. These
assignments are widely used in molecular descriptions of the narrow $P_c$
states~\cite{Liu2019HQSS,Xiao2019HQSS,Wang2020QI,Lin2023CSM}.

Because the relative spin ordering of the two heavier peaks remains
model dependent~\cite{Liu2019HQSS,PavonValderrama2019OPE,Du2021,
	Lin2023CSM}, we do not impose either ordering at this stage. Instead, both
$J^P=\tfrac12^-$ and $\tfrac32^-$ are evaluated for each peak. Hence,
Eqs.~\eqref{eq:pc4312_assignment} and~\eqref{eq:pc44_assignment} define five
state--spin cases: one for $P_c(4312)$ and two for each heavier state.

At the level of the initial molecular wave function, every case is
approximated as a pure, single-channel $\Sigma_c\bar D^{(*)}$ molecule;
coupled-channel admixtures and their dynamical feedback are outside the
baseline calculation. The spin--flavor wave functions and charge-channel
reconstruction are evaluated in the isospin-symmetric limit. Physical charged
and neutral hadron masses enter only through the explicitly stated molecular
threshold prescription and two-body kinematics; they do not alter the
isospin Clebsch--Gordan algebra.

\subsection{Isospin, spin, and allowed partial waves}
\label{sec:states_isospin}

We denote total isospin and its third component by $I$ and $I_3$, total
angular momentum and its projection by $J$ and $M_J$, and the initial (final)
relative orbital angular momentum and channel spin by $L_i,S_i$
($L_f,S_f$). We use the Condon--Shortley convention for isospin
Clebsch--Gordan coefficients and fix the charge-state phases by
\begin{equation}
	\begin{aligned}
		|\Sigma_c^{++}\rangle&=|1,+1\rangle,
		&|\Sigma_c^+\rangle&=|1,0\rangle,\\
		|\bar D^{(*)0}\rangle&=\bigl|\tfrac12,+\tfrac12\bigr\rangle,
		&|D^{(*)-}\rangle&=\bigl|\tfrac12,-\tfrac12\bigr\rangle.
	\end{aligned}
	\label{eq:isospin_phase_conventions}
\end{equation}
The positively charged $I=1/2$ initial state is then
\begin{equation}
	\begin{aligned}
		&\bigl|\Sigma_c\bar D^{(*)};
		I=\tfrac12,I_3=+\tfrac12\bigr\rangle\\
		&\quad=\sqrt{\frac{2}{3}}\,
		\bigl|\Sigma_c^{++}D^{(*)-}\bigr\rangle
		-\frac{1}{\sqrt{3}}\,
		\bigl|\Sigma_c^+\bar D^{(*)0}\bigr\rangle .
	\end{aligned}
	\label{eq:initial_isospin_state}
\end{equation}
Because $\Lambda_c$ is an isosinglet, the corresponding $I=1/2$ final state
is simply
\begin{equation}
	\bigl|\Lambda_c\bar D^{(*)};
	I=\tfrac12,I_3=+\tfrac12\bigr\rangle
	=\bigl|\Lambda_c^+\bar D^{(*)0}\bigr\rangle .
	\label{eq:final_isospin_state}
\end{equation}

The QI calculation explicitly evaluates the charged initial component
$\Sigma_c^{++}D^{(*)-}$. To show how this result determines the physical
$I=1/2$ transition, let $\mathcal A_{\rm ch}$ and
$\mathcal A_{\rm neu}$ denote, for any fixed spin and momentum arguments, the
amplitudes from the charged and neutral initial components, respectively. Let
$\mathcal A_{1/2}$ be the amplitude between the normalized $I=1/2$ states in
Eqs.~\eqref{eq:initial_isospin_state} and
\eqref{eq:final_isospin_state}. Since the strong transition operator is an
isoscalar, its $I=3/2\to I=1/2$ matrix element vanishes. The Clebsch--Gordan
projections therefore give
\begin{equation}
	\begin{aligned}
		\mathcal A_{\rm ch}
		&=\sqrt{\frac{2}{3}}\,\mathcal A_{1/2},\\
		\mathcal A_{\rm neu}
		&=-\frac{1}{\sqrt{3}}\,\mathcal A_{1/2}
		=-\frac{\mathcal A_{\rm ch}}{\sqrt{2}},\\
		\mathcal A_{1/2}
		&=\sqrt{\frac{2}{3}}\,\mathcal A_{\rm ch}
		-\frac{1}{\sqrt{3}}\,\mathcal A_{\rm neu}
		=\sqrt{\frac{3}{2}}\,\mathcal A_{\rm ch}.
	\end{aligned}
	\label{eq:isospin_amplitude_reconstruction}
\end{equation}
By linearity, Eq.~\eqref{eq:isospin_amplitude_reconstruction} applies both to
a charge-channel kernel at fixed momenta and to the amplitude obtained after
folding that kernel with the molecular wave function. Its use in constructing
the physical QI kernel is made explicit in Sec.~\ref{sec:qi}. The neutral
charge diagrams are therefore not recalculated independently with
charge-dependent masses. An overall phase of either channel is conventional,
but the relative sign in Eq.~\eqref{eq:initial_isospin_state} must be retained.
The additional relative phase required when QI and OPE are combined is
discussed in Sec.~\ref{sec:ope}.

For later use, consider an initial molecular state with relative momentum
$\mathbf p$, $p=|\mathbf p|$, and
$\widehat{\mathbf p}=\mathbf p/p$. Its spin--orbital wave function is written
as
\begin{equation}
	\begin{aligned}
		\Psi_{P_c}^{JM_J}(\mathbf p)
		={}&\sum_{m_{L_i},M_{S_i}}
		\langle L_i m_{L_i},S_i M_{S_i}\mid J M_J\rangle\\
		&\times R_{L_i}(p)Y_{L_i m_{L_i}}(\widehat{\mathbf p})
		\bigl[\chi_A\otimes\chi_B\bigr]_{S_iM_{S_i}},
	\end{aligned}
	\label{eq:pc_spin_orbital_wavefunction}
\end{equation}
where $A=\bar D^{(*)}$ is the initial anticharmed meson, $B=\Sigma_c$ is the
initial charmed baryon, $R_{L_i}$ is the radial wave function, and
$Y_{L_i m_{L_i}}$ is a spherical harmonic. The bracketed product couples the
meson and baryon spins to $S_i,M_{S_i}$. Although channel names are often
written in baryon--meson order, spin states are coupled in meson--baryon order
in both the initial and final channels; this convention fixes the associated
Clebsch--Gordan phases. The constituent-hadron spins are
$s_{\bar D}=0$, $s_{\bar D^*}=1$, and $s_{\Sigma_c}=1/2$.

For the initial $S$ waves retained here, $L_i=m_{L_i}=0$ and $S_i=J$;
$R_{L_i=0}$ is the state-dependent function $R_0$ defined in
Sec.~\ref{sec:molecular_wavefunction}. The final meson and baryon spins are
likewise coupled in the basis
$[\chi_{M_f}\otimes\chi_{B_f}]_{S_fM_{S_f}}$, where
$M_f=\bar D^{(*)}$ and $B_f=\Lambda_c$.

The intrinsic parities of $M_f=\bar D^{(*)}$ and $B_f=\Lambda_c$ are $-$ and
$+$, respectively. Their two-body parity is therefore
\begin{equation}
	P_f=P_{M_f}P_{B_f}(-1)^{L_f}=-(-1)^{L_f}.
	\label{eq:final_state_parity}
\end{equation}
Parity conservation for a negative-parity $P_c$ requires even $L_f$. For
$L_f=0$, the $\Lambda_c\bar D$ channel has $S_f=1/2$ and hence only
$J=1/2$, whereas $\Lambda_c\bar D^*$ has $S_f=1/2$ or $3/2$ and admits both
values of $J$ considered here. The lowest allowed partial waves are summarized
in Table~\ref{tab:allowed_partial_waves}.

\begin{table}[t]
	\caption{Lowest parity- and angular-momentum-allowed partial waves for the
		selected open-charm channels. The $D$-wave channel is physically allowed
		but is not evaluated in this work.}
	\label{tab:allowed_partial_waves}
	\begin{ruledtabular}
		\begin{tabular}{lccc}
			Initial configuration & $J^P$ & $\Lambda_c\bar D$ &
			$\Lambda_c\bar D^*$ \\
			\colrule
			$\Sigma_c\bar D$   & $\tfrac12^-$ & $S$ wave & $S$ wave \\
			$\Sigma_c\bar D^*$ & $\tfrac12^-$ & $S$ wave & $S$ wave \\
			$\Sigma_c\bar D^*$ & $\tfrac32^-$ & $D$ wave & $S$ wave
		\end{tabular}
	\end{ruledtabular}
\end{table}

The $\Lambda_c\bar D$ entry in the last row is not a selection-rule zero:
$P_c(3/2^-)\to\Lambda_c\bar D$ begins at $L_f=2$. Our calculation retains
only the diagonal $L_i=L_f=0$ projection and does not evaluate the
$L_i=0\to L_f=2$ amplitude generated by noncentral rank-two terms. Thus, this
decay is uncalculated rather than forbidden. A rough order-of-magnitude comparison with calculations
that retain the required rank-two transition operators is
given in Sec.~\ref{sec:results_comparison}. The five state--spin cases and
the selected final channels consequently give eight calculated $S$-wave
widths.

\subsection{Molecular wave function and binding prescription}
\label{sec:molecular_wavefunction}

Following the pure-molecule scenario of Wang \textit{et al.}
\cite{Wang2020QI}, we retain only the dominant initial
$\Sigma_c\bar D^{(*)}$ component. Its relative $S$-wave momentum
distribution is approximated by the normalized Gaussian
\begin{equation}
	\begin{aligned}
		\psi_{P_c}(\mathbf p)&=R_0(p)Y_{00}(\widehat{\mathbf p}),\\
		R_0(p)&=\frac{2}{\pi^{1/4}\beta_{P_c}^{3/2}}
		\exp\!\left(-\frac{p^2}{2\beta_{P_c}^2}\right),
	\end{aligned}
	\label{eq:molecular_gaussian}
\end{equation}
where $Y_{00}=1/\sqrt{4\pi}$ and $\beta_{P_c}$ is the molecular momentum
width. The normalization is
\begin{equation}
	\int_0^\infty p^2\,dp\,|R_0(p)|^2=1,
	\qquad
	\int d^3p\,|\psi_{P_c}(\mathbf p)|^2=1.
	\label{eq:molecular_normalization}
\end{equation}
The peak dependence of $R_0$ enters through $\beta_{P_c}$ and is suppressed
in the notation.

For a selected pair of initial-hadron masses $M_A$ and $M_B$, define the
positive binding energy and reduced mass by
\begin{equation}
	E_{\rm bind}=M_A+M_B-M_{P_c}>0,
	\qquad
	\mu_{AB}=\frac{M_AM_B}{M_A+M_B}.
	\label{eq:binding_energy_reduced_mass}
\end{equation}
The Gaussian root-mean-square radius and the characteristic molecular size
used in Ref.~\cite{Wang2020QI} are
\begin{equation}
	r_{\rm rms}^{(G)}=\sqrt{\frac{3}{2}}\,\frac{1}{\beta_{P_c}},
	\qquad
	r_{\rm typ}=\frac{1}{\sqrt{2\mu_{AB}E_{\rm bind}}}.
	\label{eq:gaussian_size_matching}
\end{equation}
Equating these two scales gives the working prescription
\begin{equation}
	\beta_{P_c}=\sqrt{3\mu_{AB}E_{\rm bind}}.
	\label{eq:beta_binding}
\end{equation}
Equation~\eqref{eq:beta_binding} is a phenomenological Gaussian-size
prescription; it is not the universal zero-range expression for a bound-state
root-mean-square radius.

For the reference calculation, $M_A$ and $M_B$ in
Eq.~\eqref{eq:binding_energy_reduced_mass} are set equal to the arithmetic
charge averages
\begin{equation}
	\begin{aligned}
		M_A^{\rm av}
		&=\frac12\left(M_{D^{(*)-}}+M_{\bar D^{(*)0}}\right),\\
		M_B^{\rm av}
		&=\frac12\left(M_{\Sigma_c^{++}}+M_{\Sigma_c^+}\right),\\
		\overline M_{\rm th}^{\,D^{(*)}}
		&\equiv M_A^{\rm av}+M_B^{\rm av},\\
		\mu_{AB}^{\rm av}
		&\equiv
		\frac{M_A^{\rm av}M_B^{\rm av}}
		{M_A^{\rm av}+M_B^{\rm av}}.
	\end{aligned}
	\label{eq:average_threshold_masses}
\end{equation}
Thus, the central values use
$E_{\rm bind}=\overline M_{\rm th}^{\,D^{(*)}}-M_{P_c}$ and
$\mu_{AB}=\mu_{AB}^{\rm av}$. The averaged threshold is the arithmetic mean
of the $\Sigma_c^{++}D^{(*)-}$ and
$\Sigma_c^+\bar D^{(*)0}$ thresholds, not a Clebsch--Gordan-weighted average.
The two physical thresholds and direct variations of $\beta_{P_c}$ are later
used as sensitivity tests; their numerical values are given in
Sec.~\ref{sec:validation}.

\subsection{Two-body kinematics, amplitudes, and widths}
\label{sec:kinematics_width}

For a fixed transition, let $i$ denote the initial $P_c$ channel and
$f=M_fB_f$ the final two-hadron channel, with $M_f=\bar D^{(*)}$ and
$B_f=\Lambda_c$. A suffix $fi$ is always ordered as final, initial, in
accordance with a matrix element $\langle f|V|i\rangle$. In the $P_c$ rest
frame, the four-momenta are
\begin{equation}
	\begin{aligned}
		P^\mu&=(M_{P_c},\mathbf 0),& k&=|\mathbf k|,\\
		p_M^\mu&=(E_{M_f},\mathbf k),&
		p_B^\mu&=(E_{B_f},-\mathbf k).
	\end{aligned}
	\label{eq:rest_frame_momenta}
\end{equation}
The final relative-momentum magnitude is
\begin{equation}
	\begin{aligned}
		k&=\frac{\lambda_{\rm K}^{1/2}
			(M_{P_c}^2,M_{M_f}^2,M_{B_f}^2)}{2M_{P_c}},\\
		\lambda_{\rm K}(x,y,z)
		&=x^2+y^2+z^2-2xy-2xz-2yz,
	\end{aligned}
	\label{eq:two_body_momentum}
\end{equation}
where $M_{M_f}$ and $M_{B_f}$ are the physical final-hadron masses. Their
energies are
\begin{equation}
	E_{M_f}=\sqrt{M_{M_f}^2+k^2},
	\qquad
	E_{B_f}=\sqrt{M_{B_f}^2+k^2}.
	\label{eq:final_energies}
\end{equation}

We use the delta-normalized scattering states and partial-wave convention of
Ref.~\cite{Wang2020QI},
$\langle\mathbf p'|\mathbf p\rangle=\delta^3(\mathbf p'-\mathbf p)$.
The transition dynamics is first represented by the spin-coupled, but
molecularly unfolded, kernel
$V_{\mathrm{eff};fi}^{(J)}(k,p,u)$, where
\begin{equation*}
	u=\widehat{\mathbf k}\!\cdot\!\widehat{\mathbf p}
\end{equation*}
is the cosine of the angle between the final and initial relative momenta. Here \(\bf{p}\) is the initial relative momentum as mentioned earlier. 
The kernel is evaluated between normalized spin-coupled initial and final
channel states with the same $J,M_J$. For the central $L_i=L_f=0$ sector
retained here, rotational invariance makes the projected scalar amplitude
independent of the particular value of $M_J$.
For $L_i=L_f=0$, the zeroth Legendre polynomial is $P_0(u)=1$. Folding the
corresponding $S$-wave kernel with the initial radial wave function therefore
defines
\begin{equation}
	\begin{aligned}
		T_{00;fi}^{(J)}(k)
		={}&\int_0^\infty p^2\,dp\,R_0(p)\\
		&\times\int_{-1}^{1}du\,
		V_{\mathrm{eff};fi}^{(J)}(k,p,u).
	\end{aligned}
	\label{eq:t00_projection}
\end{equation}
The first and second zero in $T_{00;fi}^{(J)}$ denote $L_f=0$ and $L_i=0$,
respectively. The superscript $(J)$ labels the conserved total-angular-momentum
channel.

Let $s_M,m_M$ and $s_B,m_B$ be the spins and canonical spin projections of
$M_f$ and $B_f$, defined along a common quantization axis. Since $L_f=0$,
$S_f=J$, and the polarization-resolved nonrelativistic amplitude is
\begin{equation}
	\begin{aligned}
		\mathcal T_{fi;M_J;m_Mm_B}(\mathbf k)
		={}&\frac{T_{00;fi}^{(J)}(k)}{(2\pi)^2}
		Y_{00}(\widehat{\mathbf k})\\
		&\times
		\langle s_Mm_M,s_Bm_B\mid JM_J\rangle.
	\end{aligned}
	\label{eq:T_s_wave_expansion}
\end{equation}
The Clebsch--Gordan coefficient converts the final state from the coupled-spin
basis to definite meson and baryon spin projections. All dynamical spin
dependence remains in $T_{00;fi}^{(J)}$.

Writing $\widehat V_{\rm eff}$ for the transition operator whose unfolded
kernel appears in Eq.~\eqref{eq:t00_projection}, and abbreviating the complete
final state as
$|f;\mathbf k,m_Mm_B\rangle\equiv
|M_f(\mathbf k,m_M)B_f(-\mathbf k,m_B)\rangle$, the same amplitude is related
to the relativistically normalized decay amplitude by
\begin{equation}
	\begin{aligned}
		\mathcal T_{fi;M_J;m_Mm_B}(\mathbf k)
		&\equiv\langle f;\mathbf k,m_Mm_B|
		\widehat V_{\rm eff}|P_c(JM_J)\rangle\\
		&=-\frac{\mathcal M_{fi;M_J;m_Mm_B}}{\mathcal N_f},\\
		\mathcal N_f
		&=(2\pi)^{3/2}\sqrt{2M_{P_c}}
		\sqrt{2E_{M_f}}\sqrt{2E_{B_f}}.
	\end{aligned}
	\label{eq:T_to_invariant_amplitude}
\end{equation}
Thus, $V_{\mathrm{eff};fi}^{(J)}$, $T_{00;fi}^{(J)}$, $\mathcal T_{fi}$, and
$\mathcal M_{fi}$ denote, respectively, the unfolded kernel, its molecularly
folded $S$-wave amplitude, the polarization-resolved nonrelativistic matrix
element, and the relativistic decay amplitude.

In terms of $\mathcal M_{fi}$, the standard two-body decay formula is
\begin{equation}
	\begin{aligned}
		\Gamma_{fi}^{(J)}
		={}&\frac{k}{32\pi^2M_{P_c}^2}
		\frac{1}{2J+1}\\
		&\times\sum_{M_J,m_M,m_B}
		\int d\Omega_k\,
		\bigl|\mathcal M_{fi;M_J;m_Mm_B}\bigr|^2.
	\end{aligned}
	\label{eq:invariant_two_body_width}
\end{equation}
For the QI contribution, explicit spin matrix elements are evaluated at
$M_J=+1/2$; the OPE coefficients are already projected onto definite-$J$
channels. The quantity $T_{00;fi}^{(J)}$ is independent of $M_J$ by rotational
invariance, but it is not a Wigner--Eckart reduced matrix element. The
Clebsch--Gordan normalization gives
\begin{equation}
	\begin{aligned}
		&\frac{1}{2J+1}
		\sum_{M_J,m_M,m_B}
		\Bigl|
		\langle s_Mm_M,s_Bm_B\mid JM_J\rangle
		T_{00;fi}^{(J)}(k)
		\Bigr|^2\\
		&\hspace{5em}=\bigl|T_{00;fi}^{(J)}(k)\bigr|^2.
	\end{aligned}
	\label{eq:spin_average_cancellation}
\end{equation}
Using Eqs.~\eqref{eq:T_s_wave_expansion}--
\eqref{eq:spin_average_cancellation} together with
$\int d\Omega_k\,|Y_{00}|^2=1$, Eq.~\eqref{eq:invariant_two_body_width}
reduces to
\begin{equation}
	\begin{aligned}
		\Gamma_{fi}^{(J)}
		&=\rho_{fi}\,
		\bigl|T_{00;fi}^{(J)}(k)\bigr|^2,\\
		\rho_{fi}
		&\equiv\frac{kE_{M_f}E_{B_f}}
		{(2\pi)^3M_{P_c}}.
	\end{aligned}
	\label{eq:width_t00}
\end{equation}
Thus, for each retained $S$-wave channel, the initial-spin average is exactly
cancelled by the complete sum over the normalized final spin states; no
additional factor of $2J+1$ is inserted. 

In the conventions used here,
$[V_{\mathrm{eff};fi}^{(J)}]=\mathrm{GeV}^{-2}$,
$[R_0]=\mathrm{GeV}^{-3/2}$, and
$[T_{00;fi}^{(J)}]=\mathrm{GeV}^{-1/2}$. Consequently,
$[\rho_{fi}]=\mathrm{GeV}^{2}$ and Eq.~\eqref{eq:width_t00} yields a width in
GeV.

The assignments and normalized radial wave functions specified above define
the initial states. Section~\ref{sec:qi} constructs the short-range QI
contribution to $V_{\mathrm{eff};fi}^{(J)}$, while Sec.~\ref{sec:ope}
constructs the OPE contribution and converts it to the same normalization.
Only the spin--isospin-reduced OPE transition kernel is taken from
Ref.~\cite{Lin2023CSM}; its complex-scaled resonance wave functions and
$c$-product normalization are not used. Inserting either kernel into
Eq.~\eqref{eq:t00_projection} gives its contribution to
$T_{00;fi}^{(J)}$, and the corresponding partial width then follows from
Eq.~\eqref{eq:width_t00}. All QI and OPE amplitudes quoted below are expressed
in this common convention.

\section{Quark-interchange transition mechanism}
\label{sec:qi}

This section constructs the short-range quark-interchange (QI) contribution
$V_{\mathrm{QI};fi}^{(J)}$ to the generic transition kernel
$V_{\mathrm{eff};fi}^{(J)}$ introduced in
Sec.~\ref{sec:kinematics_width}. The construction proceeds in four steps. We
first identify the labeled constituent rearrangements and their prior and post
Born representations. We then specify the pairwise constituent interaction
and evaluate its color, flavor, spin, and spatial matrix elements. Finally, we
antisymmetrize the two identical light quarks and project the explicitly
calculated charge component onto the physical $I=1/2$ channel. Folding the
resulting kernel with $R_0(p)$ gives the QI contribution to
$T_{00;fi}^{(J)}$.

At Born order, a constituent interaction acts between the two incoming
color-singlet hadrons, after which their constituents are recombined into the
outgoing hadrons~\cite{BarnesSwanson1992,Swanson1992,BarnesSwanson1994,
	Wong2002,Hilbert2007,Wang2020QI}. The mechanism is short ranged at the hadronic level
because the internal wave functions of the incoming hadrons must overlap for
the constituent rearrangement to occur. Those finite wave functions also
suppress the kernel at large external momenta.

We evaluate explicitly the charged initial component appearing in
Eq.~\eqref{eq:initial_isospin_state}. Its constituents are labeled as
\begin{equation}
	A(12)=D^{(*)-}(\bar c_1d_2),\qquad
	B(345)=\Sigma_c^{++}(c_3u_4u_5),
	\label{eq:qi_initial_labels}
\end{equation}
where $A$ denotes the initial anticharmed meson and $B$ the initial charmed
baryon. Either identical $u$ quark can enter the final anticharmed meson, so
the same physical final channel has two temporarily labeled clusterings:
\begin{align}
	F_{14}&:\quad M_f(14)+B_f(325),\nonumber\\[-0.4ex]
	&\quad M_f(14)=(\bar c_1u_4),\qquad
	B_f(325)=(c_3d_2u_5),
	\label{eq:topology14}\\
	F_{15}&:\quad M_f(15)+B_f(324),\nonumber\\[-0.4ex]
	&\quad M_f(15)=(\bar c_1u_5),\qquad
	B_f(324)=(c_3d_2u_4).
	\label{eq:topology15}
\end{align}
Here $M_f=\bar D$ or $\bar D^*$ and $B_f=\Lambda_c$. The labels $14$ and
$15$ identify which $u$ quark enters $M_f$; they do not denote distinct
asymptotic decay channels. Although physical channel names are commonly
written in baryon--meson order, all spin products below retain the
meson--baryon coupling order fixed in Sec.~\ref{sec:states_isospin}.

\subsection{Born kernels and prior--post representations}
\label{sec:qi_born_prior_post}

The same constituent Hamiltonian may be partitioned with respect to either
the initial or final hadron clusters,
\begin{equation}
	\widehat H
	=\widehat H_A^0+\widehat H_B^0+\widehat V_{AB}
	=\widehat H_{M_f}^0+\widehat H_{B_f}^0
	+\widehat V_{M_fB_f}.
	\label{eq:qi_hamiltonian_partition}
\end{equation}
The residual operator $\widehat V_{AB}$ connects constituents belonging to
different initial clusters, whereas $\widehat V_{M_fB_f}$ connects
constituents belonging to different final clusters. These two partitions
define, respectively, the prior and post Born kernels. For a labeled final
clustering $F_a$, with $a\in\{14,15\}$, they are
\begin{equation}
	\begin{aligned}
		V_{\mathrm{pr},a}^{(J)}(k,p,u)
		&\equiv
		\langle F_a;JM_J|\widehat V_{AB}|AB;JM_J\rangle,\\
		V_{\mathrm{po},a}^{(J)}(k,p,u)
		&\equiv
		\langle F_a;JM_J|\widehat V_{M_fB_f}|AB;JM_J\rangle.
	\end{aligned}
	\label{eq:prior_post_kernels}
\end{equation}
Thus, Eq.~\eqref{eq:prior_post_kernels} is the constituent-level realization
of the unfolded kernel in Eq.~\eqref{eq:t00_projection}, before identical-quark
antisymmetrization and isospin projection. The external variables are the
final momentum $k$, initial momentum $p$, and
$u=\widehat{\mathbf k}\!\cdot\!\widehat{\mathbf p}$. The superscript $(J)$
labels the conserved total-angular-momentum channel. We evaluate explicit spin
matrix elements at $M_J=+1/2$; rotational invariance makes the scalar kernel
independent of this choice.

\begin{figure*}[t]
	\centering
	\includegraphics[width=\textwidth]{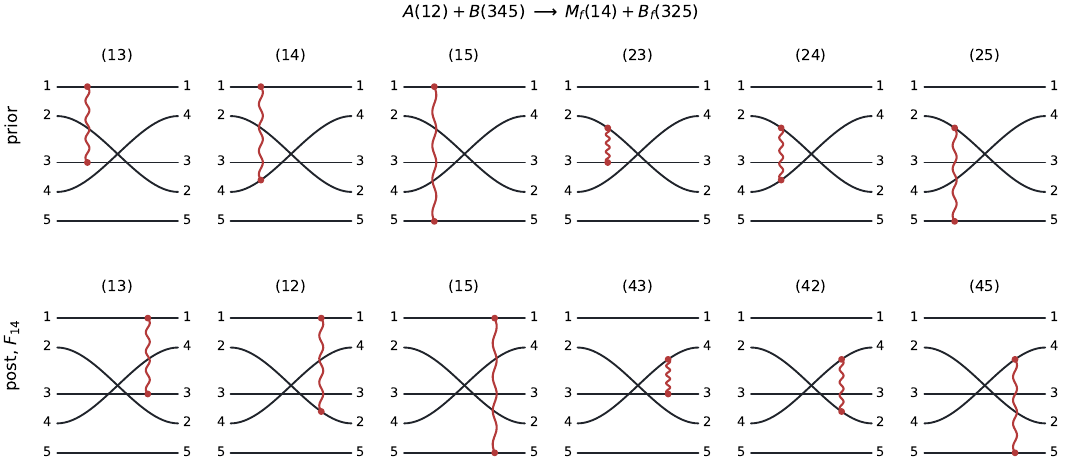}
	\caption{Born-order QI diagrams for the labeled clustering $F_{14}$,
		$A(12)+B(345)\to M_f(14)+B_f(325)$. Quark lines run from left to right.
		The red wavy line denotes one constituent interaction
		$V_{\nu\nu'}$; only its endpoints are interaction vertices, and crossings
		with spectator lines do not represent additional interactions. The upper
		and lower rows contain the complete prior and post pair sets,
		respectively. The crossing trajectories display the constituent
		interchange. The clustering $F_{15}$ is obtained by
		$4\leftrightarrow5$ and is not drawn separately.}
	\label{fig:qi_open_charm_diagrams}
\end{figure*}

Replacing the two residual operators by sums of pairwise constituent
interactions gives the ordered pair sets
\begin{align}
	\mathcal P_{\mathrm{pr},a}
	&=\{13,14,15,23,24,25\},
	\label{eq:qi_prior_pairs}\\
	\mathcal P_{\mathrm{po},14}
	&=\{13,12,15,43,42,45\},
	\label{eq:qi_post14_pairs}\\
	\mathcal P_{\mathrm{po},15}
	&=\{13,12,14,53,52,54\}.
	\label{eq:qi_post15_pairs}
\end{align}
For example, $13$ denotes an interaction between constituents 1 and 3, with
the first label fixing the direction assigned to the transferred momentum.
The prior set connects the two initial clusters; each post set connects the
two final clusters. Figure~\ref{fig:qi_open_charm_diagrams} displays the
complete prior and post sets for $F_{14}$. The $F_{15}$ diagrams follow from
the exchange $4\leftrightarrow5$.

For exact on-shell eigenstates, the prior and post representations are
equivalent. Small differences arise in practical calculations that use
variational internal hadron wave functions~\cite{Swanson1992,Wong2002,
	Wang2020QI}. Following Ref.~\cite{Wang2020QI}, we therefore use their
arithmetic average for each labeled clustering,
\begin{equation}
	V_a^{(J)}(k,p,u)
	=\frac12\left[
	V_{\mathrm{pr},a}^{(J)}(k,p,u)
	+V_{\mathrm{po},a}^{(J)}(k,p,u)\right].
	\label{eq:qi_prior_post_average}
\end{equation}
The factor $1/2$ in Eq.~\eqref{eq:qi_prior_post_average} is included once,
when the diagram coefficients are assembled below.

\subsection{Constituent interaction and internal hadron states}
\label{sec:qi_interaction_wavefunctions}

For an ordered constituent pair $\nu\nu'$, with
$\nu,\nu'\in\{1,\ldots,5\}$, let $\mathbf q_{\nu\nu'}$ be the momentum
transferred to constituent $\nu$; constituent $\nu'$ receives
$-\mathbf q_{\nu\nu'}$. We use the momentum-space interaction of
Ref.~\cite{Wang2020QI},
\begin{equation}
	\begin{aligned}
		V_{\nu\nu'}(\mathbf q_{\nu\nu'})
		={}&\mathbf t_\nu\!\cdot\!\mathbf t_{\nu'}
		\left[
		\frac{4\pi\alpha_s(\mathcal Q_{\nu\nu'}^2)}
		{q_{\nu\nu'}^2}
		+\frac{6\pi b}{q_{\nu\nu'}^4}\right.\\
		&\left.\hspace{1.6em}
		-\frac{8\pi\alpha_s(\mathcal Q_{\nu\nu'}^2)}
		{3m_\nu m_{\nu'}}
		\mathbf s_\nu\!\cdot\!\mathbf s_{\nu'}
		e^{-q_{\nu\nu'}^2/(4\tau_{\mathrm{hf}}^2)}
		\right],
	\end{aligned}
	\label{eq:qi_constituent_potential}
\end{equation}
where $q_{\nu\nu'}=|\mathbf q_{\nu\nu'}|$, $m_\nu$ is a constituent mass,
$b$ is the string tension, $\tau_{\mathrm{hf}}$ is the hyperfine smearing
scale, and $\mathbf s_\nu$ is the spin operator of constituent $\nu$. The
color generator is $\mathbf t=\boldsymbol\lambda^{\,c}/2$ for a quark and
$\mathbf t_{\bar q}=-\mathbf t_q^*$ for an antiquark, where
$\boldsymbol\lambda^{\,c}$ denotes the eight Gell-Mann matrices. The first
two terms in Eq.~\eqref{eq:qi_constituent_potential} are the Coulomb and
linear-confinement contributions; together we label them $\mathrm{CL}$. The
last term is the smeared hyperfine contribution, labeled $\mathrm{HF}$.
Tensor and spin--orbit interactions are not included in the present
$S$-wave calculation.

The running coupling is
\begin{equation}
	\alpha_s(\mathcal Q^2)=
	\frac{12\pi}
	{(33-2n_f)\ln\!\left(A_{\alpha}
		+\mathcal Q^2/B_{\alpha}^2\right)}.
	\label{eq:qi_running_coupling}
\end{equation}

We take $\mathcal Q^2$ to represent the characteristic invariant-mass
scale of the interacting constituent pair. Since the constituents are
generally off shell inside the molecular folding integral, this scale is
not uniquely fixed by on-shell kinematics. We therefore adopt the
fixed-threshold prescription
\begin{equation}
	\mathcal Q_{\nu\nu'}^2=(m_\nu+m_{\nu'})^2
	\label{eq:qi_scale_prescription}
\end{equation}
and set $n_f=4$. These choices fully specify the coupling used in the present
calculation. With the central parameters in
Table~\ref{tab:qi_model_parameters}, the resulting values
\begin{equation}
	\begin{aligned}
		\alpha_s(qq)&=0.561838,\\
		\alpha_s(qc)&=0.374074,\\
		\alpha_s(c\bar c)&=0.304590
	\end{aligned}
	\label{eq:qi_pair_couplings}
\end{equation}
provide direct numerical cross-checks; here $q$ denotes either light
constituent $u$ or $d$.

\begin{table}[t]
	\caption{Constituent-model parameters and internal Gaussian widths used in
		the central QI calculation. The potential parameters and widths follow
		Ref.~\cite{Wang2020QI}; Eq.~\eqref{eq:qi_scale_prescription} and $n_f=4$
		complete the running-coupling prescription used here.}
	\label{tab:qi_model_parameters}
	\begin{ruledtabular}
		\begin{tabular}{lcc}
			Quantity & Value & Unit \\
			\colrule
			$m_u=m_d$               & $0.334$ & $\mathrm{GeV}$ \\
			$m_c=m_{\bar c}$        & $1.776$ & $\mathrm{GeV}$ \\
			$b$                     & $0.180$ & $\mathrm{GeV}^2$ \\
			$\tau_{\mathrm{hf}}$   & $0.897$ & $\mathrm{GeV}$ \\
			$A_{\alpha}$            & $10.0$  & -- \\
			$B_{\alpha}$            & $0.310$ & $\mathrm{GeV}$ \\
			$n_f$                   & $4$     & -- \\
			$\alpha_\rho$           & $0.400$ & $\mathrm{GeV}$ \\
			$\alpha_\lambda^{(c)}$ & $0.490$ & $\mathrm{GeV}$ \\
			$\beta_D$               & $0.600$ & $\mathrm{GeV}$ \\
			$\beta_{D^*}$           & $0.520$ & $\mathrm{GeV}$ \\
		\end{tabular}
	\end{ruledtabular}
\end{table}

Each external hadron state is factorized into normalized color, flavor,
spin, and internal momentum-space factors,
\begin{equation}
	\Psi_H=\chi_H^c\chi_H^f\chi_H^s\phi_H,
\end{equation}
where \(H\in\{A,B,M_f,B_f\}\) labels any external hadron in the
transition \(A+B\to M_f+B_f\). The internal $S$-wave functions
$\phi_H$ are Gaussian. We use the same baryon widths
$(\alpha_\rho,\alpha_\lambda^{(c)})$ for $\Sigma_c$ and $\Lambda_c$, and the
meson widths in Table~\ref{tab:qi_model_parameters} for both charge states.
No parameter is fitted separately to the decay channels considered here. The
normalized meson and baryon Gaussians, their mass-weighted relative and Jacobi
momenta, and the topology-dependent baryon orderings are given in
Appendix~\ref{app:qi_internal_wavefunctions}. The internal functions $\phi_H$
must be distinguished from the molecular radial function $R_0(p)$ in
Eq.~\eqref{eq:molecular_gaussian}: the former resolve the constituent overlap
inside the Born kernel, whereas the latter folds that kernel over the initial
two-hadron molecule.

The internal and molecular Gaussians describe two distinct
levels of the cluster approximation and are not eigenfunctions
obtained from a single common Hamiltonian.  The constituent-
interaction parameters and the isolated-hadron widths in
Table~\ref{tab:qi_model_parameters} are adopted together as the QI input set.  The latter
are phenomenological single-Gaussian representations of the
hadrons: the meson widths are tied to isolated-meson size
estimates, whereas the baryon widths follow the oscillator
prescription in Eq.~\eqref{eq:app_baryon_width_relation}.  They are not exact solutions of
the pairwise interaction in Eq.~\eqref{eq:qi_constituent_potential}.  In contrast,
$\beta_{P_c}$ is fixed independently from the binding energy
through Eq.~\eqref{eq:beta_binding}.  Neither the QI rearrangement kernel nor
the OPE transition kernel is iterated to generate the
molecular bound state in the present first-Born treatment.
Possible binding-induced modifications of the internal
hadron wave functions are consequently neglected.

For the central inputs, Eq.~\eqref{eq:gaussian_size_matching} gives molecular rms
separations of approximately $1.46$, $0.90$, and $1.97$~fm
for the $P_c(4312)$, $P_c(4440)$, and $P_c(4457)$,
respectively.  Applying the corresponding coordinate-space
Gaussian conversion to the internal widths gives relative
coordinate lengths of approximately $0.40$ and $0.46$~fm
for the $\bar D$ and $\bar D^*$ mesons, and $0.49$--$0.60$~fm
for the baryon Jacobi variables.  These quantities are
Gaussian coordinate lengths rather than charge radii.  The
molecular coordinate is therefore more extended than the
internal hadron coordinates, although the separation is only
moderate for the more deeply bound $P_c(4440)$.  This
hierarchy supports the cluster factorization but should not
be interpreted as a fully self-consistent five-body solution.
The residual dependence on the two scales is probed in
Table~\ref{tab:reference_sensitivities} by varying $\beta_{P_c}$ and the internal meson and
baryon widths independently; the resulting envelopes are
reported in Table~\ref{tab:qi_central_results} and Fig.~\ref{fig:qi_sensitivity}.

\subsection{Color, flavor, and spin factors}
\label{sec:qi_cfs}

To expose how the constituent interaction enters each diagram, we write
\begin{equation}
	\begin{gathered}
		V_{\nu\nu'}
		=
		\mathbf t_\nu\!\cdot\!\mathbf t_{\nu'}
		\sum_{X\in\{\mathrm{CL},\,\mathrm{HF}\}}
		v_{\nu\nu'}^X(q_{\nu\nu'})\,\widehat O_X,
		\\[0.4ex]
		\widehat O_{\mathrm{CL}}=1,
		\qquad
		\widehat O_{\mathrm{HF}}
		=\mathbf s_\nu\!\cdot\!\mathbf s_{\nu'} .
	\end{gathered}
	\label{eq:qi_operator_decomposition}
\end{equation}
Here $v^{\mathrm{CL}}$ contains the Coulomb and confinement terms, whereas
$v^{\mathrm{HF}}$ contains the scalar coefficient and Gaussian smearing of
the hyperfine term. For representation
$r\in\{\mathrm{pr},\mathrm{po}\}$, clustering
$a\in\{14,15\}$, and pair $\nu\nu'\in\mathcal P_{r,a}$, define
\begin{align}
	I_{C,r}^{a}(\nu\nu')
	&=\langle\chi_{F_a}^{c}|
	\mathbf t_\nu\!\cdot\!\mathbf t_{\nu'}|\chi_{AB}^{c}\rangle,
	\label{eq:qi_color_factor}\\
	I_F^a
	&=\langle\chi_{F_a}^{f}|\chi_{AB}^{f}\rangle,
	\label{eq:qi_flavor_factor}\\
	I_{S,X,r}^{a,J}(\nu\nu')
	&=\langle\chi_{F_a}^{s;JM_J}|
	\widehat O_X|\chi_{AB}^{s;JM_J}\rangle.
	\label{eq:qi_spin_factor}
\end{align}
These matrix elements respectively carry the color, flavor, and spin
dependence of a diagram; its momentum dependence resides in the spatial
overlap introduced below. All states in
Eqs.~\eqref{eq:qi_color_factor}--\eqref{eq:qi_spin_factor} are normalized and
use the phase conventions of Sec.~\ref{sec:states_isospin} and
Appendix~\ref{app:qi_cfs}.

The color singlets are formed with $\delta_{ab}/\sqrt3$ for a meson and
$\epsilon_{abc}/\sqrt6$ for a baryon. The final $\Lambda_c$ contains an
antisymmetric light-flavor pair. For the labeled states in
Eqs.~\eqref{eq:topology14} and~\eqref{eq:topology15}, its overlap with the
initial charged flavor state is
\begin{equation}
	I_F^{14}=I_F^{15}=\frac{1}{\sqrt2}.
	\label{eq:qi_flavor_overlap_value}
\end{equation}
The complete color and flavor factors are listed in
Table~\ref{tab:qi_color_flavor}. Each color vector sums to zero. This
color-singlet identity removes a constant interquark interaction and is used
below in the distributional treatment of the confinement kernel.

\begin{table*}[t]
	\caption{Color and flavor factors for the labeled open-charm clusterings.
		Within each row, the entries of $\boldsymbol I_C$ follow the displayed
		ordered pair set. All entries are matrix elements of normalized states.}
	\label{tab:qi_color_flavor}
	\begin{ruledtabular}
		\begin{tabular}{lccc}
			Set & Ordered pairs & $\boldsymbol I_C$ & $I_F$ \\
			\colrule
			Prior, $F_{14}$
			& $(13,14,15,23,24,25)$
			& $(2,-4,2,-2,4,-2)/9$ & $1/\sqrt2$ \\
			Post, $F_{14}$
			& $(13,12,15,43,42,45)$
			& $(2,-4,2,-2,4,-2)/9$ & $1/\sqrt2$ \\
			Prior, $F_{15}$
			& $(13,14,15,23,24,25)$
			& $(-2,-2,4,2,2,-4)/9$ & $1/\sqrt2$ \\
			Post, $F_{15}$
			& $(13,12,14,53,52,54)$
			& $(-2,4,-2,2,-4,2)/9$ & $1/\sqrt2$
		\end{tabular}
	\end{ruledtabular}
\end{table*}

The spin states are constructed in the Condon--Shortley convention. The
$u_4u_5$ pair in the initial $\Sigma_c$ has spin 1 and is coupled with the
charm spin to $s_{\Sigma_c}=1/2$, whereas the light pair in the final
$\Lambda_c$ has spin 0. The meson and baryon spins are subsequently coupled,
in that order, to total $J$. Table~\ref{tab:qi_spin_factors} gives the
spin-independent overlap $I_{S,\mathrm{CL}}^{14,J}$ and the pair-dependent hyperfine
factors for $F_{14}$. The $F_{15}$ factors follow from
$4\leftrightarrow5$ together with the pair reordering shown in
Table~\ref{tab:qi_color_flavor}.

\begin{table*}[t]
	\caption{Spin factors for clustering $F_{14}$. The prior vectors are ordered
		as $(13,14,15,23,24,25)$ and the post vectors as
		$(13,12,15,43,42,45)$. We use $\zeta=\sqrt3/24$.
		The two heavier peaks have the same factors for a fixed
		$\Sigma_c\bar D^*$ spin assignment. The table covers every $S$-wave
		transition evaluated in this work.}
	\label{tab:qi_spin_factors}
	\begin{ruledtabular}
		\begin{tabular}{lcccl}
			Transition & $J$ & $I_{S,\mathrm{CL}}^{14,J}$ &
			$\boldsymbol I_{S,\mathrm{HF},\mathrm{pr}}^{14,J}$ &
			$\boldsymbol I_{S,\mathrm{HF},\mathrm{po}}^{14,J}$ \\
			\colrule
			$\Sigma_c\bar D\to\Lambda_c\bar D$
			& $\tfrac12$ & $0$
			& $(3\zeta,0,0,-3\zeta,0,0)$
			& $(3\zeta,0,0,-3\zeta,0,0)$ \\
			$\Sigma_c\bar D\to\Lambda_c\bar D^*$
			& $\tfrac12$ & $-\tfrac12$
			& $(\tfrac14,-\tfrac18,-\tfrac38,-\tfrac14,\tfrac18,\tfrac38)$
			& $(\tfrac14,\tfrac38,-\tfrac38,\tfrac14,\tfrac18,-\tfrac18)$ \\
			$\Sigma_c\bar D^*\to\Lambda_c\bar D$
			& $\tfrac12$ & $-\tfrac12$
			& $(-\tfrac14,\tfrac38,\tfrac18,-\tfrac14,\tfrac18,\tfrac38)$
			& $(-\tfrac14,-\tfrac18,\tfrac18,\tfrac14,\tfrac18,-\tfrac18)$ \\
			$\Sigma_c\bar D^*\to\Lambda_c\bar D^*$
			& $\tfrac12$ & $1/\sqrt3$
			& $(-\zeta,2\zeta,-2\zeta,\zeta,-2\zeta,-6\zeta)$
			& $(-\zeta,2\zeta,-2\zeta,-7\zeta,-2\zeta,2\zeta)$ \\
			$\Sigma_c\bar D^*\to\Lambda_c\bar D^*$
			& $\tfrac32$ & $-1/(2\sqrt3)$
			& $(-\zeta,-\zeta,\zeta,-5\zeta,\zeta,3\zeta)$
			& $(-\zeta,-\zeta,\zeta,-\zeta,\zeta,-\zeta)$
		\end{tabular}
	\end{ruledtabular}
\end{table*}

Together, Tables~\ref{tab:qi_color_flavor} and
\ref{tab:qi_spin_factors} provide the complete algebraic input for the
$S$-wave QI transitions. Explicit constituent states, phase conventions, and
representative contractions are collected in
Appendix~\ref{app:qi_cfs}; no channel-dependent coupling is introduced.

We now combine these algebraic factors with the spatial matrix element. To
include the prior--post average of Eq.~\eqref{eq:qi_prior_post_average} at the
coefficient level, define
\begin{equation}
	\kappa_{r,a,\nu\nu',X}^{(J)}
	=\frac12 I_{C,r}^{a}(\nu\nu')I_F^a
	I_{S,X,r}^{a,J}(\nu\nu'),
	\qquad \nu\nu'\in\mathcal P_{r,a}.
	\label{eq:qi_combined_coefficient}
\end{equation}
The kernel for one labeled clustering is then
\begin{equation}
	\begin{aligned}
		V_a^{(J)}(k,p,u)
		={}&\sum_{r\in\{\mathrm{pr},\mathrm{po}\}}
		\sum_{\nu\nu'\in\mathcal P_{r,a}}
		\sum_{X\in\{\mathrm{CL},\mathrm{HF}\}}
		\kappa_{r,a,\nu\nu',X}^{(J)}\\
		&\times
		I_{\mathrm{space},X}^{r,a,\nu\nu'}(k,p,u).
	\end{aligned}
	\label{eq:qi_kernel_assembly}
\end{equation}
Equation~\eqref{eq:qi_kernel_assembly} therefore assembles the separately
evaluated color, flavor, spin, and spatial overlaps into the constituent-level
Born kernel. Because the factor $1/2$ is already contained in
$\kappa_{r,a,\nu\nu',X}^{(J)}$, no additional prior--post averaging factor is
applied to Eq.~\eqref{eq:qi_kernel_assembly}.

For illustration, the coefficient of the prior $13$ hyperfine diagram for
$F_{14}$ in
$\Sigma_c\bar D\to\Lambda_c\bar D$ is
\begin{equation}
	\begin{aligned}
		\kappa_{\mathrm{pr},14,13,\mathrm{HF}}^{(1/2)}
		&=\frac12\left(\frac{2}{9}\right)
		\left(\frac{1}{\sqrt2}\right)
		\left(\frac{\sqrt3}{8}\right)\\
		&=\frac{\sqrt6}{144}.
	\end{aligned}
	\label{eq:qi_example_cfs_coefficient}
\end{equation}
This example also provides a direct check of the coefficient conventions and
of the single prior--post averaging factor.

\subsection{Spatial matrix elements and reduced kernel}
\label{sec:qi_spatial}

For each ordered pair $\nu\nu'$, the remaining diagram dependence is carried
by the internal momentum overlap. We choose the initial momenta
$\mathbf p_3$ and $\mathbf p_4$ of constituents 3 and 4 as independent
internal integration variables. For fixed external momenta
$\mathbf k$ and $\mathbf p$, transferred momentum
$\mathbf q_{\nu\nu'}$, ordered pair $\nu\nu'$, and clustering $a$, all other
constituent momenta are fixed by momentum conservation and the momentum-transfer
routing. Suppressing the resulting arguments of the four hadron wave
functions, the spatial overlap takes the common form
\begin{equation}
	\begin{aligned}
		I_{\mathrm{space},X}^{r,a,\nu\nu'}
		(\mathbf k,\mathbf p)
		={}&\int d^3q_{\nu\nu'}\,d^3p_3\,d^3p_4\;
		v_{\nu\nu'}^X(q_{\nu\nu'})\\
		&\times
		\phi_{M_f}^{*}\phi_{B_f}^{*}\phi_A\phi_B .
	\end{aligned}
	\label{eq:qi_general_spatial_integral}
\end{equation}
The complete constituent-momentum routing and an explicit $F_{14}$ example
for the ordered pair $13$ are given in
Appendix~\ref{app:qi_momentum_routing}. The mass-weighted Jacobi momenta
entering the hadron wave functions are defined in
Appendix~\ref{app:qi_internal_wavefunctions}. The integration measures in
Eq.~\eqref{eq:qi_general_spatial_integral} follow the convention of
Ref.~\cite{Wang2020QI}; no additional $(2\pi)^{-3}$ factor is inserted into
the internal overlap.

Because the four internal hadron functions are Gaussian, all internal
integrations for the hyperfine contribution can be performed analytically.
For the Coulomb and confinement terms, the six components associated with
$\mathbf p_3$ and $\mathbf p_4$ are integrated analytically, leaving a
three-dimensional integral over the transferred momentum. After the complete
color--flavor--spin weighted $\mathrm{CL}$ diagram sum has been assembled,
denote its reduced form factor for representation $r$ and clustering $a$ by
$\mathcal F_{r,a}^{(J)}(\mathbf q)$. The linear-confinement contribution is
then evaluated through the finite distributional combination
\begin{equation}
	\int d^3q\,
	\frac{\mathcal F_{r,a}^{(J)}(\mathbf q)
		-\mathcal F_{r,a}^{(J)}(\mathbf0)}{q^4}.
	\label{eq:qi_confinement_subtraction}
\end{equation}
The subtraction is applied only after the complete color-weighted
spin-independent sum is formed. At $\mathbf q=\mathbf0$, the pair form
factors in a given representation reduce to a common overlap, while the
corresponding color factors in Table~\ref{tab:qi_color_flavor} sum to zero.
Consequently, the subtraction in
Eq.~\eqref{eq:qi_confinement_subtraction} implements the regulated Fourier
transform of the linear potential; it is not an adjustable infrared cutoff.
The Gaussian reduction and distributional derivation are given in
Appendix~\ref{app:qi_spatial}, and the numerical convergence tests are
reported in Appendix~\ref{app:qi_numerics}.

In the normalization of Sec.~\ref{sec:kinematics_width},
$[I_{\mathrm{space},X}]=[V_a^{(J)}]=\mathrm{GeV}^{-2}$, and the subsequent
molecular fold gives
$[T_{00,\mathrm{QI};fi}^{(J)}]=\mathrm{GeV}^{-1/2}$. The instantaneous
constituent interaction and real Gaussian functions make the Born-order QI
amplitudes real in the phase convention adopted here.

\subsection{Antisymmetrization, isospin projection, and physical QI amplitude}
\label{sec:qi_antisymmetrization}

Equation~\eqref{eq:qi_kernel_assembly} still refers to one temporarily labeled
clustering. We next combine $F_{14}$ and $F_{15}$ into the normalized
asymptotic state and then reconstruct the physical $I=1/2$ kernel. The two
clusterings are labeled representations of the same final channel, not
independent final states. Fermionic antisymmetry of the identical $u$ quarks
requires
\begin{equation}
	|F_{\mathrm{AS}}\rangle
	=\frac{|F_{14}\rangle-|F_{15}\rangle}{\sqrt2}.
	\label{eq:qi_antisymmetric_final_state}
\end{equation}
Projection onto this state gives the kernel for the explicitly calculated
charged initial component,
\begin{equation}
	V_{\mathrm{ch}}^{(J)}(k,p,u)
	=\frac{V_{14}^{(J)}(k,p,u)-V_{15}^{(J)}(k,p,u)}{\sqrt2}.
	\label{eq:qi_charged_topology_sum}
\end{equation}
To see the relative sign, let $P_{45}$ exchange the identical light quarks.
Then $|F_{15}\rangle=P_{45}|F_{14}\rangle$ and
$P_{45}|AB;JM_J\rangle=-|AB;JM_J\rangle$. The isospin-symmetric
constituent interaction is covariant under the same relabeling. The explicit
color, flavor, spin, and spatial factors therefore satisfy, pointwise in
$(k,p,u)$,
\begin{equation}
	V_{15}^{(J)}=-V_{14}^{(J)},
	\qquad
	V_{\mathrm{ch}}^{(J)}=\sqrt2\,V_{14}^{(J)}.
	\label{eq:qi_topology_relation}
\end{equation}
The factor $1/\sqrt2$ in
Eq.~\eqref{eq:qi_antisymmetric_final_state} normalizes the asymptotic
fermionic state and is distinct from both the flavor overlap in
Eq.~\eqref{eq:qi_flavor_overlap_value} and the prior--post factor in
Eq.~\eqref{eq:qi_prior_post_average}. A creation-operator derivation and the
scope of the asymptotic-state normalization are given in
Appendix~\ref{app:qi_topology}.

Only
$\Sigma_c^{++}D^{(*)-}\to\Lambda_c^+\bar D^{(*)0}$ is evaluated directly.
The neutral initial component is reconstructed with the exact-isospin
relation derived in Eq.~\eqref{eq:isospin_amplitude_reconstruction}. Applied
to the unfolded QI kernels, it gives
\begin{equation}
	\begin{aligned}
		V_{\mathrm{neu}}^{(J)}(k,p,u)
		&=-\frac{V_{\mathrm{ch}}^{(J)}(k,p,u)}{\sqrt2},\\
		V_{\mathrm{QI};fi}^{(J)}(k,p,u)
		&=\sqrt{\frac32}\,V_{\mathrm{ch}}^{(J)}(k,p,u).
	\end{aligned}
	\label{eq:qi_physical_isospin_kernel}
\end{equation}
Here $V_{\mathrm{QI};fi}^{(J)}$ is the short-range kernel between normalized
physical $I=1/2$ channel states, and the suffix $fi$ retains the final--initial
ordering established in Sec.~\ref{sec:kinematics_width}. Physical
charged--neutral mass splittings enter only the molecular-scale and kinematic
variations of Sec.~\ref{sec:validation}; they do not modify the exact-isospin
projection in Eq.~\eqref{eq:qi_physical_isospin_kernel}.

The QI contribution to the folded amplitude of
Eq.~\eqref{eq:t00_projection} is therefore
\begin{equation}
	\begin{aligned}
		T_{00,\mathrm{QI};fi}^{(J)}(k)
		={}&\int_0^\infty p^2\,dp\,R_0(p)
		\int_{-1}^{1}du\,
		V_{\mathrm{QI};fi}^{(J)}(k,p,u)\\
		={}&T_{00,\mathrm{CL};fi}^{(J)}(k)
		+T_{00,\mathrm{HF};fi}^{(J)}(k).
	\end{aligned}
	\label{eq:qi_physical_t00}
\end{equation}
Both terms on the second line are in the physical $I=1/2$ normalization:
$T_{00,\mathrm{CL};fi}^{(J)}$ contains the Coulomb and confinement pieces,
and $T_{00,\mathrm{HF};fi}^{(J)}$ contains the smeared hyperfine piece. Thus,
Eq.~\eqref{eq:qi_physical_t00} completes the chain from the constituent Born
diagrams to the short-range contribution in the common amplitude convention
of Sec.~\ref{sec:kinematics_width}. Its QI-only partial width follows by
inserting $T_{00,\mathrm{QI};fi}^{(J)}$ into Eq.~\eqref{eq:width_t00}.
Section~\ref{sec:ope} next constructs the OPE contribution in the same
normalization before the two mechanisms are combined coherently.

\section{Long-range one-pion-exchange contribution}
\label{sec:ope}

The QI mechanism constructed in Sec.~\ref{sec:qi} describes a short-range
rearrangement of the constituent quarks. We supplement it with the direct
one-pion-exchange (OPE) transition
\begin{equation}
	\Sigma_c\bar D^{(*)}\longrightarrow\Lambda_c\bar D^{(*)},
	\label{eq:ope_transition}
\end{equation}
which provides the leading chiral long-distance connection between these
channels whenever both pion vertices are allowed. The OPE interaction is
evaluated once at Born order and folded with the same molecular radial wave
function $R_0(p)$ as the QI kernel. It is not iterated in a coupled-channel
equation, and the complex-scaled resonance wave functions of
Ref.~\cite{Lin2023CSM} are not imported into the present calculation.

The construction proceeds as follows. We first obtain the spin--isospin
reduced three-dimensional OPE potential from the chiral vertices and specify
its channel phases. We then define the retarded energy transfer and treat the
physical-sheet pion cut analytically. After separating the contact-like and
pion-pole pieces, we convert the potential to the $T_{00;fi}^{(J)}$
normalization of Sec.~\ref{sec:kinematics_width} and combine it coherently
with $T_{00,\mathrm{QI};fi}^{(J)}$. The QI-only widths remain the central
predictions; the OPE and QI+OPE results quantify the sensitivity to
long-distance dynamics and to the relative matching of the two descriptions.

\subsection{Chiral vertices and the direct OPE kernel}
\label{sec:ope_chiral_kernel}

We follow the heavy-hadron chiral conventions of
Ref.~\cite{Lin2023CSM}. Only the axial interactions that generate the
$\Sigma_c\Lambda_c\pi$ and $\bar D^{(*)}\bar D^{(*)}\pi$ vertices required
by Eq.~\eqref{eq:ope_transition} are needed:
\begin{equation}
	\begin{aligned}
		\mathcal L_{\pi}^{\mathrm{int}}={}&
		g_2\,\operatorname{Tr}\!\left(
		\bar\psi^{\mu}u_{\mu}B_1+\mathrm{H.c.}\right)\\
		&+g\,\bigl\langle
		\bar{\widetilde H}\,u\!\!\!/\,\gamma_5\widetilde H
		\bigr\rangle,
	\end{aligned}
	\label{eq:ope_axial_lagrangian}
\end{equation}
where
\begin{equation}
	u_\mu=-\frac{\tau^a\partial_\mu\pi^a}{2f_\pi}
	+\mathcal O(\pi^3).
	\label{eq:ope_axial_current}
\end{equation}
Here $\pi^a$ are the pion fields, $\tau^a$ are the Pauli matrices in
$SU(2)$ flavor space, $f_\pi$ is the pion decay constant, and
$u\!\!\!/\equiv\gamma^\mu u_\mu$. The symbol $\mathrm{H.c.}$ denotes the
Hermitian-conjugate interaction. The trace $\operatorname{Tr}$ acts in
flavor space, whereas $\langle\cdots\rangle$ denotes a Dirac trace.

The isosinglet field $B_1$ contains the $\Lambda_c$. The spin-$1/2$
isotriplet $B_3$ and the spin-$3/2$ isotriplet $B_3^{*\mu}$, containing the
$\Sigma_c$ and $\Sigma_c^*$ states, are combined into the heavy-quark-spin
superfield
\begin{equation}
	\psi^\mu=B_3^{*\mu}
	-\frac{1}{\sqrt3}(\gamma^\mu+v^\mu)\gamma_5B_3,
	\label{eq:ope_baryon_superfield}
\end{equation}
where $v^\mu$ is the common heavy-hadron velocity, $v^2=1$, and
$\bar\psi^\mu$ is the Dirac adjoint. The anticharmed pseudoscalar and vector
mesons are collected in
\begin{equation}
	\widetilde H=
	\left(\widetilde P^*_{\mu}\gamma^\mu
	+i\widetilde P\gamma_5\right)
	\frac{1-\not v}{2},
	\qquad
	\bar{\widetilde H}=\gamma^0\widetilde H^\dagger\gamma^0.
	\label{eq:ope_antimeson_superfield}
\end{equation}
In this convention, the $SU(2)$ antimeson doublets are
\begin{equation*}
	\widetilde P=
	\begin{pmatrix}\bar D^0\\ D^-\end{pmatrix},
	\qquad
	\widetilde P^*_{\mu}=
	\begin{pmatrix}\bar D^{*0}\\ D^{*-}\end{pmatrix}_{\!\mu}.
\end{equation*}
Expanding Eq.~\eqref{eq:ope_axial_lagrangian} produces the
$\bar D\bar D^*\pi$ and $\bar D^*\bar D^*\pi$ vertices, but no leading-order
$\bar D\bar D\pi$ axial vertex. This accounts for the vanishing
$\Sigma_c\bar D\to\Lambda_c\bar D$ entry in
Table~\ref{tab:ope_channel_coefficients}.

Let the exchanged pion carry
$q_\pi^\mu=(q_\pi^0,\mathbf q_\pi)$, with
$\mathbf q_\pi=\mathbf k-\mathbf p$. Before the central reduction, the
OPE interaction has the schematic spin--momentum structure
\begin{equation}
	V_{1\pi}\propto
	-\frac{(\mathbf S_B\!\cdot\!\mathbf q_\pi)
		(\mathbf S_M\!\cdot\!\mathbf q_\pi)}
	{(q_\pi^0)^2-\mathbf q_\pi^2-m_\pi^2+i0},
	\label{eq:ope_unprojected_structure}
\end{equation}
where $\mathbf S_B$ and $\mathbf S_M$ are the transition-spin operators at
the baryon and meson vertices. The present $S$-wave calculation retains the
trace part of the momentum tensor,
\begin{equation}
	q_{\pi i}q_{\pi j}
	=\frac{\delta_{ij}}{3}\mathbf q_\pi^2
	+\left(q_{\pi i}q_{\pi j}
	-\frac{\delta_{ij}}{3}\mathbf q_\pi^2\right)
	\longrightarrow\frac{\delta_{ij}}{3}\mathbf q_\pi^2,
	\label{eq:ope_central_reduction}
\end{equation}
and omits the traceless tensor part.

For a transition $i\to f$, let $\widehat c_{fi}^{(J)}$ denote the real,
dimensionless channel coefficient after projection onto the physical
$I=1/2$ and definite-$J$ basis. Its extraction from the coefficient matrices
of Ref.~\cite{Lin2023CSM}, including the required rephasing, is given in
Sec.~\ref{sec:ope_coefficients}. In the three-dimensional potential
normalization of that reference, the full central OPE kernel is
\begin{equation}
	\begin{aligned}
		V_{\pi,\mathrm{full};fi}^{[\mathrm{Lin}],J}(k,p,u)
		={}&\frac{|g|\,|g_2|\,\widehat c_{fi}^{(J)}}{f_\pi^2}
		\frac{\mathbf q_\pi^2}
		{\mathbf q_\pi^2+m_\pi^2-[q_\pi^0(p,k)]^2-i0}\\
		&\times F_{\Lambda_\pi}(k,p),
	\end{aligned}
	\label{eq:ope_full_kernel}
\end{equation}
where the relative-angle variable $u$ was defined in
Sec.~\ref{sec:kinematics_width}, and
\begin{equation}
	\begin{gathered}
		\mathbf q_\pi^2=k^2+p^2-2kpu,\\
		F_{\Lambda_\pi}(k,p)=
		\exp\!\left[-\frac{k^2+p^2}{\Lambda_\pi^2}\right].
	\end{gathered}
	\label{eq:ope_momentum_regulator}
\end{equation}
The cutoff $\Lambda_\pi$ distinguishes the OPE regulator from the other
momentum scales in the calculation. The factor
$F_{\Lambda_\pi}$ is the single potential-level regulator used in
Ref.~\cite{Lin2023CSM}; it is not a vertex form factor and is therefore not
squared. The coefficient $\widehat c_{fi}^{(J)}$ already includes the
component-field factors, the physical $I=1/2$ projection, the spin
recoupling, and the factor $1/3$ in
Eq.~\eqref{eq:ope_central_reduction}. No additional isospin coefficient,
spin factor, or central-reduction factor multiplies
Eq.~\eqref{eq:ope_full_kernel}.

The reference inputs are collected in Table~\ref{tab:ope_inputs}. Coupling
magnitudes are used here so that Eq.~\eqref{eq:ope_full_kernel} defines a
real reference basis before the pion cut is encountered. The remaining
relative sign between the OPE and QI bases is treated explicitly in
Sec.~\ref{sec:ope_coherent_matching}.

\begin{table}[t]
	\caption{OPE inputs used in the reference calculation. The pion mass is
		the charge average $(m_{\pi^0}+2m_{\pi^\pm})/3$. The mass
		$M_{\Sigma_c}^{\rm av}\equiv M_B^{\rm av}$ is the same arithmetic
		charge average defined in Eq.~\eqref{eq:average_threshold_masses}.}
	\label{tab:ope_inputs}
	\begin{ruledtabular}
		\begin{tabular}{lc}
			Quantity & Value \\
			\colrule
			$|g|$ & $0.59$ \\
			$|g_2|$ & $1.04$ \\
			$f_\pi$ & $0.092~\mathrm{GeV}$ \\
			$m_\pi$ & $0.138039~\mathrm{GeV}$ \\
			$\Lambda_\pi$ & $0.500~\mathrm{GeV}$ \\
			$M_{\Sigma_c}^{\rm av}$ & $2.453310~\mathrm{GeV}$
		\end{tabular}
	\end{ruledtabular}
\end{table}

\begin{figure}[t]
	\centering
	\includegraphics[width=0.98\columnwidth]{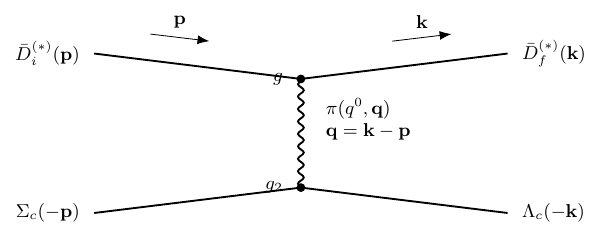}
	\caption{Direct OPE transition
		$\Sigma_c\bar D_i^{(*)}\to\Lambda_c\bar D_f^{(*)}$. The initial
		meson and baryon carry momenta $\mathbf p$ and $-\mathbf p$, while the
		final pair carries $\mathbf k$ and $-\mathbf k$. Directing the pion
		momentum from the baryon line to the meson line gives
		$q_\pi^\mu=(q_\pi^0,\mathbf k-\mathbf p)$. The physical $I=1/2$ and
		definite-$J$ projections are contained in $\widehat c_{fi}^{(J)}$.}
	\label{fig:ope_open_charm}
\end{figure}

Figure~\ref{fig:ope_open_charm} shows the corresponding hadron-level Born
topology. It is evaluated once and is distinct from the constituent
rearrangement diagrams of Sec.~\ref{sec:qi}.

\subsection{Channel coefficients and phase convention}
\label{sec:ope_coefficients}

Let $C_{fi}^{[\mathrm{Lin}],J}$ denote an entry of the OPE coefficient
matrices in Tables VI and VII of Ref.~\cite{Lin2023CSM}, with the first
suffix denoting the final channel and the second the initial channel. The
numerical channel labels used in those matrices are quoted only to make the
translation reproducible; they are unrelated to the constituent labels of
Sec.~\ref{sec:qi}.

The off-diagonal $\bar D\leftrightarrow\bar D^*$ entries in the Lin basis
contain conventional factors of $i$. We introduce a real reference basis by
\begin{equation}
	|\bar D\rangle_{\mathrm{ref}}
	=i|\bar D\rangle_{[\mathrm{Lin}]},
	\qquad
	|\bar D^*\rangle_{\mathrm{ref}}
	=|\bar D^*\rangle_{[\mathrm{Lin}]}.
	\label{eq:ope_rephasing}
\end{equation}
More generally, if
$|h\rangle_{\mathrm{ref}}=e^{i\phi_h}|h\rangle_{[\mathrm{Lin}]}$, then
\begin{equation}
	C_{fi}^{[\mathrm{ref}],J}
	=e^{-i\phi_f}C_{fi}^{[\mathrm{Lin}],J}e^{i\phi_i},
	\qquad
	\phi_{\bar D}=\frac{\pi}{2},\quad
	\phi_{\bar D^*}=0.
	\label{eq:ope_coefficient_rephasing}
\end{equation}
The dimensionless coefficient used in Eq.~\eqref{eq:ope_full_kernel} is
therefore
\begin{equation}
	\widehat c_{fi}^{(J)}
	\equiv\frac{C_{fi}^{[\mathrm{ref}],J}}{|g|\,|g_2|}.
	\label{eq:ope_coefficient_definition}
\end{equation}
This definition separates the channel-dependent spin, isospin, and phase
factor from the common positive coupling magnitude.

\begin{table*}[t]
	\caption{OPE coefficients for the selected physical $I=1/2$
		transitions. The Lin entry identifies the corresponding matrix element
		in Ref.~\cite{Lin2023CSM}. The next column gives its phase and numerical
		factor in that convention, and $\widehat c_{fi}^{(J)}$ is the real
		reference-basis coefficient of Eq.~\eqref{eq:ope_coefficient_definition}.
		The $J=3/2\to\Lambda_c\bar D$ channel is not an $S$-wave zero: its
		leading contribution has $L_f=2$ and is outside the present
		truncation.}
	\label{tab:ope_channel_coefficients}
	\scriptsize
	\begin{ruledtabular}
		\begin{tabular}{lclcccc}
			Initial & $J$ & Final & Lin entry
			& $C_{fi}^{[\mathrm{Lin}],J}/(g g_2)$
			& $\widehat c_{fi}^{(J)}$ & Treatment \\
			\colrule
			$\Sigma_c\bar D$ & $\tfrac12$ & $\Lambda_c\bar D$
			& VI: $C_{13}$ & $0$ & $0$ & no $\bar D\bar D\pi$ vertex \\
			$\Sigma_c\bar D$ & $\tfrac12$ & $\Lambda_c\bar D^*$
			& VI: $C_{23}$ & $-i\sqrt6/12$ & $+\sqrt6/12$ & retained \\
			$\Sigma_c\bar D^*$ & $\tfrac12$ & $\Lambda_c\bar D$
			& VI: $C_{14}$ & $+i\sqrt6/12$ & $+\sqrt6/12$ & retained \\
			$\Sigma_c\bar D^*$ & $\tfrac12$ & $\Lambda_c\bar D^*$
			& VI: $C_{24}$ & $-\sqrt2/6$ & $-\sqrt2/6$ & retained \\
			$\Sigma_c\bar D^*$ & $\tfrac32$ & $\Lambda_c\bar D$
			& --- & --- & --- & $L_f=2$; not calculated \\
			$\Sigma_c\bar D^*$ & $\tfrac32$ & $\Lambda_c\bar D^*$
			& VII: $C_{13}$ & $+\sqrt2/12$ & $+\sqrt2/12$ & retained
		\end{tabular}
	\end{ruledtabular}
\end{table*}

Equations~\eqref{eq:ope_rephasing}--
\eqref{eq:ope_coefficient_definition} convert both retained
$\bar D\leftrightarrow\bar D^*$ coefficients to $+\sqrt6/12$ and leave the
real diagonal-in-meson-spin entries unchanged. These coefficients already
act between normalized physical $I=1/2$ channel states. Consequently, the
$\sqrt{3/2}$ charged-to-isospin factor used for the QI kernel in
Eq.~\eqref{eq:qi_physical_isospin_kernel} is not applied again to OPE. The
remaining relative phase between this reference chiral basis and the signed
QI basis is kept explicit in the coherent matching.

\subsection{Retarded energy transfer and the pion cut}
\label{sec:ope_retardation}

In the molecular fold, $p$ is integrated over the initial wave function,
whereas $k$ is fixed by the final two-body kinematics of
Eq.~\eqref{eq:two_body_momentum}. Our reference energy transfer follows the
direct baryon-line routing in Fig.~\ref{fig:ope_open_charm}:
\begin{equation}
	q_{\pi,\mathrm{dir}}^0(p,k)
	=E_{\Sigma_c}(p)-E_{\Lambda_c}(k),
	\qquad
	E_H(x)=\sqrt{M_H^2+x^2}.
	\label{eq:ope_q0_baryon}
\end{equation}
Here $M_{\Sigma_c}=M_{\Sigma_c}^{\rm av}$ from
Table~\ref{tab:ope_inputs}, while $M_{\Lambda_c}$ is the physical final
baryon mass. Only $(q_\pi^0)^2$ enters Eq.~\eqref{eq:ope_full_kernel}; hence
reversing the overall four-momentum direction assigned to the pion does not
change the kernel.

To probe the off-shell continuation, we also consider
\begin{equation}
	\begin{aligned}
		q_{\pi,\mathrm{swap}}^0(p,k)
		&=E_{\Sigma_c}(k)-E_{\Lambda_c}(p),\\
		q_{\pi,\mathrm{stat}}^0&=0.
	\end{aligned}
	\label{eq:ope_q0_diagnostics}
\end{equation}
The first diagnostic merely interchanges $p$ and $k$ in the baryon energy
difference; it does not represent a crossed Feynman topology. These two
choices are sensitivity prescriptions, not additional decay channels or
statistical variations.

Retardation is important because the $\Sigma_c-\Lambda_c$ mass difference is
comparable to $m_\pi$. For any chosen $q_\pi^0(p,k)$, define
\begin{equation}
	\begin{aligned}
		\Delta_\pi(p,k)
		&=m_\pi^2-[q_\pi^0(p,k)]^2,\\
		D_0(p,k)&=p^2+k^2+\Delta_\pi(p,k),\\
		D_1(p,k)&=2pk.
	\end{aligned}
	\label{eq:ope_delta_ab}
\end{equation}
The propagator denominator in Eq.~\eqref{eq:ope_full_kernel} is then
$D_0-D_1u-i0$. For $D_1>0$, a pole lies inside the angular integration
interval when $|D_0|<D_1$. This is the on-shell-pion cut associated with the
allowed three-body intermediate state. On the physical sheet, the required
angular integral is
\begin{equation}
	\begin{aligned}
		\mathcal L_\pi(p,k)
		&\equiv\int_{-1}^{1}
		\frac{du}{D_0-D_1u-i0}\\
		&=\frac{1}{D_1}
		\ln\!\left|\frac{D_0+D_1}{D_0-D_1}\right|
		+i\frac{\pi}{D_1}\,
		\Theta\!\left(D_1-|D_0|\right).
	\end{aligned}
	\label{eq:ope_angular_log}
\end{equation}
Its continuous $D_1\to0$ limit is
$\mathcal L_\pi=2/(D_0-i0)$. The positive imaginary part follows from
$1/(x-i0)=\operatorname{PV}(1/x)+i\pi\delta(x)$. Equation
\eqref{eq:ope_angular_log} is used directly, so no artificial finite width is
introduced in the pion propagator.

The absorptive part generated by Eq.~\eqref{eq:ope_angular_log} is retained
in the two-body Born amplitude. Because the kernel is not iterated and the
three-body channel is not solved explicitly, this treatment is not a unitary
three-body calculation and does not furnish a separate
$\Lambda_c\bar D^{(*)}\pi$ partial width.

\subsection{Contact-like and pion-pole pieces}
\label{sec:ope_contact_subtraction}

The central numerator can be separated algebraically before the molecular
fold. For finite $\epsilon>0$,
\begin{equation}
	\frac{\mathbf q_\pi^2}
	{\mathbf q_\pi^2+\Delta_\pi-i\epsilon}
	=1-
	\frac{\Delta_\pi-i\epsilon}
	{\mathbf q_\pi^2+\Delta_\pi-i\epsilon}
	\xrightarrow[\epsilon\to0^+]{}
	1-
	\frac{\Delta_\pi}
	{\mathbf q_\pi^2+\Delta_\pi-i0},
	\label{eq:ope_contact_identity}
\end{equation}
where the last equality is understood in the distributional sense. We label
the resulting OPE prescriptions by
$\xi\in\{\mathrm{ct},\mathrm{pole},\mathrm{full}\}$ and define
\begin{align}
	V_{\pi,\mathrm{ct};fi}^{[\mathrm{Lin}],J}
	&=\frac{|g|\,|g_2|\widehat c_{fi}^{(J)}}{f_\pi^2}
	F_{\Lambda_\pi}(k,p),
	\label{eq:ope_contact_kernel}\\
	V_{\pi,\mathrm{pole};fi}^{[\mathrm{Lin}],J}
	&=-\frac{|g|\,|g_2|\widehat c_{fi}^{(J)}}{f_\pi^2}
	\frac{\Delta_\pi}
	{\mathbf q_\pi^2+\Delta_\pi-i0}
	F_{\Lambda_\pi}(k,p).
	\label{eq:ope_pole_kernel}
\end{align}
Thus
$V_{\pi,\mathrm{full};fi}^{[\mathrm{Lin}],J}
=V_{\pi,\mathrm{ct};fi}^{[\mathrm{Lin}],J}
+V_{\pi,\mathrm{pole};fi}^{[\mathrm{Lin}],J}$. After integration over
$u$, the three prescriptions are represented by
\begin{equation}
	\begin{aligned}
		\mathcal K_{\mathrm{ct}}(p,k)
		&=2,\\
		\mathcal K_{\mathrm{pole}}(p,k)
		&=-\Delta_\pi\mathcal L_\pi,\\
		\mathcal K_{\mathrm{full}}(p,k)
		&=2-\Delta_\pi\mathcal L_\pi.
	\end{aligned}
	\label{eq:ope_angular_kernels}
\end{equation}
The factor 2 in $\mathcal K_{\mathrm{ct}}$ is simply
$\int_{-1}^{1}du$.

Because the QI kernel already provides a model of the
predominantly short-distance rearrangement transition, directly
adding the full regulated OPE kernel would risk double counting
its contact-like short-range part. We therefore take the
contact-subtracted pion-pole term as the reference exploratory
OPE prescription and retain the full kernel only as a diagnostic
of short-distance matching sensitivity. This subtraction removes
the explicit momentum-independent term in Eq.~(69), but it is not
a microscopic matching condition. The QI constituent parameters
and the OPE chiral couplings are fixed independently, and no common
short-distance counterterm is fitted. Moreover, at finite
$\Lambda_\pi$ the separation is regulator dependent, and the
retained pion-pole term can still overlap with QI at intermediate
momenta. The coherent QI--OPE widths should therefore be interpreted
as regulator- and prescription-dependent matching-sensitivity
scenarios rather than unique hybrid predictions. A systematic
treatment of double counting would require a common
regulator-consistent coupled-channel framework in which the
short-distance operators and OPE are matched simultaneously,
including the counterterms associated with tensor-induced
$S$--$D$ mixing~\cite{Du2021}.

The present Born calculation retains only the central $L_i=L_f=0$
projection. Tensor OPE, OPE iteration, $S$--$D$ mixing, and the physical
$L_i=0\to L_f=2$ transition
$\Sigma_c\bar D^*(J=3/2)\to\Lambda_c\bar D$ are outside this truncation. The
last decay is therefore uncalculated rather than forbidden.

\subsection{Normalization bridge and molecular fold}
\label{sec:ope_normalization_fold}

Equations~\eqref{eq:ope_full_kernel}, \eqref{eq:ope_contact_kernel}, and
\eqref{eq:ope_pole_kernel} have so far carried the plane-wave normalization
of Ref.~\cite{Lin2023CSM}. We now convert them to the
$T_{00;fi}^{(J)}$ convention fixed in Sec.~\ref{sec:kinematics_width}. To
avoid confusing a normalization label with orbital angular momentum, we use
$[\mathrm{Lin}]$ and $[\mathrm{pres}]$ for the two conventions.

The plane-wave states satisfy
\begin{equation}
	\begin{aligned}
		{}_{[\mathrm{Lin}]}\!\langle\mathbf p'|\mathbf p
		\rangle_{[\mathrm{Lin}]}
		&=(2\pi)^3\delta^3(\mathbf p'-\mathbf p),\\
		{}_{[\mathrm{pres}]}\!\langle\mathbf p'|\mathbf p
		\rangle_{[\mathrm{pres}]}
		&=\delta^3(\mathbf p'-\mathbf p).
	\end{aligned}
	\label{eq:ope_lin_plane_wave_normalization}
\end{equation}
Consequently,
\begin{equation}
	|\mathbf p\rangle_{[\mathrm{pres}]}
	=(2\pi)^{-3/2}|\mathbf p\rangle_{[\mathrm{Lin}]},
	\qquad
	V^{[\mathrm{pres}]}=(2\pi)^{-3}V^{[\mathrm{Lin}]}.
	\label{eq:ope_plane_wave_bridge}
\end{equation}
For the same normalized state, the corresponding radial functions obey
\begin{equation}
	\int_0^\infty\frac{p^2dp}{(2\pi)^3}
	\left|\phi_\ell^{[\mathrm{Lin}]}(p)\right|^2=1,
	\qquad
	\phi_\ell^{[\mathrm{Lin}]}(p)
	=(2\pi)^{3/2}R_\ell(p).
	\label{eq:ope_radial_normalization_bridge}
\end{equation}
For the molecular $S$ wave, $R_\ell$ reduces to the $R_0$ of
Eq.~\eqref{eq:molecular_gaussian}.

For a rotationally invariant scalar potential and normalized spherical
harmonics, the Funk--Hecke relation gives
\begin{equation}
	\begin{aligned}
		&\int d\Omega_p\,
		V^{[\mathrm{Lin}]}(k,p,u)
		Y_{\ell m}(\widehat{\mathbf p})\\
		&\quad=2\pi Y_{\ell m}(\widehat{\mathbf k})
		\int_{-1}^{1}du\,P_\ell(u)
		V^{[\mathrm{Lin}]}(k,p,u).
	\end{aligned}
	\label{eq:ope_funk_hecke}
\end{equation}
Combining Eqs.~\eqref{eq:ope_plane_wave_bridge}--
\eqref{eq:ope_funk_hecke} for $\ell=0$ yields the polarization-resolved OPE
amplitude in the present convention:
\begin{equation}
	\begin{aligned}
		&\mathcal T_{\pi;fi;M_J;m_Mm_B}^{[\mathrm{pres}]}(\mathbf k)
		=\frac{Y_{00}(\widehat{\mathbf k})}{(2\pi)^2}
		\langle s_Mm_M,s_Bm_B\mid JM_J\rangle\\
		&\quad\times\Biggl\{
		\int_0^\infty p^2dp\,R_0(p)
		\int_{-1}^{1}du\,
		V_{\pi,\xi;fi}^{[\mathrm{Lin}],J}(k,p,u)
		\Biggr\}.
	\end{aligned}
	\label{eq:ope_explicit_normalization_bridge}
\end{equation}
Comparison with Eq.~\eqref{eq:T_s_wave_expansion} identifies the expression
in braces as $T_{00,\pi,\xi;fi}^{(J)}(k)$. Thus, although the state-level
matrix elements in Eq.~\eqref{eq:ope_plane_wave_bridge} differ by
$(2\pi)^{-3}$, the radial-state and partial-wave factors compensate so that
the Lin three-dimensional potential appears directly under the braces in
Eq.~\eqref{eq:ope_explicit_normalization_bridge}.

The same result can be expressed through a generic partial wave. Define
\begin{equation}
	\begin{aligned}
		G_\ell^{[\mathrm{Lin}]}(k)
		&\equiv\int_0^\infty\frac{p^2dp}{(2\pi)^3}
		V_\ell^{[\mathrm{Lin}]}(k,p)
		\phi_\ell^{[\mathrm{Lin}]}(p)\\
		&=\frac{1}{\sqrt{2\pi}}
		T_\ell^{[\mathrm{pres}]}(k),\\
		V_\ell^{[\mathrm{Lin}]}(k,p)
		&=2\pi\int_{-1}^{1}du\,P_\ell(u)
		V^{[\mathrm{Lin}]}(k,p,u).
	\end{aligned}
	\label{eq:ope_vertex_bridge}
\end{equation}
If $\mu_f$ denotes the reduced mass of the final two-hadron channel, the
nonrelativistic expression
$\mu_fk|G_\ell^{[\mathrm{Lin}]}|^2/(4\pi^2)$ becomes
$\mu_fk|T_\ell^{[\mathrm{pres}]}|^2/(2\pi)^3$. The relativistic replacement
$\mu_f\to E_{M_f}E_{B_f}/M_{P_c}$ then reproduces the width convention of
Eq.~\eqref{eq:width_t00}.

Reference~\cite{Lin2023CSM} relates its heavy-hadron chiral Feynman amplitude
to the nonrelativistic potential through
$V^{[\mathrm{Lin}]}=-\mathcal M_{\mathrm{HH}\chi\mathrm{PT}}^{[\mathrm{Lin}]}/4$.
That conversion is already contained in
Eq.~\eqref{eq:ope_full_kernel}; the factor $-1/4$ and the external
heavy-field normalizations must not be applied again. The chiral amplitude
$\mathcal M_{\mathrm{HH}\chi\mathrm{PT}}^{[\mathrm{Lin}]}$ is distinct from
the relativistic decay amplitude $\mathcal M_{fi}$ defined in
Eq.~\eqref{eq:T_to_invariant_amplitude}. Similarly,
Eqs.~\eqref{eq:ope_central_reduction},
\eqref{eq:ope_coefficient_definition}, and
\eqref{eq:ope_explicit_normalization_bridge} show why no additional factor
of $1/3$, $\sqrt{3/2}$, $2\pi$, or $(2\pi)^3$ is inserted.

Using the angular functions in Eq.~\eqref{eq:ope_angular_kernels}, the OPE
contribution in the common present convention is therefore
\begin{equation}
	\begin{aligned}
		T_{00,\pi,\xi;fi}^{(J)}(k)
		={}&\frac{|g|\,|g_2|\widehat c_{fi}^{(J)}}{f_\pi^2}
		\int_0^\infty dp\,p^2R_0(p)F_{\Lambda_\pi}(k,p)\\
		&\times\mathcal K_\xi(p,k),
		\qquad
		\xi\in\{\mathrm{ct},\mathrm{pole},\mathrm{full}\}.
	\end{aligned}
	\label{eq:ope_t00_fold}
\end{equation}
The reference-sign OPE amplitude used in the coherent matching is
\begin{equation}
	T_{00,\pi,\mathrm{ref};fi}^{(J)}(k)
	\equiv T_{00,\pi,\mathrm{pole};fi}^{(J)}(k).
	\label{eq:ope_reference_amplitude}
\end{equation}
Equation~\eqref{eq:ope_t00_fold} is the endpoint of the normalization bridge:
$T_{00,\pi,\xi;fi}^{(J)}$ now has the same normalization as the QI amplitude
and a fixed real reference-channel phase convention. Only its overall sign
relative to the independently constructed QI basis remains to be matched.

The logarithmic endpoint singularities in
Eq.~\eqref{eq:ope_angular_log} are integrable. Their numerical treatment and
the radial cutoff are specified in Sec.~\ref{sec:numerical_evaluation}, with
convergence tests in Appendix~\ref{app:qi_numerics}. In the conventions used
here, $[V_\pi]=\mathrm{GeV}^{-2}$,
$[R_0]=\mathrm{GeV}^{-3/2}$, and
$[T_{00,\pi}^{(J)}]=\mathrm{GeV}^{-1/2}$, as required by
Eq.~\eqref{eq:width_t00}. The complex-scaled channel wave functions and their
$c$-product normalization play no role in this Gaussian Born fold.

\subsection{Coherent matching to quark interchange}
\label{sec:ope_coherent_matching}

The QI and OPE mechanisms feed the same asymptotic two-body channel and must
therefore be combined at the amplitude level. The rephasing in
Sec.~\ref{sec:ope_coefficients} fixes the relative phases among the OPE
channels, but it does not by itself determine the overall phase of the
chiral heavy-hadron basis relative to the independently constructed QI
basis. We parameterize this remaining correlated ambiguity by one sign:
\begin{equation}
	\begin{aligned}
		T_{00,\mathrm{tot};fi}^{(J)}
		(k;\eta_{\mathrm{match}})
		={}&T_{00,\mathrm{QI};fi}^{(J)}(k)\\
		&+\eta_{\mathrm{match}}
		T_{00,\pi,\mathrm{ref};fi}^{(J)}(k),\\
		\eta_{\mathrm{match}}&=\pm1.
	\end{aligned}
	\label{eq:ope_hybrid_amplitude}
\end{equation}
Equation~\eqref{eq:ope_hybrid_amplitude} is a correlated matching ansatz,
not a relative sign derived from either model separately. Within a reported
branch, the same $\eta_{\mathrm{match}}$ is therefore used for every decay
channel; it is not selected channel by channel. The two branches are not
statistical limits and are not averaged.

The factor $\eta_{\rm match}$ parameterizes only the unresolved
global relative phase between the two amplitude conventions; it
does not remove or quantify the residual short-distance overlap
discussed in Sec.~IV D.

Using the phase-space factor $\rho_{fi}$ defined in
Eq.~\eqref{eq:width_t00}, the separate mechanism widths are
\begin{equation}
	\Gamma_{\mathrm{QI};fi}^{(J)}
	=\rho_{fi}\left|T_{00,\mathrm{QI};fi}^{(J)}\right|^2,
	\qquad
	\Gamma_{\pi,\mathrm{ref};fi}^{(J)}
	=\rho_{fi}\left|T_{00,\pi,\mathrm{ref};fi}^{(J)}\right|^2.
	\label{eq:ope_component_widths}
\end{equation}
The coherent total width is
\begin{equation}
	\begin{aligned}
		\Gamma_{\mathrm{tot};fi}^{(J)}(\eta_{\mathrm{match}})
		={}&\Gamma_{\mathrm{QI};fi}^{(J)}
		+\Gamma_{\pi,\mathrm{ref};fi}^{(J)}\\
		&+2\eta_{\mathrm{match}}\rho_{fi}
		\operatorname{Re}\!\left[
		T_{00,\mathrm{QI};fi}^{(J)}
		T_{00,\pi,\mathrm{ref};fi}^{(J)*}\right],\\
		\Gamma_{\pm;fi}^{(J)}
		&\equiv
		\Gamma_{\mathrm{tot};fi}^{(J)}(\eta_{\mathrm{match}}=\pm1).
	\end{aligned}
	\label{eq:ope_interference_width}
\end{equation}
Thus, the separately quoted QI and OPE widths are not added incoherently.
Fixing which correlated branch is realized requires a signed microscopic
matching of the quark-model states to the chiral heavy-hadron fields, or an
interference-sensitive observable. Until such information is available, we
report both $\Gamma_+$ and $\Gamma_-$ alongside the QI-only central results.
The numerical inputs, integration settings, and stability tests used for all
three amplitudes are given in Sec.~\ref{sec:validation}.

\section{Numerical inputs, evaluation, and validation}
\label{sec:validation}

This section completes the setup needed for the results. We first list the
physical masses and the molecular scales derived from the prescription of
Sec.~\ref{sec:molecular_wavefunction}. We then specify the numerical
integration settings, summarize the algebraic and convergence checks, and
define the model-variation scenarios used to assess the stability of the
predicted width hierarchies. Unless stated otherwise, all dimensionful inputs
are given in GeV. No parameter is fitted to the selected
$\Lambda_c\bar D^{(*)}$ partial widths. The QI and OPE interaction parameters
already given in Tables~\ref{tab:qi_model_parameters} and
\ref{tab:ope_inputs}, respectively, are not repeated here.

\subsection{Hadron masses and derived molecular scales}
\label{sec:numerical_masses}

The reference masses are collected in Table~\ref{tab:hadron_masses}. We use
the central LHCb peak masses for the three $P_c$ states~\cite{LHCb2019} and
the conventional-hadron masses from the Particle Data Group~\cite{PDG2024}.
Experimental mass uncertainties are not propagated as statistical errors,
because the uncertainty of the calculation is dominated by the molecular and
interaction-model assumptions. The effect of choosing the neutral, averaged,
or charged molecular threshold is instead examined explicitly in
Sec.~\ref{sec:numerical_sensitivities}.

\begin{table}[t]
	\caption{Hadron masses used in the reference calculation.}
	\label{tab:hadron_masses}
	\begin{ruledtabular}
		\begin{tabular}{lc}
			Hadron & $M_H$ (GeV) \\
			\colrule
			$P_c(4312)^+$       & $4.31190$ \\
			$P_c(4440)^+$       & $4.44030$ \\
			$P_c(4457)^+$       & $4.45730$ \\
			$\bar D^0$          & $1.86484$ \\
			$D^-$                & $1.86966$ \\
			$\bar D^{*0}$       & $2.00685$ \\
			$D^{*-}$             & $2.01026$ \\
			$\Sigma_c^+$        & $2.45265$ \\
			$\Sigma_c^{++}$     & $2.45397$ \\
			$\Lambda_c^+$       & $2.28646$ \\
		\end{tabular}
	\end{ruledtabular}
\end{table}

For the reference molecular calculation, the masses in
Eq.~\eqref{eq:binding_energy_reduced_mass} are the arithmetic charge averages
$M_A^{\rm av}$ and $M_B^{\rm av}$ defined in
Eq.~\eqref{eq:average_threshold_masses}. Thus
$E_{\rm bind}=\overline M_{\rm th}^{\,D^{(*)}}-M_{P_c}$,
$\mu_{AB}=\mu_{AB}^{\rm av}$, and Eq.~\eqref{eq:beta_binding} fixes
$\beta_{P_c}$. The resulting thresholds, binding energies, Gaussian widths,
and final-state momenta are listed in
Table~\ref{tab:molecular_kinematics}. The momenta are obtained from
Eq.~\eqref{eq:two_body_momentum} using the physical masses of
$\Lambda_c^+\bar D^0$ or $\Lambda_c^+\bar D^{*0}$ in the final charge
channel.

\begin{table*}[t]
	\caption{Reference molecular scales and final-state momenta. The averaged
		thresholds and reduced masses are defined in
		Eq.~\eqref{eq:average_threshold_masses}; $E_{\rm bind}$ and
		$\beta_{P_c}$ then follow from Eqs.~\eqref{eq:binding_energy_reduced_mass}
		and \eqref{eq:beta_binding}. The final-state momenta use the physical
		$\bar D^0$ and $\bar D^{*0}$ masses.}
	\label{tab:molecular_kinematics}
	\begin{ruledtabular}
		\begin{tabular}{lccccc}
			Assignment & $\overline M_{\rm th}$ (GeV) & $E_{\rm bind}$ (MeV)
			& $\beta_{P_c}$ (GeV) & $k_{\Lambda_c\bar D}$ (GeV)
			& $k_{\Lambda_c\bar D^*}$ (GeV) \\
			\colrule
			$P_c(4312):\ \Sigma_c\bar D,\ J=\tfrac12$
			& $4.320560$ & $8.660$  & $0.165969$ & $0.580128$ & $0.199561$ \\
			$P_c(4440):\ \Sigma_c\bar D^*,\ J=\tfrac12,\tfrac32$
			& $4.461865$ & $21.565$ & $0.267298$ & $0.784313$ & $0.565390$ \\
			$P_c(4457):\ \Sigma_c\bar D^*,\ J=\tfrac12,\tfrac32$
			& $4.461865$ & $4.565$  & $0.122982$ & $0.807877$ & $0.597780$ \\
		\end{tabular}
	\end{ruledtabular}
\end{table*}

At a fixed peak mass, the $J=1/2$ and $J=3/2$
$\Sigma_c\bar D^*$ assignments therefore have identical kinematics and
differ only through their spin-dependent transition kernels. The particularly
small reference binding energy of $P_c(4457)$ also explains why its molecular
width is most sensitive to the charged--neutral threshold prescription.

\subsection{Numerical evaluation and reference settings}
\label{sec:numerical_evaluation}

The QI amplitude in Eq.~\eqref{eq:qi_physical_t00} is evaluated with
Gauss--Legendre quadrature in the initial relative momentum $p$ and the
external angular variable $u=\widehat{\mathbf k}\!\cdot\!\widehat{\mathbf p}$.
After the analytic Gaussian reduction described in
Sec.~\ref{sec:qi_spatial}, the remaining Coulomb-plus-confinement integral is
evaluated in the magnitude $q=|\mathbf q|$, the polar-angle cosine
$u_q\equiv\cos\theta_q$, and the azimuthal angle $\phi_q$. We use
Gauss--Legendre quadrature for $q$ and $u_q$ and a uniform periodic sum for
$\phi_q$. The Gaussian-smeared hyperfine contribution is evaluated after its
internal momentum integrations have been carried out analytically.

For OPE, the $u$ integration is performed analytically through
Eq.~\eqref{eq:ope_angular_log}. The remaining radial domain is partitioned at
the logarithmic endpoint singularities. Each pole-separated interval is
mapped with the $\sin^2$ transformation given in
Appendix~\ref{app:ope_checks} and evaluated with Gauss--Legendre quadrature.
The endpoint conditions are first bracketed on a uniform scan and then
refined before the integration. Table~\ref{tab:numerical_settings} lists the
reference settings for both mechanisms.

\begin{table}[t]
	\caption{Reference numerical settings. The orders $N_p$ and $N_u$ refer
		to the QI molecular fold; $N_q$, $N_{u_q}$, and $N_\phi$ refer to the
		remaining Coulomb-plus-confinement integral. The order $N_\pi$ is used on
		each pole-separated OPE interval.}
	\label{tab:numerical_settings}
	\begin{ruledtabular}
		\begin{tabular}{lc}
			Setting & Reference value \\
			\colrule
			$(N_p,N_u)$                         & $(20,16)$ \\
			$p_{\max}^{\rm QI}$                 & $1.6~\mathrm{GeV}$ \\
			$(N_q,N_{u_q},N_\phi)$              & $(24,12,12)$ \\
			$q_{\max}$                          & $4.0~\mathrm{GeV}$ \\
			$p_{\max}^{\pi}$                   & $2.0~\mathrm{GeV}$ \\
			Endpoint-bracketing scan             & $2000$ subintervals \\
			$N_\pi$ per pole-separated interval & $64$ \\
		\end{tabular}
	\end{ruledtabular}
\end{table}

\subsection{Algebraic consistency and numerical convergence}
\label{sec:numerical_convergence}

We distinguish exact consistency relations from convergence of the remaining
numerical integrations. In the QI sector, the initial and final color and spin
states are normalized, both labeled rearrangement topologies have flavor
overlap $1/\sqrt2$, and the explicit color--flavor--spin construction
reproduces the coefficient $\sqrt6/144$ in
Eq.~\eqref{eq:qi_example_cfs_coefficient}. The two labeled final clusterings
satisfy Eq.~\eqref{eq:qi_topology_relation} pointwise in $(k,p,u)$ before the
molecular fold. The subsequent antisymmetrization and exact-isospin
reconstruction yield Eq.~\eqref{eq:qi_physical_isospin_kernel}; the neutral
initial charge component is therefore not evaluated independently.

In the OPE sector, the five $S$-wave coefficients used here agree with
Tables VI and VII of Ref.~\cite{Lin2023CSM} after the rephasing in
Eq.~\eqref{eq:ope_rephasing}. The contact--pole relation
\eqref{eq:ope_contact_identity} is satisfied in the distributional limit
$\epsilon\to0^+$, and an independent delta-function evaluation reproduces the
absorptive part implied by Eq.~\eqref{eq:ope_angular_log}. The same global
value of $\eta_{\rm match}$ is used in every channel, as required by
Eq.~\eqref{eq:ope_hybrid_amplitude}. These are algebraic or convention-level
checks and are independent of the quadrature orders.

The convergence tests show that the numerical uncertainty is negligible on
the scale of the model variations. Raising the QI orders from
$(N_p,N_u;N_q,N_{u_q},N_\phi)=(20,16;24,12,12)$ to
$(28,22;32,16,16)$ changes the most sensitive calculated width by only
$9.42\times10^{-5}\%$. For the reference
$P_c(4312)\to\Lambda_c\bar D^*$ pion-pole amplitude with
$q_{\pi,\rm dir}^0$ and $\Lambda_\pi=0.50~\mathrm{GeV}$, increasing
$N_\pi$ from $64$ to $128$ changes the complex amplitude by
$9.83\times10^{-8}~\mathrm{GeV}^{-1/2}$. Extending
$p_{\max}^{\pi}$ from $2$ to $3~\mathrm{GeV}$ at $N_\pi=256$ changes it by
$1.84\times10^{-10}~\mathrm{GeV}^{-1/2}$. Appendix~\ref{app:qi_numerics}
gives the channel-by-channel QI comparison, the OPE endpoint analysis, and the
complex amplitudes entering these tests.

\subsection{Reference prescriptions and model-variation scenarios}
\label{sec:numerical_sensitivities}

The convergence tests above quantify numerical errors; they do not measure
the uncertainty associated with the physical and modeling assumptions.
Table~\ref{tab:reference_sensitivities} defines the variations used for that
second purpose. In the QI sector, one of eight input groups is varied at a
time while all other inputs remain at their reference values. Each group is
evaluated at a lower, reference, and upper choice, giving the 24 entries used
to construct the scenario envelopes in Sec.~\ref{sec:results}. Repeated
reference entries are retained in this count. The variations are not assigned
probability distributions and are not combined in quadrature.

The OPE scenarios separately probe the regulator cutoff, the off-shell
energy-transfer prescription, and the treatment of the short-distance part
of the regulated interaction. Finally, $\eta_{\rm match}=+1$ and $-1$ define
two globally correlated QI--OPE matching branches. Neither branch is selected
as the preferred sign, and the sign is never chosen separately for different
channels.

\begin{table*}[t]
	\caption{Reference prescriptions and one-at-a-time model-variation
		choices. These alternatives define model scenarios rather than statistical
		confidence intervals. The two values of $\eta_{\rm match}$ are globally
		correlated matching branches.}
	\label{tab:reference_sensitivities}
	\begin{ruledtabular}
		\begin{tabular}{llll}
			Sector & Ingredient & Reference choice & Alternative choices \\
			\colrule
			QI & Molecular constituent masses
			& $(M_A^{\rm av},M_B^{\rm av})$ from
			Eq.~\eqref{eq:average_threshold_masses}
			& neutral or charged constituent masses \\
			QI & Molecular width
			& $\beta_{P_c}$ from Eq.~\eqref{eq:beta_binding}
			& $0.9\beta_{P_c}$ and $1.1\beta_{P_c}$ \\
			QI & Meson oscillator widths
			& $(\beta_D,\beta_{D^*})$ in Table~\ref{tab:qi_model_parameters}
			& common factors $0.9$ and $1.1$ \\
			QI & Baryon oscillator widths
			& $(\alpha_\rho,\alpha_\lambda^{(c)})$ in
			Table~\ref{tab:qi_model_parameters}
			& common factors $0.9$ and $1.1$ \\
			QI & Light constituent mass
			& $m_q$ in Table~\ref{tab:qi_model_parameters}
			& $0.95m_q$ and $1.05m_q$ \\
			QI & Charm constituent mass
			& $m_c$ in Table~\ref{tab:qi_model_parameters}
			& $0.95m_c$ and $1.05m_c$ \\
			QI & String tension
			& $b$ in Table~\ref{tab:qi_model_parameters}
			& $0.9b$ and $1.1b$ \\
			QI & Hyperfine smearing
			& $\tau_{\rm hf}$ in Table~\ref{tab:qi_model_parameters}
			& $0.9\tau_{\rm hf}$ and $1.1\tau_{\rm hf}$ \\
			OPE & Regulator cutoff
			& $\Lambda_\pi=0.50~\mathrm{GeV}$
			& $0.40\leq\Lambda_\pi\leq1.00~\mathrm{GeV}$ in
			$0.10~\mathrm{GeV}$ steps \\
			OPE & Energy transfer
			& $q_{\pi,\rm dir}^0$
			& $q_{\pi,\rm swap}^0$ and $q_{\pi,\rm stat}^0$ \\
			OPE & Short-distance treatment
			& $\xi=\mathrm{pole}$
			& $\xi=\mathrm{full}$ \\
			QI--OPE & Relative matching sign
			& no preferred branch
			& $\eta_{\rm match}=+1$ and $-1$, globally \\
		\end{tabular}
	\end{ruledtabular}
\end{table*}

Finite $\Sigma_c$ and $\bar D^*$ widths are not included. Their possible
smearing of the pion cut, together with tensor OPE, OPE iteration, and
$S$--$D$ mixing, lies beyond the present first-Born, central-$S$-wave
treatment. With the reference inputs, numerical accuracy, and model scenarios
now specified, Sec.~\ref{sec:results} presents the central QI predictions and
the exploratory coherent QI--OPE branches.

\section{Results and discussion}
\label{sec:results}

With the transition amplitudes, numerical settings, and model-variation
scenarios now specified, we turn to the predicted decay patterns. We first
present the QI amplitudes and partial widths, which constitute the central
results of this work. We then test the stability of their channel and spin
hierarchies, examine OPE through the matching scenarios defined in
Secs.~\ref{sec:ope_coherent_matching} and \ref{sec:numerical_sensitivities},
and compare the resulting patterns with other open-charm calculations.

All amplitudes quoted below are the physical $I=1/2$, molecularly folded
$S$-wave amplitudes of Sec.~\ref{sec:kinematics_width}. The channel labels and
on-shell momentum argument are suppressed in tables only after the relevant
initial and final states have been specified. For
$J=3/2\to\Lambda_c\bar D$, angular-momentum and parity conservation require
$L_f=2$. Those two $D$-wave transitions are outside the central-$S$-wave
treatment adopted here; an entry marked ``not calculated'' must therefore not
be interpreted as a vanishing width.

\subsection{Central quark-interchange results}
\label{sec:results_qi_central}

The reference QI amplitudes and partial widths are collected in
Table~\ref{tab:qi_central_results}. For each row,
$T_{00,\mathrm{CL}}^{(J)}$ and $T_{00,\mathrm{HF}}^{(J)}$ abbreviate the
channel-specific amplitudes $T_{00,\mathrm{CL};fi}^{(J)}(k)$ and
$T_{00,\mathrm{HF};fi}^{(J)}(k)$ defined in
Eq.~\eqref{eq:qi_physical_t00}; hence
$T_{00,\mathrm{QI}}^{(J)}=T_{00,\mathrm{CL}}^{(J)}+
T_{00,\mathrm{HF}}^{(J)}$. All three amplitudes are in the common convention
of Sec.~\ref{sec:kinematics_width} and are real in the adopted phase
convention. The range accompanying each width is the minimum-to-maximum
envelope of the 24 QI scenarios defined in
Table~\ref{tab:reference_sensitivities}, not a statistical confidence
interval.

\begin{table*}[t]
	\caption{Reference QI amplitudes and partial widths obtained with the
		averaged molecular thresholds and the inputs of
		Sec.~\ref{sec:validation}. Amplitudes are in $\mathrm{GeV}^{-1/2}$ and
		widths are in MeV. The parenthetical interval in the last column is the
		minimum-to-maximum envelope of the 24 one-at-a-time QI scenarios.}
	\label{tab:qi_central_results}
	\begin{ruledtabular}
		\begin{tabular}{lclrrrr}
			Initial state & $J$ & Final state
			& $T_{00,\mathrm{CL}}^{(J)}$
			& $T_{00,\mathrm{HF}}^{(J)}$
			& $T_{00,\mathrm{QI}}^{(J)}$
			& $\Gamma_{\mathrm{QI}}^{(J)}$ (scenario range) \\
			\colrule
			$P_c(4312)$ & $\tfrac12$ & $\Lambda_c\bar D$
			& $0$ & $-0.033759$ & $-0.033759$
			& $0.002848\;(0.001537\text{--}0.004318)$ \\
			$P_c(4312)$ & $\tfrac12$ & $\Lambda_c\bar D^*$
			& $1.067412$ & $-0.140435$ & $0.926978$
			& $0.742111\;(0.398344\text{--}1.345232)$ \\
			$P_c(4440)$ & $\tfrac12$ & $\Lambda_c\bar D$
			& $0.215196$ & $-0.113425$ & $0.101771$
			& $0.036067\;(0.012826\text{--}0.065088)$ \\
			$P_c(4440)$ & $\tfrac12$ & $\Lambda_c\bar D^*$
			& $-0.916089$ & $0.472811$ & $-0.443278$
			& $0.495335\;(0.212903\text{--}0.899050)$ \\
			$P_c(4440)$ & $\tfrac32$ & $\Lambda_c\bar D$
			& \multicolumn{4}{c}{$D$ wave not calculated} \\
			$P_c(4440)$ & $\tfrac32$ & $\Lambda_c\bar D^*$
			& $0.458044$ & $-0.324335$ & $0.133709$
			& $0.045068\;(0.010919\text{--}0.103664)$ \\
			$P_c(4457)$ & $\tfrac12$ & $\Lambda_c\bar D$
			& $0.053088$ & $-0.035964$ & $0.017124$
			& $0.001056\;(0.000016\text{--}0.004277)$ \\
			$P_c(4457)$ & $\tfrac12$ & $\Lambda_c\bar D^*$
			& $-0.341078$ & $0.167280$ & $-0.173798$
			& $0.080820\;(0.030208\text{--}0.141703)$ \\
			$P_c(4457)$ & $\tfrac32$ & $\Lambda_c\bar D$
			& \multicolumn{4}{c}{$D$ wave not calculated} \\
			$P_c(4457)$ & $\tfrac32$ & $\Lambda_c\bar D^*$
			& $0.170539$ & $-0.114254$ & $0.056285$
			& $0.008477\;(0.002569\text{--}0.017729)$ \\
		\end{tabular}
	\end{ruledtabular}
\end{table*}

The two calculated $P_c(4312)$ channels illustrate the origin of the main
hierarchy. In the transition
$\Sigma_c\bar D\to\Lambda_c\bar D$, the Coulomb-plus-confinement contribution
vanishes, leaving the small width to be generated entirely by the hyperfine
interaction. For $\Sigma_c\bar D\to\Lambda_c\bar D^*$, the
Coulomb-plus-confinement amplitude is large and is only moderately reduced by
destructive interference with the hyperfine term. Consequently,
\begin{equation}
	\frac{\Gamma_{\mathrm{QI}}
		[P_c(4312)\to\Lambda_c\bar D^*]}
	{\Gamma_{\mathrm{QI}}
		[P_c(4312)\to\Lambda_c\bar D]}
	=260.6.
	\label{eq:qi_4312_ratio}
\end{equation}
For the $J=1/2$ $\Sigma_c\bar D^*$ assignments, the analogous
$\Lambda_c\bar D^*$-to-$\Lambda_c\bar D$ ratios are $13.73$ at the
$P_c(4440)$ mass and $76.53$ at the $P_c(4457)$ mass. The latter value is
enhanced by the cancellation
$0.05309-0.03596=0.01712$ in the $\Lambda_c\bar D$ amplitude and should not
be interpreted as a symmetry-protected prediction.

This distinction becomes transparent in the heavy-quark spin symmetry
(HQSS) limit. Before phase-space corrections, Voloshin obtained the
spin-recoupling relation~\cite{Voloshin2019}
\begin{equation}
	\frac{\Gamma(\Sigma_c\bar D^*,J=1/2
		\to\Lambda_c\bar D^*)}
	{\Gamma(\Sigma_c\bar D^*,J=1/2
		\to\Lambda_c\bar D)}
	=\frac{4}{3}.
	\label{eq:hqss_open_charm_ratio}
\end{equation}
As an internal limit check, making the $D$ and $D^*$ masses and Gaussian
widths degenerate and removing the spin-dependent interaction gives
\begin{equation*}
	\left|
	\frac{T_{00,\mathrm{CL}}(\Lambda_c\bar D^*)}
	{T_{00,\mathrm{CL}}(\Lambda_c\bar D)}
	\right|^2=\frac{4}{3}.
\end{equation*}
The substantially larger physical ratios in
Table~\ref{tab:qi_central_results} therefore arise from unequal spatial
overlaps, the hyperfine contribution, and channel-dependent cancellations.
Likewise, the small but nonzero
$P_c(4312)\to\Lambda_c\bar D$ width is a spin-dependent correction to the
leading HQSS suppression.

For the three calculated $J=1/2$ assignments, the relative share of the two
selected open-charm modes can be summarized by the conditional fraction
\begin{equation}
	\mathcal F_{\bar D^*}^{(\bar D,\bar D^*)}
	\equiv
	\frac{\Gamma_{\mathrm{QI}}(\Lambda_c\bar D^*)}
	{\Gamma_{\mathrm{QI}}(\Lambda_c\bar D)
		+\Gamma_{\mathrm{QI}}(\Lambda_c\bar D^*)}.
	\label{eq:selected_channel_fraction}
\end{equation}
The reference values are $99.62\%$, $93.21\%$, and $98.71\%$ for
$P_c(4312)$, $P_c(4440,J=1/2)$, and $P_c(4457,J=1/2)$, respectively. These
are conditional two-channel fractions, not inclusive branching fractions:
hidden-charm, $\Sigma_c^{(*)}\bar D^{(*)}$, and three-body modes are not
included in the denominator.

\subsection{Stability of the QI hierarchies}
\label{sec:results_qi_sensitivity}

\begin{figure*}[t]
	\centering
	\includegraphics[width=\textwidth]{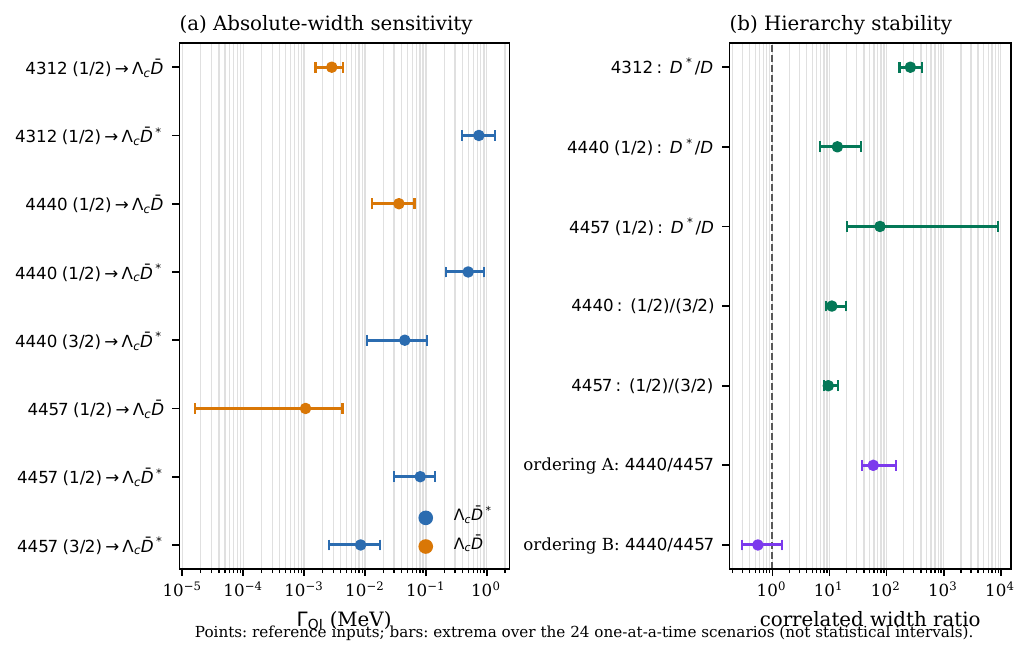}
	\caption{Stability of the QI results over the 24 one-at-a-time scenarios
		of Table~\ref{tab:reference_sensitivities}. Points show the reference
		inputs, and horizontal bars span the corresponding minimum and maximum.
		Panel (a) gives the eight calculated $S$-wave widths. Panel (b) gives
		correlated width ratios, with the vertical dashed line marking unity.
		The bars are model-scenario envelopes, not statistical error intervals.}
	\label{fig:qi_sensitivity}
\end{figure*}

Figure~\ref{fig:qi_sensitivity} summarizes the 24 QI scenarios. Absolute
widths vary appreciably, particularly when the reference amplitude is small
because of cancellation. To display the correlated ratios compactly, define
\begin{equation*}
	\Gamma_{n,J}^{\mathrm{QI}}(f)
	\equiv
	\Gamma_{\mathrm{QI};fi}^{(J)}
	\quad\text{for}\quad i=P_c(n),
\end{equation*}
where $n\in\{4312,4440,4457\}$ and $f$ is the final channel named in the
argument. Within each one-at-a-time scenario, the same varied input is used
in the numerator and denominator. The channel ratios then satisfy
\begin{align}
	168.0&\leq
	\frac{\Gamma_{4312,1/2}^{\mathrm{QI}}(\Lambda_c\bar D^*)}
	{\Gamma_{4312,1/2}^{\mathrm{QI}}(\Lambda_c\bar D)}
	\leq421.6,\nonumber\\
	6.90&\leq
	\frac{\Gamma_{4440,1/2}^{\mathrm{QI}}(\Lambda_c\bar D^*)}
	{\Gamma_{4440,1/2}^{\mathrm{QI}}(\Lambda_c\bar D)}
	\leq36.15,\nonumber\\
	20.61&\leq
	\frac{\Gamma_{4457,1/2}^{\mathrm{QI}}(\Lambda_c\bar D^*)}
	{\Gamma_{4457,1/2}^{\mathrm{QI}}(\Lambda_c\bar D)}
	\leq8912.4.
	\label{eq:qi_channel_ratio_ranges}
\end{align}
The large upper endpoint in the final line results from an accidental
near-zero in the $\Lambda_c\bar D$ denominator. It is therefore more useful
as evidence for a stable hierarchy than as a precise enhancement factor.

At each heavier peak mass, the calculated $\Lambda_c\bar D^*$ width also
shows a stable dependence on the assigned spin:
\begin{align}
	8.67&\leq
	\frac{\Gamma_{4440,1/2}^{\mathrm{QI}}(\Lambda_c\bar D^*)}
	{\Gamma_{4440,3/2}^{\mathrm{QI}}(\Lambda_c\bar D^*)}
	\leq19.50,\nonumber\\
	7.99&\leq
	\frac{\Gamma_{4457,1/2}^{\mathrm{QI}}(\Lambda_c\bar D^*)}
	{\Gamma_{4457,3/2}^{\mathrm{QI}}(\Lambda_c\bar D^*)}
	\leq14.22.
	\label{eq:qi_spin_ratio_ranges}
\end{align}
Thus, within all tested QI scenarios, the selected
$\Lambda_c\bar D^*$ width is larger for $J=1/2$ than for $J=3/2$ at the
same peak mass. The size of the enhancement remains model dependent.

The two possible assignments of the heavier peaks can now be compared without
reusing the QI cluster labels $A$ and $B$. We call
\begin{align*}
	\text{spin ordering I:}\quad
	&J[P_c(4440)]=\tfrac12,\qquad
	J[P_c(4457)]=\tfrac32,\\
	\text{spin ordering II:}\quad
	&J[P_c(4440)]=\tfrac32,\qquad
	J[P_c(4457)]=\tfrac12.
\end{align*}
For spin ordering I, the reference peak-to-peak ratio and its scenario
envelope are
\begin{equation}
	\begin{aligned}
		R_{\mathrm I}^{\bar D^*}
		&\equiv
		\frac{\Gamma_{4440,1/2}^{\mathrm{QI}}(\Lambda_c\bar D^*)}
		{\Gamma_{4457,3/2}^{\mathrm{QI}}(\Lambda_c\bar D^*)}
		=58.4,\\
		R_{\mathrm I}^{\bar D^*}&\in[37.5,143.4].
	\end{aligned}
	\label{eq:qi_ordering_a_ratio}
\end{equation}
The reversed spin ordering II gives
\begin{equation}
	\begin{aligned}
		R_{\mathrm{II}}^{\bar D^*}
		&\equiv
		\frac{\Gamma_{4440,3/2}^{\mathrm{QI}}(\Lambda_c\bar D^*)}
		{\Gamma_{4457,1/2}^{\mathrm{QI}}(\Lambda_c\bar D^*)}
		=0.558,\\
		R_{\mathrm{II}}^{\bar D^*}&\in[0.299,1.495].
	\end{aligned}
	\label{eq:qi_ordering_b_ratio}
\end{equation}
Accordingly, a much larger $P_c(4440)$ partial width in this channel would be
characteristic of spin ordering I within the present QI model. A ratio of
order unity would remain compatible with spin ordering II, but would not by
itself establish it because the corresponding scenario envelope crosses
unity. In applying either test, an experimental amplitude analysis must
separate resonance couplings from production effects and interference with
nonresonant contributions.

\subsection{Exploratory one-pion-exchange scenarios}
\label{sec:results_ope}

We next examine how the long-range contribution modifies the central QI
pattern. The reference OPE scenario uses the direct baryon-line energy
transfer $q_{\pi,\mathrm{dir}}^0$, the cutoff
$\Lambda_\pi=0.50~\mathrm{GeV}$, and the pion-pole prescription
$\xi=\mathrm{pole}$. Table~\ref{tab:ope_reference_results} lists the
corresponding amplitude $T_{00,\pi,\mathrm{ref};fi}^{(J)}$ of
Eq.~\eqref{eq:ope_reference_amplitude}, its magnitude relative to QI, the
OPE-only width, and the two coherent widths $\Gamma_+$ and $\Gamma_-$ defined
in Eq.~\eqref{eq:ope_interference_width}. The amplitude itself does not
contain $\eta_{\mathrm{match}}$; the same global sign is used in every row of
each coherent branch. Imaginary parts generated by the on-shell pion cut are
retained.

\begin{table*}[t]
	\caption{Reference OPE matching scenario. For each row, the $fi$ and $k$
		labels of the amplitudes and widths are suppressed. Amplitudes are in
		$\mathrm{GeV}^{-1/2}$ and widths are in MeV. The OPE-only width is
		independent of $\eta_{\mathrm{match}}$; $\Gamma_+$ and $\Gamma_-$ use the
		globally correlated choices $\eta_{\mathrm{match}}=+1$ and $-1$,
		respectively.}
	\label{tab:ope_reference_results}
	\begin{ruledtabular}
		\begin{tabular}{lclrrrr}
			Initial state & $J$ & Final state
			& $T_{00,\pi,\mathrm{ref}}^{(J)}$
			& $\left|T_{00,\pi,\mathrm{ref}}^{(J)}/
			T_{00,\mathrm{QI}}^{(J)}\right|$
			& $\Gamma_{\pi,\mathrm{ref}}^{(J)}$
			& $\Gamma_+^{(J)}\ ;\ \Gamma_-^{(J)}$ \\
			\colrule
			$P_c(4312)$ & $\tfrac12$ & $\Lambda_c\bar D^*$
			& $0.463698+0.274212\,i$ & $0.581$ & $0.250634$
			& $1.735189\ ;\ 0.250299$ \\
			$P_c(4440)$ & $\tfrac12$ & $\Lambda_c\bar D$
			& $-0.008156+0.000021\,i$ & $0.080$ & $0.000232$
			& $0.030518\ ;\ 0.042079$ \\
			$P_c(4440)$ & $\tfrac12$ & $\Lambda_c\bar D^*$
			& $0.010338-0.002907\,i$ & $0.024$ & $0.000291$
			& $0.472523\ ;\ 0.518729$ \\
			$P_c(4440)$ & $\tfrac32$ & $\Lambda_c\bar D^*$
			& $-0.005169+0.001453\,i$ & $0.040$ & $0.000073$
			& $0.041656\ ;\ 0.048625$ \\
			$P_c(4457)$ & $\tfrac12$ & $\Lambda_c\bar D$
			& $-0.003996$ & $0.233$ & $0.000057$
			& $0.000621\ ;\ 0.001606$ \\
			$P_c(4457)$ & $\tfrac12$ & $\Lambda_c\bar D^*$
			& $0.014743$ & $0.085$ & $0.000582$
			& $0.067689\ ;\ 0.095113$ \\
			$P_c(4457)$ & $\tfrac32$ & $\Lambda_c\bar D^*$
			& $-0.007372$ & $0.131$ & $0.000145$
			& $0.006402\ ;\ 0.010842$ \\
		\end{tabular}
	\end{ruledtabular}
\end{table*}

The reference OPE amplitude is largest relative to QI for
$P_c(4312)\to\Lambda_c\bar D^*$. Its QI width of $0.742~\mathrm{MeV}$
becomes $1.735~\mathrm{MeV}$ in the $\eta_{\mathrm{match}}=+1$ branch and
$0.250~\mathrm{MeV}$ in the $\eta_{\mathrm{match}}=-1$ branch. The leading
OPE amplitude for $P_c(4312)\to\Lambda_c\bar D$ is exactly zero because no
$\bar D\bar D\pi$ axial vertex exists; that channel is therefore omitted
from Table~\ref{tab:ope_reference_results}, and its small QI width is
unchanged. For the other listed transitions,
$|T_{00,\pi,\mathrm{ref}}/T_{00,\mathrm{QI}}|$ ranges from $0.024$ to
$0.233$.

At the reference point, the spin-discriminating pattern survives in both
globally correlated branches. For the peak-to-peak
$\Lambda_c\bar D^*$ ratio, spin ordering I gives $73.8$ for
$\eta_{\mathrm{match}}=+1$ and $47.8$ for
$\eta_{\mathrm{match}}=-1$, whereas spin ordering II gives $0.615$ and
$0.511$, respectively. Thus the reference OPE contribution changes the
absolute widths but does not reverse the QI separation between the two spin
orderings.

\begin{figure*}[t]
	\centering
	\includegraphics[width=\textwidth]{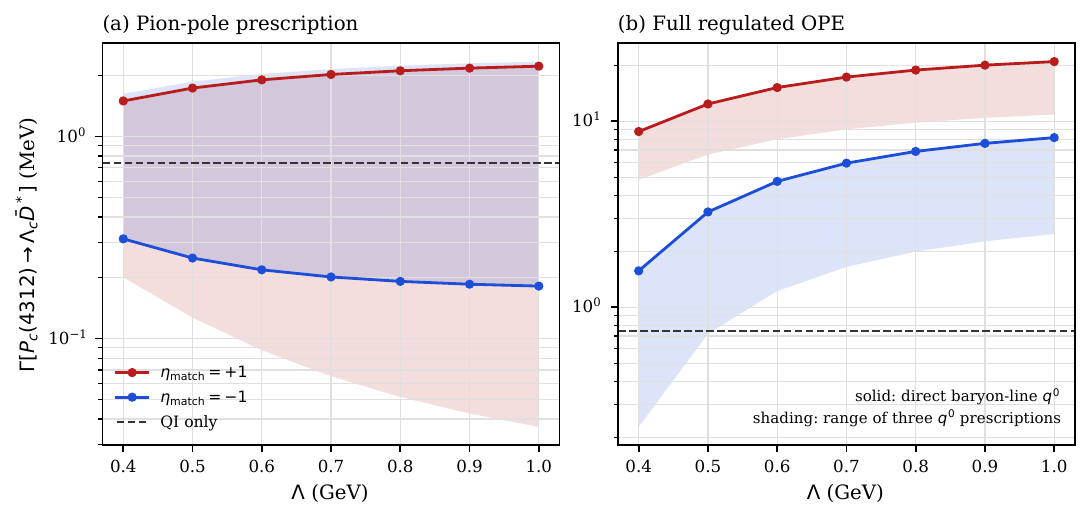}
	\caption{Exploratory cutoff dependence of
		$P_c(4312)\to\Lambda_c\bar D^*$. Solid curves use
		$q_{\pi,\mathrm{dir}}^0$, the shaded regions span
		$q_{\pi,\mathrm{dir}}^0$, $q_{\pi,\mathrm{swap}}^0$, and
		$q_{\pi,\mathrm{stat}}^0$, and the dashed line is the QI-only width.
		Panel (a) uses the pion-pole prescription; panel (b) retains the full
		regulated OPE kernel. The two values of $\eta_{\mathrm{match}}$ are
		globally correlated matching alternatives. Neither the shading nor the
		separation of the two branches is a probability interval.}
	\label{fig:ope_cutoff_pc4312}
\end{figure*}

The reference numbers should not, however, be promoted to unique QI+OPE
predictions. Figure~\ref{fig:ope_cutoff_pc4312} displays the regulator
dependence in the channel for which OPE is most important. Even the pion-pole
prescription depends appreciably on the off-shell continuation chosen for
$q_\pi^0$. Retaining the full regulated kernel produces a much stronger
dependence because its contact-like part has not been matched to the
short-distance QI interaction. For a general diagnostic prescription
$\xi$, we use the coherent formula of Eq.~\eqref{eq:ope_hybrid_amplitude}
with $T_{00,\pi,\mathrm{ref}}^{(J)}$ replaced by
$T_{00,\pi,\xi}^{(J)}$ and denote the resulting branches by
$\Gamma_{\mathrm{tot},\xi,+}$ and $\Gamma_{\mathrm{tot},\xi,-}$. For example,
the full kernel with $q_{\pi,\mathrm{dir}}^0$ and
$\Lambda_\pi=0.50~\mathrm{GeV}$ gives
\begin{equation}
	\begin{aligned}
		\Gamma_{\pi,\mathrm{full}}&=7.06~\mathrm{MeV},\\
		\Gamma_{\mathrm{tot},\mathrm{full},+}&=12.36~\mathrm{MeV},\\
		\Gamma_{\mathrm{tot},\mathrm{full},-}&=3.25~\mathrm{MeV}
	\end{aligned}
	\label{eq:ope_full_pc4312_example}
\end{equation}
for $P_c(4312)\to\Lambda_c\bar D^*$. Across the full cutoff and energy-transfer scan, several pion-pole scenarios
make the OPE amplitude comparable to the QI amplitude, while the
corresponding full-kernel ratios can be considerably larger. This spread diagnoses regulator,
energy-transfer, and short-distance-matching sensitivity; it is not a
statistical uncertainty and does not justify choosing an extreme scenario as
a prediction.

This interpretation is consistent with the general observation that the
short-range part of OPE cannot be separated from contact or quark-exchange
dynamics in a model-independent manner. Regulator-dependent spin assignments
in OPE-based descriptions have been emphasized in
Refs.~\cite{PavonValderrama2019OPE,Du2021}, while a recent quark-exchange study
shows explicitly that the treatment of short-range spin-dependent OPE can
alter the molecular interaction~\cite{Xu2025QuarkExchange}. The pion-pole
subtraction adopted here is therefore a stated matching prescription, not a
derived QI--OPE identity.

\subsection{Comparison with other approaches and experimental implications}
\label{sec:results_comparison}

Wang \textit{et al.} applied the same prior--post QI framework and amplitude
normalization to the hidden-charm $J/\psi p$ and $\eta_c p$ modes
~\cite{Wang2020QI}. Their calculation fixes the constituent interaction and
normalization adopted here, but it does not provide an open-charm numerical
benchmark. We therefore compare with studies that explicitly evaluated the
$\Lambda_c\bar D^{(*)}$ final states, keeping in mind that their dynamics and
normalization prescriptions differ.

\begin{table*}[t]
	\caption{Selected results for
		$P_c(4312)\to\Lambda_c\bar D^{(*)}$. The first four rows quote
		absolute partial widths. The final row gives the restricted lower-channel
		two-body fractions reported in the complex-scaling analysis and must not
		be interpreted as a width in MeV.}
	\label{tab:pc4312_literature_comparison}
	\begin{ruledtabular}
		\begin{tabular}{lccc}
			Approach & $\Lambda_c\bar D$ & $\Lambda_c\bar D^*$
			& Reported observable \\
			\colrule
			Present reference QI & $0.00285$ & $0.742$ & partial width (MeV) \\
			Dong \textit{et al.}~\cite{Dong2020}
			& $0$ & $8.36$ & partial width (MeV) \\
			Lin and Zou~\cite{LinZou2019}
			& $0.06$--$0.30$ & $3.8$--$10.7$ & partial width (MeV) \\
			Li \textit{et al.}~\cite{Li2023BS}
			& $0.026$ & $8.8$ & partial width (MeV) \\
			Lin \textit{et al.}~\cite{Lin2023CSM}
			& $0.04\%$ & $99.96\%$ & restricted two-body fraction \\
		\end{tabular}
	\end{ruledtabular}
\end{table*}

The uncalculated $J^P=3/2^-\to\Lambda_c\bar D$
	transition can be placed on a rough scale by comparison
	with calculations that retain the required rank-two
	dynamics.  Although a $D$-wave width has the threshold
	behavior $\Gamma_D\propto k^5$, its normalization cannot
	be inferred from the present $S$-wave amplitudes because
	the corresponding tensor operator is absent from our QI
	interaction and from the central OPE projection.  Moreover,
	the final momenta, $k=0.784$ and $0.808$~GeV at the
	$P_c(4440)$ and $P_c(4457)$ masses, respectively, are not
	small enough to ensure a strong centrifugal suppression.

The pion-exchange triangle calculation of
	Ref.~\cite{Yang2025}, which explicitly separates the
	final $S$- and $D$-wave components, gives
	\[
	\Gamma_D[P_c(4440),J=3/2\to\Lambda_c\bar D]
	=1.06-1.79~\mathrm{MeV},
	\]
	\[
	\Gamma_D[P_c(4457),J=3/2\to\Lambda_c\bar D]
	=0.51-0.90~\mathrm{MeV}.
	\]
	The Bethe--Salpeter calculation of Ref.~\cite{Li2025BS}
	instead obtains $20.0$ and $16.7$~MeV for the corresponding
	assignments.  These values use different molecular wave
	functions, regulators, and transition mechanisms and should
	not be interpreted as an uncertainty interval for the present
	calculation.  They nevertheless indicate external model estimates spanning approximately $0.5-20$~MeV.  The
	$J^P=3/2^-\to\Lambda_c\bar D$ mode therefore need not be
	negligible, but a quantitative prediction within our
	framework requires inclusion of the tensor QI and OPE
	operators and a complete $L_i=0\to L_f=2$ projection.

The reference QI widths are smaller than the constituent-quark,
effective-Lagrangian, and Bethe--Salpeter values in
Table~\ref{tab:pc4312_literature_comparison}. The pronounced preference for
$\Lambda_c\bar D^*$ is nevertheless common to these calculations. This is
qualitative agreement on the channel hierarchy, not agreement on the
absolute normalization. For example, Li \textit{et al.} obtain $8.8$ and
$0.026~\mathrm{MeV}$ for the $\Lambda_c\bar D^*$ and
$\Lambda_c\bar D$ modes, respectively~\cite{Li2023BS}, compared with the
present reference QI values $0.742$ and $0.00285~\mathrm{MeV}$.

The percentages of Lin \textit{et al.} require a separate interpretation.
Their complex-scaling calculation includes contact interactions, retarded
OPE, and coupled channels, with low-energy constants fitted to the observed
pole properties~\cite{Lin2023CSM}. On-shell three-body effects are partly
included in the total pole width but excluded from the tabulated
lower-channel two-body fractions. Multiplying those percentages by the total
pole width would therefore not yield partial widths directly comparable with
the quantities calculated here. We use them only to compare channel
hierarchies. Their $P_c(4440,J=1/2)$ fractions---$10.8\%$, $38.4\%$, and
$50.9\%$ for $\Lambda_c\bar D$, $\Lambda_c\bar D^*$, and
$\Sigma_c\bar D$, respectively---also illustrate why the conditional
quantity in Eq.~\eqref{eq:selected_channel_fraction} is not an inclusive
branching fraction.

Independent calculations find that the
$\bar D^{(*)}\Lambda_c\pi$ three-body modes can themselves be sizable
~\cite{LiuPanGeng2024}; these channels lie outside the selected two-body
widths considered here.

Several analyses find that open-charm decays can be strongly spin
dependent~\cite{LinZou2019,BurnsSwanson2022,Yang2025,Li2025BS,
	Brambilla2025BOEFT}, but no unique assignment of
the two heavier peaks has emerged. Under the assumption of relatively weak
off-diagonal contact interactions, Lin \textit{et al.}, together with the
triangle-diagram analysis of Yang \textit{et al.}, favor spin ordering I
~\cite{Lin2023CSM,Yang2025}. Other coupled-channel and Bethe--Salpeter
studies favor spin ordering II~\cite{Du2021,Li2025BS,Deng2026}. The present
calculation does not infer the spin ordering from the peak masses. Instead,
Eqs.~\eqref{eq:qi_ordering_a_ratio} and
\eqref{eq:qi_ordering_b_ratio} give distinct conditional decay patterns for
the two hypotheses.

The relevant final states can be studied experimentally in
$\Lambda_b^0\to\Lambda_c^+\bar D^{(*)0}K^-$ decays
~\cite{LHCb2024OpenCharm}. The two channels offer complementary experimental advantages. In the
LHCb study cited above, the
$\Lambda_b^0\to\Lambda_c^+\bar D^0K^-$ mode was fully reconstructed
from charged particles, whereas the $\bar D^{*0}$ mode was treated as
partially reconstructed because the neutral particle in
$\bar D^{*0}\to\bar D^0\pi^0$ or $\bar D^0\gamma$ was not reconstructed.
Full $\bar D^{*0}$ reconstruction can therefore incur an efficiency
penalty from the additional soft neutral particle, while partial
reconstruction provides less complete kinematic information and
requires a more involved signal description.  The
$\Lambda_c\bar D$ mode may consequently provide a cleaner experimental
spectrum even when its calculated $J=1/2$ partial width is smaller.
Conversely, the larger calculated $\Lambda_c\bar D^*$ widths and their
pronounced spin dependence make that mode the sharper discriminator
within the present model.  The actual experimental sensitivity depends
also on production rates, reconstruction efficiencies, resolution,
backgrounds, and interference, and cannot be inferred from the partial
widths alone.

Searches in the two modes are therefore complementary.  Direct tests
would include a search for the $P_c(4312)$ in
$\Lambda_c^+\bar D^{*0}$, searches for the two heavier peaks in both
$\Lambda_c^+\bar D^0$ and $\Lambda_c^+\bar D^{*0}$, and an amplitude
analysis comparing their resonant $\Lambda_c^+\bar D^{*0}$ strengths
while separating nonresonant backgrounds and production-dependent
interference.  These observables probe the stable QI hierarchies
without requiring a preferred value of $\eta_{\rm match}$.

Taken together, the results identify the $\Lambda_c\bar D^*$ hierarchy and
its spin dependence as the robust predictions of the present QI calculation.
The absolute widths remain model dependent, especially where the QI pieces
cancel. The reference OPE contribution preserves the central spin pattern,
whereas the broader OPE scan reveals unresolved cutoff, energy-transfer,
short-distance, and matching-sign dependence. The following section
summarizes these conclusions and their scope.

\section{Summary and conclusions}
\label{sec:summary}

We have studied selected open-charm decays of the $P_c(4312)$,
$P_c(4440)$, and $P_c(4457)$ as predominantly $S$-wave
$\Sigma_c\bar D^{(*)}$ molecules. The $P_c(4312)$ was assigned
$J^P=1/2^-$, while both $J^P=1/2^-$ and $3/2^-$ $\Sigma_c\bar D^*$
assignments were evaluated at each heavier peak mass. Our central results
were obtained for the $S$-wave transitions
$\Sigma_c\bar D^{(*)}\to\Lambda_c\bar D^{(*)}$ at Born order in the QI
model. The prior and post rearrangement diagrams were evaluated with their
color, flavor, spin, and spatial matrix elements, projected onto the physical
$I=1/2$ channels, and folded with the molecular wave functions.

The most robust QI result is the preference for $\Lambda_c\bar D^*$ over
$\Lambda_c\bar D$ for every $J^P=1/2^-$ assignment considered. The central
$\Lambda_c\bar D^*/\Lambda_c\bar D$ width ratios are $260.6$, $13.73$, and
$76.53$ for the $P_c(4312)$, $P_c(4440)$, and $P_c(4457)$, respectively,
and the hierarchy survives all 24 one-at-a-time input variations. These
ratios compare only the two calculated open-charm modes and are not inclusive
branching fractions. Their large values are also not universal HQSS
predictions: they reflect unequal spatial overlaps, hyperfine effects, phase
space, and channel-dependent interference. In particular, the suppressed
$P_c(4312)\to\Lambda_c\bar D$ width is generated by the hyperfine
interaction, whereas the small $P_c(4457,J=1/2)\to\Lambda_c\bar D$ width
results from a model-dependent cancellation.

The $\Lambda_c\bar D^*$ mode also shows a stable spin dependence. At a fixed
heavier peak mass, its central QI width for $J^P=1/2^-$ exceeds that for
$J^P=3/2^-$ by factors of approximately $11.0$ at $4440~\mathrm{MeV}$ and
$9.5$ at $4457~\mathrm{MeV}$; the separation persists throughout the input
scan. Consequently, the two possible spin orderings predict distinct
conditional patterns for the resonant $\Lambda_c\bar D^*$ strengths. A much
larger $P_c(4440)$ partial width is characteristic of spin ordering I,
$J[P_c(4440)]=1/2$ and $J[P_c(4457)]=3/2$, within the present model. For the
reverse ordering, the two strengths can be comparable. This provides a
decay-based discriminator but does not determine the ordering from the peak
masses alone. The $J^P=3/2^-\to\Lambda_c\bar D$ transition begins in a
$D$ wave and was not calculated; it is not predicted to have zero width.

We also explored a direct retarded OPE transition at first Born order. The
contact-subtracted pion-pole term was used as the reference long-distance
prescription, while the full regulated kernel and alternative energy-transfer
choices served as diagnostics. The
$P_c(4312)\to\Lambda_c\bar D^*$ amplitude is the most sensitive to OPE. At
the reference point, both globally correlated matching-sign branches preserve
the principal QI channel and spin hierarchies. The broader scan nevertheless
reveals substantial dependence on the cutoff, energy-transfer prescription,
treatment of the contact-like term, and unresolved relative QI--OPE sign.
The coherent widths are therefore matching-sensitivity scenarios, not unique
predictions or statistical uncertainty bands. A predictive synthesis will
require QI and pion exchange to be matched in a common regulator-consistent
framework that controls double counting and treats rescattering and
three-body cuts consistently.

The $\Lambda_c^+\bar D^0$ and $\Lambda_c^+\bar D^{*0}$ modes are
experimentally complementary: the former can provide a cleaner fully
reconstructed final state, whereas the latter has the larger calculated
$J^P=1/2^-$ widths and the stronger spin sensitivity within the present
model. The most informative experimental program would therefore examine
both final states, with a search for the $P_c(4312)$ in
$\Lambda_c^+\bar D^{*0}$ and a comparison of the resonant
$\Lambda_c^+\bar D^{*0}$ strengths of the two heavier peaks providing
direct tests of the predicted hierarchy and spin dependence. Such
measurements require amplitude analyses that separate resonance couplings
from nonresonant backgrounds, production effects, and interference. The
present conclusions remain conditional on pure-channel Gaussian molecular
states, exact-isospin reconstruction, and a first-Born treatment. Coupled
channels and final-state rescattering, $D$-wave transitions, tensor-induced
$S$--$D$ mixing, finite $\Sigma_c$ and $\bar D^*$ widths, and unitary three-body
dynamics lie beyond the present scope. Within these limitations, the stable
$\Lambda_c\bar D^*$ hierarchy and its pronounced spin dependence make the
selected open-charm modes useful probes of the molecular interpretation of
the narrow $P_c$ states.

\begin{acknowledgments}

Vandan Patel acknowledges the financial assistance by University Grants Commission (UGC) under the CSIR-UGC Junior Research Fellow (JRF) scheme with Ref No. 231610186052.

\end{acknowledgments}

\appendix

\section{Constituent-state and internal-wave-function conventions}
\label{app:qi_cfs}

This appendix specifies the normalized constituent states and phase
conventions used to obtain the color, flavor, and spin factors in
Sec.~\ref{sec:qi}.  We also collect here the internal spatial wave functions
of the external mesons and baryons.  The constituent labels are those of
Eqs.~\eqref{eq:qi_initial_labels}--\eqref{eq:topology15}: the initial state is
$A(12)=D^{(*)-}(\bar c_1d_2)$ and
$B(345)=\Sigma_c^{++}(c_3u_4u_5)$, while the two labelled final
clusterings are $F_{14}=M_f(14)B_f(325)$ and
$F_{15}=M_f(15)B_f(324)$.

\subsection{Internal spatial wave functions}
\label{app:qi_internal_wavefunctions}

For each external hadron, the internal wave function is factorized as
\begin{equation}
	\Psi_H=\chi_H^c\chi_H^f\chi_H^s\phi_H,
	\label{eq:app_hadron_factorization}
\end{equation}
where $\chi_H^c$, $\chi_H^f$, and $\chi_H^s$ denote the color, flavor, and
spin factors, respectively, and $\phi_H$ is the momentum-space spatial wave
function.  Following Ref.~\cite{Wang2020QI}, we use normalized Gaussian
functions for the ground-state mesons and baryons.

For a meson $M(\bar q_\nu q_{\nu'})$ with total momentum
$\mathbf P_M=\mathbf p_\nu+\mathbf p_{\nu'}$, the mass-weighted relative
momentum is
\begin{equation}
	\begin{aligned}
		\mathbf p_{\mathrm{rel}}^{(\nu\nu')}
		&=\frac{m_\nu\mathbf p_{\nu'}-m_{\nu'}\mathbf p_\nu}
		{m_\nu+m_{\nu'}}\\
		&=\mathbf p_{\nu'}-\frac{m_{\nu'}}{m_\nu+m_{\nu'}}\mathbf P_M .
	\end{aligned}
	\label{eq:app_meson_relative_momentum}
\end{equation}
Equivalently,
$\mathbf p_{\mathrm{rel}}=(m_{\bar q}\mathbf p_q
-m_q\mathbf p_{\bar q})/(m_q+m_{\bar q})$.  The normalized $S$-wave
Gaussian is
\begin{equation}
	\phi_M\!\left(\mathbf p_{\mathrm{rel}}\right)
	=\frac{1}{\pi^{3/4}\beta_M^{3/2}}
	\exp\!\left[-\frac{p_{\mathrm{rel}}^2}{2\beta_M^2}\right],
	\label{eq:app_meson_gaussian}
\end{equation}
with
\begin{equation}
	\int d^3p_{\mathrm{rel}}\,
	\left|\phi_M\!\left(\mathbf p_{\mathrm{rel}}\right)\right|^2=1.
	\label{eq:app_meson_normalization}
\end{equation}
Equation~\eqref{eq:app_meson_relative_momentum} is used for both the initial
$D^{(*)-}(12)$ and the final $\bar D^{(*)0}(14)$ or
$\bar D^{(*)0}(15)$, with the appropriate constituent labels and meson width
$\beta_M$.

For an ordered baryon $B(\nu_1\nu_2\nu_3)$, with constituent $\nu_1$
chosen as the first entry, define
$m_{\nu_1\nu_2\nu_3}\equiv
m_{\nu_1}+m_{\nu_2}+m_{\nu_3}$. The mass-weighted Jacobi momenta are
\begin{equation}
	\begin{aligned}
		\mathbf p_\rho^{(\nu_1\nu_2\nu_3)}={}&
		\frac{1}{\sqrt2\,m_{\nu_1\nu_2\nu_3}}
		\Big[(m_{\nu_1}+2m_{\nu_3})\mathbf p_{\nu_2}\\
		&\quad -(m_{\nu_1}+2m_{\nu_2})\mathbf p_{\nu_3}
		+(m_{\nu_3}-m_{\nu_2})\mathbf p_{\nu_1}\Big],\\
		\mathbf p_\lambda^{(\nu_1\nu_2\nu_3)}={}&
		\sqrt{\frac32}\,\frac{1}{m_{\nu_1\nu_2\nu_3}}
		\Big[m_{\nu_1}(\mathbf p_{\nu_2}+\mathbf p_{\nu_3})\\
		&\quad -(m_{\nu_2}+m_{\nu_3})\mathbf p_{\nu_1}\Big].
	\end{aligned}
	\label{eq:app_baryon_jacobi_momenta}
\end{equation}
The corresponding normalized $S$-wave baryon Gaussian is
\begin{equation}
	\begin{aligned}
		\phi_B(\mathbf p_{\nu_1},\mathbf p_{\nu_2},\mathbf p_{\nu_3})
		={}&\frac{3^{3/4}}
		{\pi^{3/2}\alpha_\rho^{3/2}\alpha_\lambda^{3/2}}\\
		&\times\exp\!\left[-\frac{(p_\rho^{(\nu_1\nu_2\nu_3)})^2}{2\alpha_\rho^2}
		-\frac{(p_\lambda^{(\nu_1\nu_2\nu_3)})^2}{2\alpha_\lambda^2}
		\right].
	\end{aligned}
	\label{eq:app_baryon_gaussian}
\end{equation}
At fixed baryon momentum $\mathbf P_B$, its normalization may be written as
\begin{equation}
	\int d^3p_{\nu_1}\,d^3p_{\nu_2}\,
	\left|\phi_B\!\left(\mathbf p_{\nu_1},\mathbf p_{\nu_2},
	\mathbf P_B-\mathbf p_{\nu_1}-\mathbf p_{\nu_2}\right)\right|^2=1.
	\label{eq:app_baryon_normalization}
\end{equation}
At fixed $\mathbf P_B$, the six-dimensional Jacobian associated with
Eq.~\eqref{eq:app_baryon_jacobi_momenta} is
\begin{equation}
	\left|
	\frac{\partial(\mathbf p_\rho,\mathbf p_\lambda)}
	{\partial(\mathbf p_{\nu_1},\mathbf p_{\nu_2})}
	\right|_{\mathbf P_B}
	=3^{3/2},
	\qquad
	d^3p_\rho\,d^3p_\lambda
	=3^{3/2}d^3p_{\nu_1}\,d^3p_{\nu_2} .
	\label{eq:app_baryon_jacobian}
\end{equation}
The factor $3^{3/4}$ in Eq.~\eqref{eq:app_baryon_gaussian} is the corresponding
wave-function normalization factor.  For equal
light-quark masses, $m_{\nu_2}=m_{\nu_3}=m_q$, one has
$\mathbf p_\rho=(\mathbf p_{\nu_2}-\mathbf p_{\nu_3})/\sqrt2$, while
$\mathbf p_\lambda$ describes the motion of the heavy quark relative to the
light pair.

The ordered triplets used here are
\begin{equation}
	\begin{aligned}
		\Sigma_c^{++}:&\quad(\nu_1,\nu_2,\nu_3)=(3,4,5),\\
		\Lambda_c(F_{14}):&\quad(\nu_1,\nu_2,\nu_3)=(3,2,5),\\
		\Lambda_c(F_{15}):&\quad(\nu_1,\nu_2,\nu_3)=(3,2,4).
	\end{aligned}
	\label{eq:app_baryon_orderings}
\end{equation}
Thus the same definition, Eq.~\eqref{eq:app_baryon_jacobi_momenta}, is used
for the initial and final baryons after replacing the ordered constituent
labels.  The central charmed-baryon widths quoted in
Table~\ref{tab:qi_model_parameters} follow the parametrization of
Ref.~\cite{Wang2020QI}; in particular, its central-width prescription is
\begin{equation}
	\alpha_\lambda^2
	=\left(\frac{3m_c}{2m_q+m_c}\right)^{1/2}\alpha_\rho^2,
	\label{eq:app_baryon_width_relation}
\end{equation}
up to the rounding of the tabulated widths.  The internal hadron Gaussians in
Eqs.~\eqref{eq:app_meson_gaussian} and~\eqref{eq:app_baryon_gaussian} are
distinct from the molecular radial wave function $R_0(p)$ in
Eq.~\eqref{eq:molecular_gaussian}.  Diagram-dependent momentum routing and
the reduction of products of four such Gaussians are given in
Appendix~\ref{app:qi_spatial}.

\subsection{Color states}
\label{app:qi_color_states}

Let $a_i\in\{1,2,3\}$ denote the color carried by constituent $i$.  In the
ordered five-constituent basis
$|\bar a_1a_2a_3a_4a_5\rangle$, the initial color state is
\begin{equation}
	\chi_{AB}^c
	=\frac{1}{\sqrt{18}}
	\delta_{a_1a_2}\epsilon_{a_3a_4a_5}
	|\bar a_1a_2a_3a_4a_5\rangle,
	\label{eq:app_initial_color_state}
\end{equation}
where repeated color indices are summed.  The two labelled final color states
are
\begin{align}
	\chi_{F_{14}}^c
	&=\frac{1}{\sqrt{18}}
	\delta_{a_1a_4}\epsilon_{a_3a_2a_5}
	|\bar a_1a_2a_3a_4a_5\rangle,
	\label{eq:app_final14_color_state}\\
	\chi_{F_{15}}^c
	&=\frac{1}{\sqrt{18}}
	\delta_{a_1a_5}\epsilon_{a_3a_2a_4}
	|\bar a_1a_2a_3a_4a_5\rangle.
	\label{eq:app_final15_color_state}
\end{align}
Each state is a product of a normalized meson singlet and a normalized baryon
singlet, and therefore
\begin{equation}
	\langle\chi_{AB}^c|\chi_{AB}^c\rangle
	=\langle\chi_{F_{14}}^c|\chi_{F_{14}}^c\rangle
	=\langle\chi_{F_{15}}^c|\chi_{F_{15}}^c\rangle=1.
	\label{eq:app_color_normalizations}
\end{equation}

For a quark, $t^A=\lambda^A/2$, whereas the generator acting on constituent
1 is the antiquark generator
\begin{equation}
	t_{\bar q}^A=-\left(t_q^A\right)^*.
	\label{eq:app_antiquark_generator}
\end{equation}
The contractions can be reduced with the fundamental $SU(3)$ identity
\begin{equation}
	\sum_{A=1}^{8}(t^A)_{ab}(t^A)_{cd}
	=\frac12\left(\delta_{ad}\delta_{bc}
	-\frac13\delta_{ab}\delta_{cd}\right).
	\label{eq:app_color_fierz}
\end{equation}
For example, inserting Eqs.~\eqref{eq:app_initial_color_state},
\eqref{eq:app_final14_color_state}, and~\eqref{eq:app_antiquark_generator}
gives
\begin{equation}
	\begin{aligned}
		I_{C,\mathrm{pr}}^{14}(13)
		&=\langle\chi_{F_{14}}^c|
		\mathbf t_1\!\cdot\!\mathbf t_3|\chi_{AB}^c\rangle\\
		&=\frac{2}{9}.
	\end{aligned}
	\label{eq:app_example_color_factor}
\end{equation}
Applying the same contraction to the remaining prior and post pairs gives the
color vectors in Table~\ref{tab:qi_color_flavor}.  For every representation
and labelled topology they satisfy
\begin{equation}
	\sum_{\nu\nu'\in\mathcal P_{r,a}}
	I_{C,r}^{a}(\nu\nu')=0,
	\qquad r\in\{\mathrm{pr},\mathrm{po}\},\quad a\in\{14,15\}.
	\label{eq:app_zero_color_sum}
\end{equation}
This identity is the color-singlet cancellation used in the distributional
treatment of the linear-confinement term.

\subsection{Flavor states}
\label{app:qi_flavor_states}

The constituent interaction is diagonal in flavor and does not change the
flavor carried by a labelled constituent line.  The charged initial component
therefore has the normalized flavor state
\begin{equation}
	\chi_{AB}^f=|\bar c_1d_2c_3u_4u_5\rangle.
	\label{eq:app_initial_flavor_state}
\end{equation}
For topology $F_{14}$, the final $\bar D^{(*)0}$ contains labels $(1,4)$ and
the antisymmetric light pair in the $\Lambda_c$ contains labels $(2,5)$.
For $F_{15}$, the corresponding pairs are $(1,5)$ and $(2,4)$.  In the same
ordered five-particle basis, the final flavor states are therefore
\begin{align}
	\chi_{F_{14}}^f
	&=\frac{1}{\sqrt2}\Bigl(
	|\bar c_1d_2c_3u_4u_5\rangle
	-|\bar c_1u_2c_3u_4d_5\rangle\Bigr),
	\label{eq:app_final14_flavor_state}\\
	\chi_{F_{15}}^f
	&=\frac{1}{\sqrt2}\Bigl(
	|\bar c_1d_2c_3u_4u_5\rangle
	-|\bar c_1u_2c_3d_4u_5\rangle\Bigr).
	\label{eq:app_final15_flavor_state}
\end{align}
The relative minus sign in each of
Eqs.~\eqref{eq:app_final14_flavor_state} and
\eqref{eq:app_final15_flavor_state}, together with the sign chosen for its
first ket, fixes the $\Lambda_c$ flavor-state phase convention used here.
A simultaneous overall rephasing of the $\Lambda_c$ state is conventional
and leaves QI-only widths unchanged, but the same phase convention must be
used when the QI amplitude is combined coherently with another transition
mechanism.
The two kets in either final state are orthogonal.  Since only the first ket
in Eqs.~\eqref{eq:app_final14_flavor_state} and
\eqref{eq:app_final15_flavor_state} overlaps with
Eq.~\eqref{eq:app_initial_flavor_state}, one obtains
\begin{equation}
	I_F^{14}=\langle\chi_{F_{14}}^f|\chi_{AB}^f\rangle
	=I_F^{15}=\langle\chi_{F_{15}}^f|\chi_{AB}^f\rangle
	=\frac{1}{\sqrt2}.
	\label{eq:app_flavor_overlap}
\end{equation}
The subsequent antisymmetrization of the two labelled clusterings is a
separate normalization issue and is discussed in
Appendix~\ref{app:qi_topology}.

\subsection{Spin states and recoupling}
\label{app:qi_spin_states}

All spin states are defined in the Condon--Shortley convention. For two
spin-$1/2$ constituents $\nu$ and $\nu'$, we use
\begin{align}
	\chi_{00}^{(\nu\nu')}
	&=\frac{1}{\sqrt2}
	\left(|\uparrow_\nu\downarrow_{\nu'}\rangle
	-|\downarrow_\nu\uparrow_{\nu'}\rangle\right),
	\label{eq:app_spin_singlet}\\
	\chi_{11}^{(\nu\nu')}&=|\uparrow_\nu\uparrow_{\nu'}\rangle,\nonumber\\
	\chi_{10}^{(\nu\nu')}
	&=\frac{1}{\sqrt2}
	\left(|\uparrow_\nu\downarrow_{\nu'}\rangle
	+|\downarrow_\nu\uparrow_{\nu'}\rangle\right),\nonumber\\
	\chi_{1,-1}^{(\nu\nu')}&=|\downarrow_\nu\downarrow_{\nu'}\rangle.
	\label{eq:app_spin_triplet}
\end{align}
Thus
\begin{equation}
	\chi_{\bar D}^{s}(\nu\nu')=\chi_{00}^{(\nu\nu')},\qquad
	\chi_{\bar D^*,m}^{s}(\nu\nu')=\chi_{1m}^{(\nu\nu')}.
	\label{eq:app_meson_spin_states}
\end{equation}

The light pair $(4,5)$ in the initial $\Sigma_c$ is in a spin-1 state and is
coupled with the charm-quark spin to $s_{\Sigma_c}=1/2$:
\begin{equation}
	\chi_{\Sigma_c,m_\Sigma}^{s}(345)
	=\sum_{m_{45},m_3}
	\langle1m_{45},\tfrac12m_3|\tfrac12m_\Sigma\rangle
	\chi_{1m_{45}}^{(45)}\chi_{\frac12m_3}^{(3)}.
	\label{eq:app_sigma_spin_coupling}
\end{equation}
For later reference, the two projections are
\begin{align}
	\chi_{\Sigma_c,+\frac12}^{s}
	&=\sqrt{\frac23}\,\chi_{11}^{(45)}\chi_{\frac12,-\frac12}^{(3)}
	-\frac{1}{\sqrt3}\,\chi_{10}^{(45)}\chi_{\frac12,+\frac12}^{(3)},
	\label{eq:app_sigma_spin_up}\\
	\chi_{\Sigma_c,-\frac12}^{s}
	&=\frac{1}{\sqrt3}\,\chi_{10}^{(45)}\chi_{\frac12,-\frac12}^{(3)}
	-\sqrt{\frac23}\,\chi_{1,-1}^{(45)}\chi_{\frac12,+\frac12}^{(3)}.
	\label{eq:app_sigma_spin_down}
\end{align}
The final $\Lambda_c$ has a spin-zero light pair, so that
\begin{align}
	\chi_{\Lambda_c,m_\Lambda}^{s}(325)
	&=\chi_{00}^{(25)}\chi_{\frac12m_\Lambda}^{(3)},
	\label{eq:app_lambda14_spin}\\
	\chi_{\Lambda_c,m_\Lambda}^{s}(324)
	&=\chi_{00}^{(24)}\chi_{\frac12m_\Lambda}^{(3)}.
	\label{eq:app_lambda15_spin}
\end{align}

The meson and baryon spins are coupled in that order.  If $s_A=0$ or 1 is
the spin of the initial $D^{(*)-}$, the initial five-constituent spin state is
\begin{equation}
	\begin{aligned}
		\chi_{AB}^{s;JM_J}(s_A)
		={}&\sum_{m_A,m_\Sigma}
		\langle s_Am_A,\tfrac12m_\Sigma|JM_J\rangle\\
		&\times\chi_{s_Am_A}^{(12)}
		\chi_{\Sigma_c,m_\Sigma}^{s}(345).
	\end{aligned}
	\label{eq:app_initial_total_spin}
\end{equation}
For $a=14,15$, define
\begin{equation}
	(\nu_{14}^{(M)},\nu_{14}^{(B)})=(4,5),\qquad
	(\nu_{15}^{(M)},\nu_{15}^{(B)})=(5,4).
	\label{eq:app_topology_light_labels}
\end{equation}
The corresponding final spin state, with final-meson spin
$s_{M_f}=0$ or 1, is
\begin{equation}
	\begin{aligned}
		\chi_{F_a}^{s;JM_J}(s_{M_f})
		={}&\sum_{m_M,m_\Lambda}
		\langle s_{M_f}m_M,\tfrac12m_\Lambda|JM_J\rangle\\
		&\times\chi_{s_{M_f}m_M}^{(1\nu_a^{(M)})}
		\chi_{\Lambda_c,m_\Lambda}^{s}(32\nu_a^{(B)}).
	\end{aligned}
	\label{eq:app_final_total_spin}
\end{equation}

The change from the initial to the final coupling tree can be organized with
Wigner symbols rather than evaluated projection by projection. We write
$\widehat j\equiv\sqrt{2j+1}$. The elementary three-angular-momentum
recoupling relation is
\begin{equation}
	\begin{aligned}
		&|((j_1j_2)j_{12},j_3)JM_J\rangle\\
		&\quad=\sum_{j_{23}}(-1)^{j_1+j_2+j_3+J}
		\widehat j_{12}\widehat j_{23}\\
		&\qquad\times
		\begin{Bmatrix}
			j_1&j_2&j_{12}\\
			j_3&J&j_{23}
		\end{Bmatrix}
		|(j_1,(j_2j_3)j_{23})JM_J\rangle,
	\end{aligned}
	\label{eq:app_wigner_6j_recoupling}
\end{equation}
while the four-spin change of pairings is
\begin{equation}
	\begin{aligned}
		&\langle[(j_1j_3)j_{13}(j_2j_4)j_{24}]KM_K|
		[(j_1j_2)j_{12}(j_3j_4)j_{34}]KM_K\rangle\\
		&\quad=\widehat j_{12}\widehat j_{34}
		\widehat j_{13}\widehat j_{24}\\
		&\qquad\times
		\begin{Bmatrix}
			j_1&j_2&j_{12}\\
			j_3&j_4&j_{34}\\
			j_{13}&j_{24}&K
		\end{Bmatrix}.
	\end{aligned}
	\label{eq:app_wigner_9j_recoupling}
\end{equation}
For $F_{14}$, Eq.~\eqref{eq:app_wigner_9j_recoupling} is applied to the
four-spin rearrangement
$(12)(45)\to(14)(25)$; for $F_{15}$ it is applied after
$4\leftrightarrow5$. Because the charm spin $s_3$ is coupled inside both
baryons, the complete open-charm overlap is a five-spin recoupling: one uses
Eq.~\eqref{eq:app_wigner_6j_recoupling} to couple $s_3$ last, performs the
four-spin transformation in Eq.~\eqref{eq:app_wigner_9j_recoupling}, and
recouples $s_3$ to the final baryon. Thus the result is a product and sum of
$6j$ and $9j$ coefficients, not a single four-spin $9j$ coefficient. The
explicit states above fix all pair-ordering phases and provide a direct
check of the Wigner-symbol evaluation.

For a spin-1 meson and a spin-$1/2$ baryon, the two couplings needed in this
work may be displayed at $M_J=+1/2$ as
\begin{align}
	[\chi_1\chi_{\frac12}]_{\frac12,+\frac12}
	&=\sqrt{\frac23}\,\chi_{11}\chi_{\frac12,-\frac12}
	-\frac{1}{\sqrt3}\,\chi_{10}\chi_{\frac12,+\frac12},
	\label{eq:app_channel_spin_half}\\
	[\chi_1\chi_{\frac12}]_{\frac32,+\frac12}
	&=\frac{1}{\sqrt3}\,\chi_{11}\chi_{\frac12,-\frac12}
	+\sqrt{\frac23}\,\chi_{10}\chi_{\frac12,+\frac12}.
	\label{eq:app_channel_spin_threehalf}
\end{align}
Equations~\eqref{eq:app_initial_total_spin} and
\eqref{eq:app_final_total_spin} define normalized states for any $M_J$. We use
$M_J=+1/2$ to evaluate the scalar overlaps; rotational invariance makes the
result independent of this convenient choice.

\subsection{Representative spin and combined factors}
\label{app:qi_algebraic_factors}

For the $\mathrm{CL}$ interaction, $\widehat O_{\mathrm{CL}}=1$, so the
spin factor is the direct overlap of Eqs.~\eqref{eq:app_initial_total_spin}
and~\eqref{eq:app_final_total_spin}. For the $\mathrm{HF}$ term, we use
\begin{equation}
	\mathbf s_\nu\!\cdot\!\mathbf s_{\nu'}
	=s_\nu^zs_{\nu'}^z
	+\frac12\left(s_\nu^+s_{\nu'}^-+s_\nu^-s_{\nu'}^+\right).
	\label{eq:app_spin_dot_operator}
\end{equation}
Equivalently, in a basis of definite pair spin $s_{\nu\nu'}$,
\begin{equation}
	\mathbf s_\nu\!\cdot\!\mathbf s_{\nu'}
	=\frac12\left[s_{\nu\nu'}(s_{\nu\nu'}+1)-\frac32\right],
	\label{eq:app_spin_dot_eigenvalue}
\end{equation}
with eigenvalues $-3/4$ and $+1/4$ for $s_{\nu\nu'}=0$ and 1,
respectively. Here $\mathbf s_\nu$ is the spin-$1/2$ operator used in
Eq.~\eqref{eq:qi_constituent_potential}; it is related to the Pauli matrix by
$\boldsymbol\sigma_\nu=2\mathbf s_\nu$. Therefore no additional factor of
4 multiplies the hyperfine entries below. For a chosen interacting pair,
Eq.~\eqref{eq:app_wigner_6j_recoupling} may be used to expose
$s_{\nu\nu'}$ before applying Eq.~\eqref{eq:app_spin_dot_eigenvalue}.

As a compact phase-convention check, the spin-independent overlaps for
topology $F_{14}$ are
\begin{equation}
	\begin{array}{c|c|c}
		\text{Transition} & J & I_{S,\mathrm{CL}}^{14,J} \\
		\hline
		\Sigma_c\bar D\to\Lambda_c\bar D
		& \tfrac12 & 0 \\
		\Sigma_c\bar D\to\Lambda_c\bar D^*
		& \tfrac12 & -\tfrac12 \\
		\Sigma_c\bar D^*\to\Lambda_c\bar D
		& \tfrac12 & -\tfrac12 \\
		\Sigma_c\bar D^*\to\Lambda_c\bar D^*
		& \tfrac12 & 1/\sqrt3 \\
		\Sigma_c\bar D^*\to\Lambda_c\bar D^*
		& \tfrac32 & -1/(2\sqrt3)
	\end{array}
	\label{eq:app_spin_independent_overlaps}
\end{equation}
These are the identity-operator entries of
Table~\ref{tab:qi_spin_factors}.  The hyperfine entries follow by inserting
Eq.~\eqref{eq:app_spin_dot_operator} for each ordered interacting pair.  For
example, for
$\Sigma_c\bar D\to\Lambda_c\bar D$ in topology $F_{14}$,
\begin{equation}
	I_{S,\mathrm{HF},\mathrm{pr}}^{14,1/2}(13)
	=\langle\chi_{F_{14}}^{s;1/2,1/2}|
	\mathbf s_1\!\cdot\!\mathbf s_3
	|\chi_{AB}^{s;1/2,1/2}\rangle
	=\frac{\sqrt3}{8}.
	\label{eq:app_example_spin_factor}
\end{equation}

Combining Eqs.~\eqref{eq:app_example_color_factor},
\eqref{eq:app_flavor_overlap}, and~\eqref{eq:app_example_spin_factor} with
the single prior--post averaging factor gives
\begin{equation}
	\begin{aligned}
		\kappa_{\mathrm{pr},14,13,\mathrm{HF}}^{(1/2)}
		&=\frac12\left(\frac{2}{9}\right)
		\left(\frac{1}{\sqrt2}\right)
		\left(\frac{\sqrt3}{8}\right)\\
		&=\frac{\sqrt6}{144}.
	\end{aligned}
	\label{eq:app_example_combined_factor}
\end{equation}
The factor $1/2$ in Eq.~\eqref{eq:app_example_combined_factor} is the
arithmetic prior--post average and is included exactly once.  The remaining
entries in Tables~\ref{tab:qi_color_flavor} and
\ref{tab:qi_spin_factors} follow from the same normalized states and
contractions.  More explicitly, let $P_{45}$ interchange the constituent
labels $4$ and $5$ in an ordered interacting pair.  The factors of topology
$F_{15}$ are then related to those of $F_{14}$ by
\begin{align}
	I_{C,r}^{15}(\nu\nu')
	&=-I_{C,r}^{14}\!\left(P_{45}(\nu\nu')\right),
	\label{eq:app_f15_color_mapping}\\
	I_F^{15}&=I_F^{14},
	\label{eq:app_f15_flavor_mapping}\\
	I_{S,X,r}^{15,J}(\nu\nu')
	&=I_{S,X,r}^{14,J}\!\left(P_{45}(\nu\nu')\right),
	\label{eq:app_f15_spin_mapping}
\end{align}
where $r\in\{\mathrm{pr},\mathrm{po}\}$ and
$X\in\{\mathrm{CL},\mathrm{HF}\}$. Together with the corresponding
spatial mapping in Appendix~\ref{app:qi_spatial}, these relations generate
all $F_{15}$ entries from the $F_{14}$ calculation.  The normalization and
relative sign with which the two labelled clusterings form the physical
identical-quark final state are a separate issue and are derived in
Appendix~\ref{app:qi_topology}.

\section{Momentum routing and Gaussian spatial matrix elements}
\label{app:qi_spatial}

This appendix gives the momentum routing and analytic Gaussian reduction used
for the spatial factors in the quark-interchange (QI) kernel.  The normalized
meson and baryon wave functions, including the mass-weighted Jacobi momenta,
are defined in Appendix~\ref{app:qi_internal_wavefunctions}.  Here we adapt the
momentum-space construction of Ref.~\cite{Wang2020QI} to the open-charm
clusterings
$A(12)+B(345)\to M_f(1\nu_a^{(M)})+B_f(32\nu_a^{(B)})$, with
\begin{equation}
	(\nu_{14}^{(M)},\nu_{14}^{(B)})=(4,5),\qquad
	(\nu_{15}^{(M)},\nu_{15}^{(B)})=(5,4).
	\label{eq:appB_topology_labels}
\end{equation}
The derivation applies to both $M_f=\bar D$ and $\bar D^*$ because their
spatial wave functions differ only through the Gaussian width.

\subsection{External momenta and constituent routing}
\label{app:qi_momentum_routing}

We work in the two-hadron center-of-mass frame and assign
\begin{equation}
	\begin{aligned}
		\mathbf P_A&=\mathbf p,&
		\mathbf P_B&=-\mathbf p,\\
		\mathbf P_{M_f}&=\mathbf k,&
		\mathbf P_{B_f}&=-\mathbf k.
	\end{aligned}
	\label{eq:appB_external_momenta}
\end{equation}
For an ordered interaction $\nu\nu'$, the transferred momentum
$\mathbf q_{\nu\nu'}$ is delivered to constituent $\nu$ and removed from
constituent $\nu'$:
\begin{equation}
	\delta\mathbf p_n^{(\nu\nu')}
	=\mathbf q_{\nu\nu'}
	(\delta_{n\nu}-\delta_{n\nu'}),
	\qquad
	\mathbf p_n'=\mathbf p_n+\delta\mathbf p_n^{(\nu\nu')}.
	\label{eq:appB_momentum_shift}
\end{equation}
Thus the order of the labels matters; for example, $43$ means
$\delta\mathbf p_4=+\mathbf q_{43}$ and
$\delta\mathbf p_3=-\mathbf q_{43}$. The ordered pair sets are
\begin{align}
	\mathcal P_{\mathrm{pr},14}=\mathcal P_{\mathrm{pr},15}
	&=\{13,14,15,23,24,25\},
	\label{eq:appB_prior_pairs}\\
	\mathcal P_{\mathrm{po},14}
	&=\{13,12,15,43,42,45\},
	\label{eq:appB_post14_pairs}\\
	\mathcal P_{\mathrm{po},15}
	&=\{13,12,14,53,52,54\}.
	\label{eq:appB_post15_pairs}
\end{align}
Equations~\eqref{eq:appB_momentum_shift}--\eqref{eq:appB_post15_pairs}
therefore specify the momentum transfer in every prior and post diagram.

It is convenient to use $\mathbf p_3$ and $\mathbf p_4$ as independent
internal momenta.  Momentum conservation in the initial baryon gives
\begin{equation}
	\mathbf p_5=-\mathbf p-\mathbf p_3-\mathbf p_4.
	\label{eq:appB_p5_routing}
\end{equation}
The condition
$\mathbf p_1'+\mathbf p_{\nu_a^{(M)}}'=\mathbf k$ for the final meson then fixes
the remaining initial momenta:
\begin{equation}
	\begin{aligned}
		\mathbf p_1={}&\mathbf k-\mathbf p_{\nu_a^{(M)}}
		-\delta\mathbf p_1^{(\nu\nu')}
		-\delta\mathbf p_{\nu_a^{(M)}}^{(\nu\nu')},\\
		\mathbf p_2={}&\mathbf p-\mathbf p_1.
	\end{aligned}
	\label{eq:appB_general_routing}
\end{equation}
These assignments satisfy
\begin{equation}
	\begin{gathered}
		\mathbf p_1+\mathbf p_2=\mathbf p,
		\qquad
		\mathbf p_3+\mathbf p_4+\mathbf p_5=-\mathbf p,\\
		\mathbf p_1'+\mathbf p_{\nu_a^{(M)}}'=\mathbf k,
		\qquad
		\mathbf p_3'+\mathbf p_2'+\mathbf p_{\nu_a^{(B)}}'=-\mathbf k,
	\end{gathered}
	\label{eq:appB_cluster_momentum_checks}
\end{equation}
for every pair in Eqs.~\eqref{eq:appB_prior_pairs}--
\eqref{eq:appB_post15_pairs}.  Consequently, no diagram-specific momentum
constraint is needed beyond the ordered shift rule in
Eq.~\eqref{eq:appB_momentum_shift}.

The meson relative momenta following from
Eq.~\eqref{eq:app_meson_relative_momentum} are
\begin{equation}
	\begin{aligned}
		\mathbf a_A&=\mathbf p_2-\xi_A\mathbf p,
		&\xi_A&=\frac{m_2}{m_1+m_2},\\
		\mathbf a_f&=\mathbf p_{\nu_a^{(M)}}'-\xi_f\mathbf k,
		&\xi_f&=\frac{m_{\nu_a^{(M)}}}{m_1+m_{\nu_a^{(M)}}}.
	\end{aligned}
	\label{eq:appB_meson_arguments}
\end{equation}
Equivalently, $1-\xi_A=m_1/(m_1+m_2)$ is the mass fraction denoted by
$f_A$ in Ref.~\cite{Wang2020QI}.  The initial and final baryon arguments are
the ordered triplets $(345)$ and $(32\nu_a^{(B)})$, respectively, as fixed in
Eq.~\eqref{eq:app_baryon_orderings}.

With these conventions, the spatial factor associated with the scalar
potential $v_{\nu\nu'}^{X}(q_{\nu\nu'})$, where
$X\in\{\mathrm{CL},\mathrm{HF}\}$, is
\begin{equation}
	\begin{aligned}
		I_{\mathrm{space},X}^{a,\nu\nu'}(\mathbf k,\mathbf p)
		={}&\int d^3q_{\nu\nu'}\,d^3p_3\,d^3p_4\;
		v_{\nu\nu'}^{X}(q_{\nu\nu'})\,
		\phi_A(\mathbf a_A)\\
		&\times\phi_B(\mathbf p_3,\mathbf p_4,\mathbf p_5)
		\phi_{M_f}^{*}(\mathbf a_f)\\
		&\times\phi_{B_f}^{*}
		(\mathbf p_3',\mathbf p_2',\mathbf p_{\nu_a^{(B)}}').
	\end{aligned}
	\label{eq:appB_general_spatial_integral}
\end{equation}
The representation label $r\in\{\mathrm{pr},\mathrm{po}\}$ retained in
Sec.~\ref{sec:qi} is a bookkeeping label.  For
$\nu\nu'\in\mathcal P_{r,a}$, we use the identification
\begin{equation}
	I_{\mathrm{space},X}^{r,a,\nu\nu'}
	\equiv I_{\mathrm{space},X}^{a,\nu\nu'}.
	\label{eq:appB_representation_bookkeeping}
\end{equation}
The spatial matrix element is fixed by the topology and the ordered
interacting pair.  The prior--post distinction enters through the pair set
and through the coefficient $\kappa_{r,a,\nu\nu',X}^{(J)}$.
The integration measure in Eq.~\eqref{eq:appB_general_spatial_integral}
follows the convention of Ref.~\cite{Wang2020QI}; no additional
$(2\pi)^{-3}$ factor is included in the internal overlap.

As a worked example, consider the ordered pair $13$ in topology $F_{14}$.
In this case
$\delta\mathbf p_1=\mathbf q_{13}$,
$\delta\mathbf p_3=-\mathbf q_{13}$, and all other shifts vanish. Equations
\eqref{eq:appB_p5_routing} and~\eqref{eq:appB_general_routing} give
\begin{equation}
	\begin{aligned}
		\mathbf p_1&=\mathbf k-\mathbf p_4-\mathbf q_{13},\\
		\mathbf p_2&=\mathbf p-\mathbf k+\mathbf p_4+\mathbf q_{13},\\
		\mathbf p_3'&=\mathbf p_3-\mathbf q_{13},
		&\mathbf p_1'&=\mathbf k-\mathbf p_4,\\
		\mathbf p_2'&=\mathbf p_2,
		&\mathbf p_4'&=\mathbf p_4,
		\qquad \mathbf p_5'=\mathbf p_5.
	\end{aligned}
	\label{eq:appB_example13_momenta}
\end{equation}
The corresponding overlap is
\begin{equation}
	\begin{aligned}
		I_{\mathrm{space},X}^{14,13}
		={}&\int d^3q_{13}\,d^3p_3\,d^3p_4\;
		v_{13}^{X}(q_{13})\\
		&\times\phi_A\!\left[(1-\xi_A)\mathbf p-\mathbf k
		+\mathbf p_4+\mathbf q_{13}\right]\\
		&\times\phi_B\!\left(\mathbf p_3,\mathbf p_4,
		-\mathbf p-\mathbf p_3-\mathbf p_4\right)\\
		&\times\phi_{M_f}^{*}\!\left(\mathbf p_4-\xi_f\mathbf k\right)\\
		&\times\phi_{B_f}^{*}\!\left(
		\mathbf p_3-\mathbf q_{13},
		\mathbf p-\mathbf k+\mathbf p_4+\mathbf q_{13},
		\mathbf p_5\right).
	\end{aligned}
	\label{eq:appB_example13_integral}
\end{equation}
All remaining overlaps follow by changing the ordered pair in
Eq.~\eqref{eq:appB_momentum_shift}; the second topology follows at the same
time from $\nu_{14}^{(M)}\leftrightarrow\nu_{15}^{(M)}$ and
$\nu_{14}^{(B)}\leftrightarrow\nu_{15}^{(B)}$.

\subsection{Quadratic form of the four-hadron overlap}
\label{app:qi_gaussian_reduction}

For the Gaussian reduction, the ordered pair is held fixed and we write
$\mathbf q\equiv\mathbf q_{\nu\nu'}$ to avoid repeating its suffix. The
product in Eq.~\eqref{eq:appB_general_spatial_integral} contains six
three-vector Gaussian arguments. We collect them as
\begin{equation}
	\begin{aligned}
		\mathbf z_1&=\mathbf a_A,&
		\mathbf z_2&=\mathbf p_{\rho,B},&
		\mathbf z_3&=\mathbf p_{\lambda,B},\\
		\mathbf z_4&=\mathbf a_f,&
		\mathbf z_5&=\mathbf p_{\rho,B_f},&
		\mathbf z_6&=\mathbf p_{\lambda,B_f},
	\end{aligned}
	\label{eq:appB_six_gaussian_arguments}
\end{equation}
with the associated widths
\begin{equation}
	(\omega_1,\ldots,\omega_6)
	=\left(\beta_A,\alpha_{\rho,B},\alpha_{\lambda,B},
	\beta_f,\alpha_{\rho,B_f},\alpha_{\lambda,B_f}\right).
	\label{eq:appB_six_widths}
\end{equation}
Every argument is an affine function of the nine integration components.
Introducing
\begin{equation}
	\mathbf x=
	\begin{pmatrix}
		\mathbf q\\ \mathbf p_3\\ \mathbf p_4
	\end{pmatrix},
	\qquad
	\mathbf z_h={\sf L}_h\mathbf x+\mathbf d_h,
	\qquad h=1,\ldots,6,
	\label{eq:appB_affine_arguments}
\end{equation}
the $3\times9$ matrices ${\sf L}_h$ follow directly from
Eqs.~\eqref{eq:appB_momentum_shift}--\eqref{eq:appB_meson_arguments} and the
Jacobi definitions in Eq.~\eqref{eq:app_baryon_jacobi_momenta}. The vectors
$\mathbf d_h$ contain only the fixed external momenta $\mathbf p$ and
$\mathbf k$.  Thus the product of the four normalized hadron wave functions
can be written as
\begin{equation}
	\begin{aligned}
		\Phi_{a,\nu\nu'}(\mathbf x)
		&\equiv\phi_A\phi_B\phi_{M_f}^{*}\phi_{B_f}^{*}\\
		&=\mathcal N_4
		\exp\!\left[-\frac12\mathbf x^{T}{\sf A}\mathbf x
		+\boldsymbol\eta^{T}\mathbf x-\frac12C\right],
	\end{aligned}
	\label{eq:appB_gaussian_product}
\end{equation}
where $\mathcal N_4$ is the product of the two meson and two baryon
normalization constants, and
\begin{equation}
	\begin{aligned}
		{\sf A}&=\sum_{h=1}^{6}
		\frac{{\sf L}_h^{T}{\sf L}_h}{\omega_h^2},\\
		\boldsymbol\eta&=-\sum_{h=1}^{6}
		\frac{{\sf L}_h^{T}\mathbf d_h}{\omega_h^2},\\
		C&=\sum_{h=1}^{6}\frac{\mathbf d_h^2}{\omega_h^2}.
	\end{aligned}
	\label{eq:appB_gaussian_matrix_definitions}
\end{equation}
The topology and pair labels on ${\sf A}$, $\boldsymbol\eta$, and $C$ are
suppressed in Eqs.~\eqref{eq:appB_gaussian_product} and
\eqref{eq:appB_gaussian_matrix_definitions}; $\boldsymbol\eta$ and $C$ also
depend on the external momenta.  For the routings used here, the matrix
obtained by stacking ${\sf L}_1,\ldots,{\sf L}_6$ has full column rank.
Consequently, positive Gaussian widths make ${\sf A}$ real, symmetric, and
positive definite.

For the explicit $F_{14}$, $13$ example, the six affine arguments entering
Eq.~\eqref{eq:appB_affine_arguments} are obtained from
\begin{equation}
	\begin{gathered}
		\mathbf a_A=(1-\xi_A)\mathbf p-\mathbf k
		+\mathbf p_4+\mathbf q_{13},
		\qquad
		\mathbf a_f=\mathbf p_4-\xi_f\mathbf k,\\
		(\mathbf p_3,\mathbf p_4,\mathbf p_5)_B,
		\qquad
		(\mathbf p_3-\mathbf q_{13},\mathbf p_2,\mathbf p_5)_{B_f},
	\end{gathered}
	\label{eq:appB_example_affine_arguments}
\end{equation}
where the last two ordered triplets are converted to their $\rho$ and
$\lambda$ momenta with Eq.~\eqref{eq:app_baryon_jacobi_momenta}.  This gives
the matrices in Eq.~\eqref{eq:appB_gaussian_matrix_definitions} without any
additional momentum convention.

\subsection{Analytic hyperfine overlap}
\label{app:qi_hyperfine_gaussian}

The scalar hyperfine potential for the fixed pair is
\begin{equation}
	\begin{aligned}
		v_{\nu\nu'}^{\mathrm{HF}}(q)
		&=-\frac{8\pi\alpha_{\nu\nu'}}{3m_\nu m_{\nu'}}
		\exp\!\left(-\frac{q^2}{4\tau_{\mathrm{hf}}^2}\right),\\
		\alpha_{\nu\nu'}&\equiv
		\alpha_s\!\left(\mathcal Q_{\nu\nu'}^2\right),
		\qquad
		\mathcal Q_{\nu\nu'}^2=(m_\nu+m_{\nu'})^2.
	\end{aligned}
	\label{eq:appB_hyperfine_scalar_potential}
\end{equation}
Let
\begin{equation}
	{\sf P}_q=
	\begin{pmatrix}
		{\sf 1}_3&0\\0&0_6
	\end{pmatrix},
	\qquad
	\widetilde{\sf A}
	={\sf A}+\frac{{\sf P}_q}{2\tau_{\mathrm{hf}}^2}.
	\label{eq:appB_hyperfine_matrix}
\end{equation}
The smearing exponential is then absorbed into the same quadratic form, and
the complete nine-dimensional integral is analytic:
\begin{equation}
	\begin{aligned}
		I_{\mathrm{space},\mathrm{HF}}^{a,\nu\nu'}
		={}&-\frac{8\pi\alpha_{\nu\nu'}}{3m_\nu m_{\nu'}}\,
		\mathcal N_4
		\frac{(2\pi)^{9/2}}{\sqrt{\det\widetilde{\sf A}}}\\
		&\times\exp\!\left[-\frac12C
		+\frac12\boldsymbol\eta^{T}
		\widetilde{\sf A}^{-1}\boldsymbol\eta\right].
	\end{aligned}
	\label{eq:appB_hyperfine_closed_form}
\end{equation}
The color, flavor, and spin factors are not included in
Eq.~\eqref{eq:appB_hyperfine_closed_form}; they multiply this spatial result
through $\kappa_{r,a,\nu\nu',\mathrm{HF}}^{(J)}$ defined in
Eq.~\eqref{eq:qi_combined_coefficient}.

\subsection{Schur reduction of the central terms}
\label{app:qi_central_gaussian}

For the Coulomb and linear-confinement terms, it is useful to retain
$\mathbf q$ and integrate analytically over
$\mathbf y=(\mathbf p_3,\mathbf p_4)^T$.  Partition the matrix and vector in
Eq.~\eqref{eq:appB_gaussian_product} as
\begin{equation}
	{\sf A}=
	\begin{pmatrix}
		{\sf A}_{qq}&{\sf A}_{qy}\\
		{\sf A}_{yq}&{\sf A}_{yy}
	\end{pmatrix},
	\qquad
	\boldsymbol\eta=
	\begin{pmatrix}
		\boldsymbol\eta_q\\ \boldsymbol\eta_y
	\end{pmatrix},
	\label{eq:appB_block_partition}
\end{equation}
where the $q$ block has dimension three and the $y$ block has dimension six.
The Gaussian integration over $\mathbf y$ gives the reduced form factor
\begin{equation}
	\mathcal G_{a,\nu\nu'}(\mathbf q;\mathbf k,\mathbf p)
	=\mathcal C_{a,\nu\nu'}
	\exp\!\left[-\frac12\mathbf q^{T}{\sf G}_{a,\nu\nu'}\mathbf q
	+\mathbf h_{a,\nu\nu'}^{T}\mathbf q\right],
	\label{eq:appB_reduced_form_factor}
\end{equation}
with the Schur complement
\begin{equation}
	\begin{aligned}
		{\sf G}_{a,\nu\nu'}
		&={\sf A}_{qq}
		-{\sf A}_{qy}{\sf A}_{yy}^{-1}{\sf A}_{yq},\\
		\mathbf h_{a,\nu\nu'}
		&=\boldsymbol\eta_q
		-{\sf A}_{qy}{\sf A}_{yy}^{-1}\boldsymbol\eta_y,
	\end{aligned}
	\label{eq:appB_schur_parameters}
\end{equation}
and
\begin{equation}
	\begin{aligned}
		\mathcal C_{a,\nu\nu'}
		={}&\mathcal N_4\,
		\frac{(2\pi)^3}{\sqrt{\det{\sf A}_{yy}}}\\
		&\times\exp\!\left[-\frac12C
		+\frac12\boldsymbol\eta_y^{T}
		{\sf A}_{yy}^{-1}\boldsymbol\eta_y\right].
	\end{aligned}
	\label{eq:appB_reduced_prefactor}
\end{equation}
The individual Coulomb spatial integral is therefore
\begin{equation}
	I_{\mathrm{space},\mathrm{C}}^{a,\nu\nu'}
	=4\pi\alpha_{\nu\nu'}
	\int d^3q\,\frac{\mathcal G_{a,\nu\nu'}(\mathbf q)}{q^2}.
	\label{eq:appB_coulomb_reduced_integral}
\end{equation}
For the angular integrations one may choose
\begin{equation}
	\mathbf k=(0,0,k),\qquad
	\mathbf p=\bigl(p\sqrt{1-u^2},0,pu\bigr),
	\qquad u=\widehat{\mathbf k}\!\cdot\!\widehat{\mathbf p}.
	\label{eq:appB_external_axis_choice}
\end{equation}
The remaining measure is
$d^3q=q^2dq\,d(\cos\theta_q)\,d\phi_q$.  Hence the Coulomb singularity is
integrable at the origin, while the Gaussian form factor suppresses the
large-$q$ region.  The radial and angular ranges used in the evaluation are
specified together with the convergence tests in
Sec.~\ref{sec:validation} and Appendix~\ref{app:qi_numerics}.

\subsection{Distributional treatment of linear confinement}
\label{app:qi_confinement_distribution}

The factor $1/q^4$ is a distribution rather than an ordinary integrable
function in three dimensions.  Its regulated action on a smooth test function
$F(\mathbf q)$ may be written as
\begin{equation}
	\begin{aligned}
		\left\langle\left[\frac{1}{q^4}\right]_{\!R},F\right\rangle
		\equiv{}&\lim_{\epsilon\to0^+}\left[
		\int d^3q\,\frac{F(\mathbf q)}{(q^2+\epsilon^2)^2}
		-\frac{\pi^2}{\epsilon}F(\mathbf0)\right]\\
		={}&\int d^3q\,
		\frac{F(\mathbf q)-F(\mathbf0)}{q^4},
	\end{aligned}
	\label{eq:appB_distribution_definition}
\end{equation}
where the last expression is understood with symmetric angular integration.
Indeed, its angular average obeys
\begin{equation}
	\frac{1}{4\pi}\int d\Omega_q
	\bigl[F(q\widehat{\mathbf q})-F(\mathbf0)\bigr]
	=\mathcal O(q^2),
	\label{eq:appB_small_q_angular_average}
\end{equation}
which makes the radial integral finite at $q=0$.

In the QI amplitude, we apply the distribution after forming the complete
spin-independent, color-weighted diagram sum.  This organization makes the
color-neutrality cancellation explicit and is numerically stable.  Applying
the same regulated distribution to the separate diagram contributions is
algebraically equivalent, provided that the subtraction is retained
consistently.  For representation $r$ and topology $a$, define
\begin{equation}
	\begin{aligned}
		\mathcal F_{r,a}^{(J)}(\mathbf q)
		&=\sum_{\nu\nu'\in\mathcal P_{r,a}}
		\kappa_{r,a,\nu\nu',\mathrm{CL}}^{(J)}
		\mathcal G_{a,\nu\nu'}(\mathbf q),\\
		\mathcal F_{r,a}^{\alpha,(J)}(\mathbf q)
		&=\sum_{\nu\nu'\in\mathcal P_{r,a}}
		\kappa_{r,a,\nu\nu',\mathrm{CL}}^{(J)}
		\alpha_{\nu\nu'}
		\mathcal G_{a,\nu\nu'}(\mathbf q).
	\end{aligned}
	\label{eq:appB_weighted_form_factors}
\end{equation}
At $\mathbf q=\mathbf0$, the momentum shifts vanish, so the individual
overlaps become independent of the interacting pair within a fixed topology:
\begin{equation}
	\mathcal G_{a,\nu\nu'}(\mathbf0)
	=\mathcal G_a^{(0)}
	\qquad (\nu\nu'\in\mathcal P_{r,a}).
	\label{eq:appB_common_zero_transfer_overlap}
\end{equation}
Using Eq.~\eqref{eq:qi_combined_coefficient}, one then obtains
\begin{equation}
	\begin{aligned}
		\mathcal F_{r,a}^{(J)}(\mathbf0)
		&=\mathcal G_a^{(0)}
		\sum_{\nu\nu'\in\mathcal P_{r,a}}
		\kappa_{r,a,\nu\nu',\mathrm{CL}}^{(J)}\\
		&=\frac12\mathcal G_a^{(0)}I_F^a I_{S,\mathrm{CL}}^{a,J}
		\sum_{\nu\nu'\in\mathcal P_{r,a}}
		I_{C,r}^{a}(\nu\nu')=0.
	\end{aligned}
	\label{eq:appB_zero_transfer_cancellation}
\end{equation}
The final equality is the color-singlet zero-sum identity in
Eq.~\eqref{eq:app_zero_color_sum}.  It holds separately in the prior and post
representations and separately for $F_{14}$ and $F_{15}$.  The prior and post
terms of a fixed topology may therefore be combined before the
transferred-momentum integration.  This gives the same result as treating the
two representations separately in the formulas below.

Since the coefficient $\kappa$ already contains the single prior--post factor
$1/2$, the central contribution carried by representation $r$ can be written
compactly as
\begin{equation}
	\begin{aligned}
		\mathcal V_{r,a,\mathrm{CL}}^{(J)}
		={}&4\pi\int d^3q\,
		\frac{\mathcal F_{r,a}^{\alpha,(J)}(\mathbf q)}{q^2}\\
		&+6\pi b\int d^3q\,
		\frac{\mathcal F_{r,a}^{(J)}(\mathbf q)
			-\mathcal F_{r,a}^{(J)}(\mathbf0)}{q^4}.
	\end{aligned}
	\label{eq:appB_complete_central_kernel}
\end{equation}
Although Eq.~\eqref{eq:appB_zero_transfer_cancellation} makes the subtracted
constant vanish analytically, retaining the difference in
Eq.~\eqref{eq:appB_complete_central_kernel} enforces the distributional
definition and keeps the small-$q$ evaluation stable.  The subtraction itself
follows from the distributional definition in
Eq.~\eqref{eq:appB_distribution_definition}.  Color neutrality makes the
zero-transfer constant vanish in the complete intercluster sum, so no
adjustable infrared regulator or arbitrary additive confinement constant is
introduced.

The hyperfine contribution from the same representation is
\begin{equation}
	\mathcal V_{r,a,\mathrm{HF}}^{(J)}
	=\sum_{\nu\nu'\in\mathcal P_{r,a}}
	\kappa_{r,a,\nu\nu',\mathrm{HF}}^{(J)}
	I_{\mathrm{space},\mathrm{HF}}^{r,a,\nu\nu'},
	\label{eq:appB_complete_hyperfine_kernel}
\end{equation}
and the labelled topology kernel is
\begin{equation}
	V_a^{(J)}
	=\sum_{r\in\{\mathrm{pr},\mathrm{po}\}}
	\left(\mathcal V_{r,a,\mathrm{CL}}^{(J)}
	+\mathcal V_{r,a,\mathrm{HF}}^{(J)}\right).
	\label{eq:appB_complete_labelled_kernel}
\end{equation}
No additional prior--post averaging factor is to be applied to
Eq.~\eqref{eq:appB_complete_labelled_kernel}.  The combination of the two
labelled topologies into the normalized identical-quark final state is derived
separately in Appendix~\ref{app:qi_topology}.

Finally, the dimensions provide a direct normalization check.  With the
normalizations of Appendix~\ref{app:qi_internal_wavefunctions},
\begin{equation}
	\begin{aligned}
		{}[\phi_M]&=\mathrm{GeV}^{-3/2},&
		[\phi_B]&=\mathrm{GeV}^{-3},\\
		[v_{\nu\nu'}^{X}]&=\mathrm{GeV}^{-2}.&&
	\end{aligned}
	\label{eq:appB_wavefunction_dimensions}
\end{equation}
The intermediate Gaussian quantities have dimensions
\begin{equation}
	\begin{aligned}
		\bigl[\mathsf A\bigr]=\bigl[\mathsf G\bigr]&=\mathrm{GeV}^{-2},&
		\bigl[\boldsymbol\eta\bigr]=\bigl[\mathbf h\bigr]&=\mathrm{GeV}^{-1},\\
		\bigl[C\bigr]&=1,&
		\bigl[\mathcal C_{a,\nu\nu'}\bigr]&=\mathrm{GeV}^{-3}.
	\end{aligned}
	\label{eq:appB_intermediate_dimensions}
\end{equation}
The nine-dimensional measure in
Eq.~\eqref{eq:appB_general_spatial_integral} therefore gives
\begin{equation}
	[I_{\mathrm{space},X}]=[V_a^{(J)}]=\mathrm{GeV}^{-2}.
	\label{eq:appB_kernel_dimensions}
\end{equation}
After the molecular folding in Eq.~\eqref{eq:qi_physical_t00},
$[R_0]=\mathrm{GeV}^{-3/2}$ implies
$[T_{00,\mathrm{QI};fi}^{(J)}]=\mathrm{GeV}^{-1/2}$, in agreement with the width
normalization of Eq.~\eqref{eq:width_t00}.

\section{Identical-quark antisymmetrization and topology normalization}
\label{app:qi_topology}

This appendix derives the relative sign and normalization with which the two
labelled open-charm clusterings, $F_{14}$ and $F_{15}$, enter the physical
charged-component kernel.  The derivation uses the constituent labels and
state phases fixed in Appendix~\ref{app:qi_cfs}.  It also clarifies the
separate exact-isospin projection used to obtain the physical $I=1/2$
amplitude.

\subsection{Asymptotic state of the two identical light quarks}
\label{app:qi_asymptotic_antisymmetry}

Let $P_{45}$ interchange all one-particle variables carried by the two
identical $u$ quarks.  In the temporarily labelled first-quantized space,
\begin{equation}
	P_{45}^{\dagger}=P_{45},\qquad
	P_{45}^{2}=1,
	\label{eq:appC_exchange_operator}
\end{equation}
and the projector onto the antisymmetric two-$u$ subspace is
\begin{equation}
	\Pi_{45}^{(-)}=\frac{1-P_{45}}{2}.
	\label{eq:appC_antisymmetric_projector}
\end{equation}
The two labelled final clusterings introduced in
Eqs.~\eqref{eq:topology14} and~\eqref{eq:topology15} obey
\begin{equation}
	|F_{15}\rangle=P_{45}|F_{14}\rangle.
	\label{eq:appC_labelled_state_exchange}
\end{equation}

The minus sign required between them can also be seen directly from fermion
creation operators.  For a normalized one-particle wave packet $f$, define
the smeared operator
\begin{equation}
	u^{\dagger}[f]=\sum_{\alpha}f_{\alpha}u_{\alpha}^{\dagger},
	\label{eq:appC_smeared_creation_operator}
\end{equation}
where $\alpha$ collectively denotes momentum, spin, and color.  The canonical
anticommutation relations give
\begin{equation}
	\{u[f],u^{\dagger}[g]\}=\langle f|g\rangle,
	\qquad
	\{u^{\dagger}[f],u^{\dagger}[g]\}=0.
	\label{eq:appC_creation_anticommutators}
\end{equation}
If $f_M$ and $f_B$ are the asymptotic modes associated with the final meson
and baryon, respectively, then
\begin{equation}
	u^{\dagger}[f_M]u^{\dagger}[f_B]|0\rangle
	\ \longleftrightarrow\ 
	\frac{|f_M\rangle_4|f_B\rangle_5
		-|f_B\rangle_4|f_M\rangle_5}{\sqrt2}.
	\label{eq:appC_two_u_slater_state}
\end{equation}
The spectator constituents and the sums over the internal cluster variables
have been suppressed in Eq.~\eqref{eq:appC_two_u_slater_state}.  Restoring
them identifies the first term with $F_{14}$ and the second with $F_{15}$.
Thus the relative minus sign is a consequence of the anticommutation of the
two identical-$u$ creation operators; it is not a phase chosen separately for
the decay amplitude.

For asymptotically separated meson and baryon wave packets,
$\langle f_M|f_B\rangle=0$.  Equivalently, the labelled cluster states obey
\begin{equation}
	\langle F_{14}|F_{14}\rangle
	=\langle F_{15}|F_{15}\rangle=1,
	\qquad
	\langle F_{14}|F_{15}\rangle=0.
	\label{eq:appC_asymptotic_labelled_norms}
\end{equation}
The normalized antisymmetric final state associated with the explicitly
evaluated charged initial component is therefore
\begin{equation}
	\begin{aligned}
		|F_{\mathrm{AS}}\rangle
		&=\sqrt2\,\Pi_{45}^{(-)}|F_{14}\rangle\\
		&=\frac{|F_{14}\rangle-|F_{15}\rangle}{\sqrt2},
	\end{aligned}
	\label{eq:appC_normalized_physical_final_state}
\end{equation}
which reproduces Eq.~\eqref{eq:qi_antisymmetric_final_state}.

\subsection{Exchange parity of the initial and final cluster states}
\label{app:qi_exchange_parities}

The exchange property of the initial state follows separately from its
internal factors.  For the $u_4u_5$ pair in the ground-state
$\Sigma_c^{++}$,
\begin{equation}
	\begin{aligned}
		P_{45}\chi_{AB}^c&=-\chi_{AB}^c,&
		P_{45}\chi_{AB}^f&=+\chi_{AB}^f,\\
		P_{45}\chi_{AB}^s&=+\chi_{AB}^s,&
		P_{45}\phi_{AB}&=+\phi_{AB}.
	\end{aligned}
	\label{eq:appC_initial_factor_parities}
\end{equation}
The color sign follows from
$\epsilon_{a_3a_4a_5}=-\epsilon_{a_3a_5a_4}$; the flavor state $u_4u_5$,
the light-pair spin triplet, and the ground-state spatial wave function are
symmetric.  Consequently, the complete initial state has the required
fermionic exchange parity,
\begin{equation}
	P_{45}|AB;JM_J\rangle=-|AB;JM_J\rangle.
	\label{eq:appC_initial_exchange_parity}
\end{equation}

For the final labelled state, $P_{45}$ moves the light quark in the meson to
the baryon and the light quark in the baryon to the meson.  With the common
cluster ordering $M_fB_f$ and the phases of Appendix~\ref{app:qi_cfs}, this
maps
\begin{equation}
	\begin{aligned}
		M_f(14)B_f(325)&\xrightarrow{\ P_{45}\ }
		M_f(15)B_f(324),\\
		\chi_{F_{14}}^c\chi_{F_{14}}^f\chi_{F_{14}}^s\phi_{F_{14}}
		&\xrightarrow{\ P_{45}\ }
		\chi_{F_{15}}^c\chi_{F_{15}}^f\chi_{F_{15}}^s\phi_{F_{15}},
	\end{aligned}
	\label{eq:appC_final_factor_mapping}
\end{equation}
which is the component form of
Eq.~\eqref{eq:appC_labelled_state_exchange}.

\subsection{Relation between the two labelled kernels}
\label{app:qi_labelled_kernel_relation}

For $r\in\{\mathrm{pr},\mathrm{po}\}$, let
$\widehat{\mathcal V}_{r,a}$ denote the complete pair-interaction sum for
representation $r$ and labelled topology $a$.  Exact isospin symmetry gives
$m_4=m_5$, and the two $u$ quarks use the same internal widths and
constituent interaction.  Hence
\begin{equation}
	P_{45}\widehat{\mathcal V}_{r,14}P_{45}
	=\widehat{\mathcal V}_{r,15}.
	\label{eq:appC_interaction_exchange_covariance}
\end{equation}
At the diagram level, the required pair mappings are
\begin{align}
	&(13,14,15,23,24,25)
	\xrightarrow{4\leftrightarrow5}
	(13,15,14,23,25,24),
	\label{eq:appC_prior_pair_mapping}\\
	&(13,12,15,43,42,45)
	\xrightarrow{4\leftrightarrow5}
	(13,12,14,53,52,54).
	\label{eq:appC_post_pair_mapping}
\end{align}
The first line is the prior set in a permuted order, while the second maps the
$F_{14}$ post set into the $F_{15}$ post set.  Under the same exchange, the
spin and spatial arguments are relabelled as in Appendices~\ref{app:qi_cfs}
and~\ref{app:qi_spatial}.  Equation~\eqref{eq:appC_interaction_exchange_covariance}
therefore applies to the complete color--flavor--spin weighted sum, including
both the $\mathrm{CL}$ and $\mathrm{HF}$ interactions.

Using Eqs.~\eqref{eq:appC_labelled_state_exchange},
\eqref{eq:appC_initial_exchange_parity}, and
\eqref{eq:appC_interaction_exchange_covariance}, one obtains
\begin{equation}
	\begin{aligned}
		V_{r,15}^{(J)}
		&=\langle F_{15};JM_J|
		\widehat{\mathcal V}_{r,15}|AB;JM_J\rangle\\
		&=\langle F_{14};JM_J|P_{45}
		\bigl(P_{45}\widehat{\mathcal V}_{r,14}P_{45}\bigr)
		|AB;JM_J\rangle\\
		&=-\langle F_{14};JM_J|
		\widehat{\mathcal V}_{r,14}|AB;JM_J\rangle\\
		&=-V_{r,14}^{(J)}.
	\end{aligned}
	\label{eq:appC_prior_post_topology_sign}
\end{equation}
The relation holds separately for the prior and post representations.  It is
therefore preserved by their arithmetic average,
\begin{equation}
	V_{15}^{(J)}=-V_{14}^{(J)},
	\label{eq:appC_averaged_topology_sign}
\end{equation}
and holds before the molecular radial and partial-wave integrations.  In
particular, it is valid pointwise in the external variables $(k,p,u)$.

Projecting the transition onto the normalized state in
Eq.~\eqref{eq:appC_normalized_physical_final_state} gives
\begin{equation}
	\begin{aligned}
		V_{\mathrm{ch}}^{(J)}
		&=\langle F_{\mathrm{AS}};JM_J|\widehat{\mathcal V}|AB;JM_J\rangle\\
		&=\frac{V_{14}^{(J)}-V_{15}^{(J)}}{\sqrt2}
		=\sqrt2\,V_{14}^{(J)}.
	\end{aligned}
	\label{eq:appC_charged_kernel_normalization}
\end{equation}
The $1/\sqrt2$ in the normalized external state is distinct from the
fixed-topology flavor overlap
$I_F^{14}=I_F^{15}=1/\sqrt2$ derived in
Eq.~\eqref{eq:app_flavor_overlap}.  Both are also distinct from the factor
$1/2$ used once in the prior--post average.  Equation~\eqref{eq:appC_charged_kernel_normalization}
contains no additional prior--post factor.

The factor $\sqrt2$ is a coherent quark-line normalization effect, rather
than a fitted enhancement or a final-state phase-space multiplicity.  The
asymptotic hadrons $\Lambda_c$ and $\bar D^{(*)}$ are distinguishable, so no
$1/2!$ identical-particle factor is introduced in the two-body width.

\subsection{Exact-isospin reconstruction}
\label{app:qi_isospin_reconstruction}

For completeness, the topology normalization can be combined with the
charge-state projection of Sec.~\ref{sec:states_isospin}.  Define
\begin{equation}
	\begin{aligned}
		V_{\mathrm{ch}}^{(J)}&=
		\langle\Lambda_c^+\bar D^{(*)0}|\widehat{\mathcal V}|
		\Sigma_c^{++}D^{(*)-}\rangle,\\
		V_{\mathrm{neu}}^{(J)}&=
		\langle\Lambda_c^+\bar D^{(*)0}|\widehat{\mathcal V}|
		\Sigma_c^+\bar D^{(*)0}\rangle.
	\end{aligned}
	\label{eq:appC_charge_amplitudes}
\end{equation}
In the Condon--Shortley convention of
Eq.~\eqref{eq:initial_isospin_state}, the orthogonal $I=3/2$ state is
\begin{equation}
	\left|\frac32,\frac12\right\rangle
	=\frac{1}{\sqrt3}|\Sigma_c^{++}D^{(*)-}\rangle
	+\sqrt{\frac23}|\Sigma_c^+\bar D^{(*)0}\rangle.
	\label{eq:appC_isospin_threehalf_state}
\end{equation}
An isospin-conserving interaction cannot connect this state to the
$I=1/2$ final channel.  Therefore,
\begin{equation}
	\frac{1}{\sqrt3}V_{\mathrm{ch}}^{(J)}
	+\sqrt{\frac23}V_{\mathrm{neu}}^{(J)}=0,
	\qquad
	V_{\mathrm{neu}}^{(J)}=-\frac{V_{\mathrm{ch}}^{(J)}}{\sqrt2}.
	\label{eq:appC_neutral_charged_relation}
\end{equation}
Projection of the initial $I=1/2$ state then gives
\begin{equation}
	\begin{aligned}
		V_{\mathrm{QI};fi}^{(J)}
		&=\sqrt{\frac23}V_{\mathrm{ch}}^{(J)}
		-\frac{1}{\sqrt3}V_{\mathrm{neu}}^{(J)}\\
		&=\sqrt{\frac32}V_{\mathrm{ch}}^{(J)}
		=\sqrt3\,V_{14}^{(J)}.
	\end{aligned}
	\label{eq:appC_complete_normalization_chain}
\end{equation}
The last equality combines two logically separate operations: antisymmetrizing
the identical $u$ quarks in the charged component and projecting that charged
component onto the exact $I=1/2$ molecular state.  Physical charge splittings
used in thresholds and final-state kinematics do not alter this amplitude
normalization convention.

\subsection{Finite exchange overlap}
\label{app:qi_finite_exchange_overlap}

Equation~\eqref{eq:appC_asymptotic_labelled_norms} is an asymptotic
external-state statement.  At finite cluster separation, quark exchange can
produce a nonzero overlap between the two labelled clusterings.  For
normalized wave packets, define
\begin{equation}
	S_{45}=\langle F_{14}|P_{45}|F_{14}\rangle
	=\langle F_{14}|F_{15}\rangle.
	\label{eq:appC_finite_exchange_overlap}
\end{equation}
Since $P_{45}$ is Hermitian, $S_{45}$ is real, and the normalized
antisymmetric combination would contain
\begin{equation}
	|F_{\mathrm{AS}}\rangle_{\mathrm{finite}}
	=\frac{|F_{14}\rangle-|F_{15}\rangle}
	{\sqrt{2(1-S_{45})}}.
	\label{eq:appC_finite_overlap_normalization}
\end{equation}
For continuum cluster states, the corresponding exchange contribution is a
nonlocal norm kernel,
\begin{equation}
	\mathcal N_{45}(\mathbf k',\mathbf k)
	=\langle F_{14};\mathbf k'|P_{45}|
	F_{14};\mathbf k\rangle.
	\label{eq:appC_rgm_norm_kernel}
\end{equation}
A finite-overlap treatment would require simultaneous Hamiltonian and norm
kernels, with an orthogonalized kernel of the schematic form
$\mathcal N^{-1/2}\mathcal H\mathcal N^{-1/2}$.  Such a
resonating-group treatment is not part of the present first-Born
calculation.  We instead use normalized asymptotic hadron states, for which
$S_{45}\to0$ and Eq.~\eqref{eq:appC_normalized_physical_final_state}
follows.  The simple $\sqrt2$ in
Eq.~\eqref{eq:appC_charged_kernel_normalization} should therefore not be
interpreted as an estimate of finite-separation exchange dynamics.

\section{Additional analytic and numerical consistency checks}
\label{app:qi_numerics}

This appendix supplements the compact validation summary in
Sec.~\ref{sec:validation}.  We first document the pointwise identities and
the higher-order quadrature comparison for the quark-interchange (QI)
amplitudes.  We then give the endpoint treatment of the retarded
one-pion-exchange (OPE) fold, an independent expression for its absorptive
part, and representative contact--pole and coherent-width identities.  The
parameter variations defined in Sec.~\ref{sec:numerical_sensitivities} test
physical and model assumptions and are distinct from the numerical
comparisons collected here.

\subsection{QI identities and quadrature refinement}
\label{app:qi_detailed_numerics}

The internal-state conventions of Appendix~\ref{app:qi_cfs} imply the exact
normalizations
\begin{equation}
	\begin{gathered}
		\langle\chi_{AB}^c|\chi_{AB}^c\rangle
		=\langle\chi_{F_{14}}^c|\chi_{F_{14}}^c\rangle
		=\langle\chi_{F_{15}}^c|\chi_{F_{15}}^c\rangle=1,\\
		I_F^{14}=I_F^{15}=\frac{1}{\sqrt2},
		\qquad
		\sum_{\nu\nu'\in\mathcal P_{r,a}}
		I_{C,r}^{a}(\nu\nu')=0.
	\end{gathered}
	\label{eq:appD_qi_state_identities}
\end{equation}
The last relation removes the zero-transfer constant from the
distributional confinement kernel, as derived in
Appendix~\ref{app:qi_spatial}.  The representative color--flavor--spin
coefficient also reproduces
\begin{equation}
	\kappa_{\mathrm{pr},14,13,\mathrm{HF}}^{(1/2)}
	=\frac{\sqrt6}{144},
	\label{eq:appD_representative_cfs_check}
\end{equation}
in agreement with Eq.~\eqref{eq:qi_example_cfs_coefficient}.

It is useful to state the topology check at the level at which the molecular
fold is performed.  Let $V_{r,a,X}^{(J)}(k,p,u)$ denote the kernel for prior
or post representation $r$, labelled topology $a$, and interaction part
$X\in\{\mathrm{CL},\mathrm{HF}\}$.  The exchange argument of
Appendix~\ref{app:qi_topology} gives
\begin{equation}
	\mathcal R_{r,X}^{(J)}(k,p,u)
	\equiv V_{r,14,X}^{(J)}(k,p,u)
	+V_{r,15,X}^{(J)}(k,p,u)=0.
	\label{eq:appD_pointwise_topology_identity}
\end{equation}
This relation is satisfied separately by the spin-independent and hyperfine
parts, for both prior and post representations, before the integrations over
$p$ and $u$.  The subsequent normalization chain is therefore
\begin{equation}
	V_{\mathrm{ch}}^{(J)}=\sqrt2\,V_{14}^{(J)},
	\qquad
	V_{\mathrm{QI};fi}^{(J)}=\sqrt{\frac32}\,V_{\mathrm{ch}}^{(J)}
	=\sqrt3\,V_{14}^{(J)},
	\label{eq:appD_qi_normalization_chain}
\end{equation}
with no additional topology, flavor, or prior--post factor.

To isolate quadrature effects, all orders are raised simultaneously from
\begin{equation}
	(N_p,N_u;N_q,N_{u_q},N_\phi)
	=(20,16;24,12,12)
	\label{eq:appD_qi_reference_grid}
\end{equation}
to
\begin{equation}
	(N_p,N_u;N_q,N_{u_q},N_\phi)
	=(28,22;32,16,16),
	\label{eq:appD_qi_higher_grid}
\end{equation}
while keeping $p_{\max}^{\rm QI}=1.6~\mathrm{GeV}$ and
$q_{\max}=4.0~\mathrm{GeV}$ fixed.  Table~\ref{tab:appD_qi_convergence}
lists the result for every calculated $S$-wave channel.  We define
\begin{equation}
	\delta_\Gamma
	=100\left|\frac{\Gamma_{\rm high}-\Gamma_{\rm ref}}
	{\Gamma_{\rm ref}}\right|.
	\label{eq:appD_relative_width_change}
\end{equation}

\begin{table*}[t]
	\caption{QI quadrature refinement for all eight calculated $S$-wave
		channels.  Widths are in MeV.  The two
		$J=3/2\to\Lambda_c\bar D$ modes are $D$ waves and are not part of this
		comparison.}
	\label{tab:appD_qi_convergence}
	\small
	\begin{ruledtabular}
		\begin{tabular}{lclccc}
			Initial state & $J$ & Final state & $\Gamma_{\rm ref}$
			& $\Gamma_{\rm high}$ & $\delta_\Gamma$ (\%) \\
			\colrule
			$P_c(4312)$ & $\tfrac12$ & $\Lambda_c\bar D$
			& $0.0028477648$ & $0.0028477648$ & $<10^{-6}$ \\
			$P_c(4312)$ & $\tfrac12$ & $\Lambda_c\bar D^*$
			& $0.7421105620$ & $0.7421105626$ & $7.19\times10^{-8}$ \\
			$P_c(4440)$ & $\tfrac12$ & $\Lambda_c\bar D$
			& $0.0360671554$ & $0.0360671894$ & $9.42\times10^{-5}$ \\
			$P_c(4440)$ & $\tfrac12$ & $\Lambda_c\bar D^*$
			& $0.4953352423$ & $0.4953354563$ & $4.32\times10^{-5}$ \\
			$P_c(4440)$ & $\tfrac32$ & $\Lambda_c\bar D^*$
			& $0.0450681927$ & $0.0450682250$ & $7.16\times10^{-5}$ \\
			$P_c(4457)$ & $\tfrac12$ & $\Lambda_c\bar D$
			& $0.0010560078$ & $0.0010560077$ & $8.63\times10^{-6}$ \\
			$P_c(4457)$ & $\tfrac12$ & $\Lambda_c\bar D^*$
			& $0.0808195485$ & $0.0808195502$ & $2.01\times10^{-6}$ \\
			$P_c(4457)$ & $\tfrac32$ & $\Lambda_c\bar D^*$
			& $0.0084765093$ & $0.0084765095$ & $2.72\times10^{-6}$ \\
		\end{tabular}
	\end{ruledtabular}
\end{table*}

The largest change is the
$P_c(4440),J=1/2\to\Lambda_c\bar D$ entry,
$0.0360671554\to0.0360671894~\mathrm{MeV}$, corresponding to
$9.42\times10^{-5}\%$.  Thus the numerical integration uncertainty of the
reference grid is negligible compared with the one-at-a-time variations of
the molecular and constituent-model inputs.

\subsection{Pion-pole endpoints and interval mapping}
\label{app:ope_checks}

For fixed final momentum $k$, recall from Eq.~\eqref{eq:ope_delta_ab} that
the angular denominator is $D_0-D_1u-i0$. It is useful to define the two
endpoint functions
\begin{equation}
	D_\pm(p,k)\equiv D_0(p,k)\pm D_1(p,k),
	\qquad D_1(p,k)=2pk.
	\label{eq:appD_ope_ab}
\end{equation}
The endpoints of the on-shell angular interval occur when the pole reaches
$u=\pm1$.  They are therefore the positive roots of
\begin{equation}
	\begin{aligned}
		D_-(p,k)&=0,\\
		D_+(p,k)&=0.
	\end{aligned}
	\label{eq:appD_ope_endpoint_equations}
\end{equation}
All roots in $0<p<p_{\max}^{\pi}$ are bracketed and then refined.  Together
with $0$ and $p_{\max}^{\pi}$, they define an ordered partition
\begin{equation}
	0=p_0<p_1<\cdots<p_n=p_{\max}^{\pi}.
	\label{eq:appD_ope_interval_partition}
\end{equation}

On each interval $[p_j,p_{j+1}]$, we use $x\in[-1,1]$ and the mapping
\begin{equation}
	\begin{aligned}
		p(x)={}&p_j+(p_{j+1}-p_j)
		\sin^2\!\left[\frac{\pi}{4}(x+1)\right],\\
		\frac{dp}{dx}={}&\frac{\pi}{4}(p_{j+1}-p_j)
		\sin\!\left[\frac{\pi}{2}(x+1)\right].
	\end{aligned}
	\label{eq:appD_sine_squared_mapping}
\end{equation}
The vanishing Jacobian at both ends clusters the integration points near an
endpoint cusp and controls the integrable logarithms in
Eq.~\eqref{eq:ope_angular_log}.  For the reference
$P_c(4312)\to\Lambda_c\bar D^*$ case with the direct baryon-line energy
transfer and $\Lambda_\pi=0.50~\mathrm{GeV}$, one finds
\begin{equation}
	\begin{aligned}
		k&=0.199560876~\mathrm{GeV},\\
		p_1&=0.116819670~\mathrm{GeV},&
		p_2&=0.311762532~\mathrm{GeV}.
	\end{aligned}
	\label{eq:appD_reference_pole_endpoints}
\end{equation}
Both internal endpoints in this case are roots of $D_-(p,k)$, and the
on-shell angular region lies between them.

\subsection{Independent absorptive-part relation}
\label{app:ope_absorptive_check}

The physical-sheet identity
\begin{equation}
	\frac{1}{x-i0}=\operatorname{PV}\frac{1}{x}+i\pi\delta(x)
	\label{eq:appD_distribution_identity}
\end{equation}
gives an expression for the imaginary part that does not require evaluating
the real logarithm.  Defining
\begin{equation}
	\Omega_\pi(k)=\left\{p\in[0,p_{\max}^{\pi}]:
	|D_0(p,k)|<D_1(p,k)\right\},
	\label{eq:appD_on_shell_radial_set}
\end{equation}
one has
\begin{equation}
	\operatorname{Im}\mathcal L_\pi(p,k)
	=\frac{\pi}{D_1(p,k)}\,
	\Theta\!\bigl(D_1-|D_0|\bigr).
	\label{eq:appD_imaginary_angular_log}
\end{equation}
Consequently, the absorptive part of the pion-pole amplitude is
\begin{equation}
	\begin{aligned}
		\operatorname{Im}T_{00,\pi,\mathrm{pole};fi}^{(J)}
		={}&-\frac{|g|\,|g_2|\widehat c_{fi}^{(J)}}{f_\pi^2}
		\int_{\Omega_\pi(k)}dp\,p^2R_0(p)F_{\Lambda_\pi}(k,p)\\
		&\times\frac{\pi\Delta_\pi(p,k)}{D_1(p,k)}.
	\end{aligned}
	\label{eq:appD_independent_absorptive_part}
\end{equation}
The contact-like contribution is real, so the full and pion-pole kernels
have the same imaginary part.  For the reference channel of
Eq.~\eqref{eq:appD_reference_pole_endpoints}, evaluating
Eq.~\eqref{eq:appD_independent_absorptive_part} gives
\begin{equation}
	\operatorname{Im}T_{00,\pi,\mathrm{pole}}^{(1/2)}
	=0.274212282530~\mathrm{GeV}^{-1/2},
	\label{eq:appD_absorptive_numerical_value}
\end{equation}
identical in the displayed digits to the result obtained from the complex
logarithm in Eq.~\eqref{eq:ope_angular_log}.  This check fixes the sign of the
physical-sheet imaginary part.  It does not turn the first-Born treatment
into a unitary three-body calculation or define a separate
$\Lambda_c\bar D^{(*)}\pi$ partial width.

\subsection{Contact--pole and coherent-width identities}
\label{app:ope_amplitude_identities}

The distributional decomposition in
Eq.~\eqref{eq:ope_contact_identity} requires the folded amplitudes to obey
\begin{equation}
	T_{00,\pi,\mathrm{full};fi}^{(J)}
	=T_{00,\pi,\mathrm{ct};fi}^{(J)}
	+T_{00,\pi,\mathrm{pole};fi}^{(J)}.
	\label{eq:appD_folded_contact_identity}
\end{equation}
For the same reference $P_c(4312)\to\Lambda_c\bar D^*$ transition,
\begin{equation}
	\begin{aligned}
		T_{00,\pi,\mathrm{ct}}^{(1/2)}
		&=2.383195854,\\
		T_{00,\pi,\mathrm{pole}}^{(1/2)}
		&=0.463697755+0.274212283i,\\
		T_{00,\pi,\mathrm{full}}^{(1/2)}
		&=2.846893609+0.274212283i,
	\end{aligned}
	\qquad [\mathrm{GeV}^{-1/2}].
	\label{eq:appD_reference_contact_decomposition}
\end{equation}
The residual of Eq.~\eqref{eq:appD_folded_contact_identity} is below
$3\times10^{-15}~\mathrm{GeV}^{-1/2}$.  This is an internal identity, not a
preference for the full OPE kernel.  As explained in
Sec.~\ref{sec:ope_contact_subtraction}, the full kernel is retained only as a
short-distance matching diagnostic.

The coherent-width relation supplies a second independent check.  Writing
\begin{equation}
	\Gamma_{\mathrm{int};fi}^{(J)}
	=2\rho_{fi}\operatorname{Re}\!\left[
	T_{00,\mathrm{QI};fi}^{(J)}
	T_{00,\pi,\mathrm{ref};fi}^{(J)*}\right],
	\label{eq:appD_interference_definition}
\end{equation}
Eq.~\eqref{eq:ope_interference_width} becomes
\begin{equation}
	\Gamma_{\mathrm{tot};fi}^{(J)}(\eta_{\mathrm{match}})
	=\Gamma_{\mathrm{QI};fi}^{(J)}
	+\Gamma_{\pi,\mathrm{ref};fi}^{(J)}
	+\eta_{\mathrm{match}}\Gamma_{\mathrm{int};fi}^{(J)}.
	\label{eq:appD_coherent_width_identity}
\end{equation}
For the reference $P_c(4312)\to\Lambda_c\bar D^*$ case, the individual
terms are
\begin{equation}
	\begin{aligned}
		\Gamma_{\mathrm{QI};fi}^{(1/2)}&=0.742110562~\mathrm{MeV},\\
		\Gamma_{\pi,\mathrm{ref};fi}^{(1/2)}&=0.250633751~\mathrm{MeV},\\
		\Gamma_{\mathrm{int};fi}^{(1/2)}&=0.742445183~\mathrm{MeV},
	\end{aligned}
	\label{eq:appD_reference_width_components}
\end{equation}
which yield
\begin{equation}
	\Gamma_{+}=1.735189496~\mathrm{MeV},
	\qquad
	\Gamma_{-}=0.250299130~\mathrm{MeV}.
	\label{eq:appD_reference_coherent_widths}
\end{equation}
The direct evaluation of
$\rho_{fi}|T_{00,\mathrm{tot};fi}^{(J)}(\eta_{\mathrm{match}})|^2$ agrees with
Eq.~\eqref{eq:appD_coherent_width_identity} to better than
$5\times10^{-16}~\mathrm{MeV}$.  The same global
$\eta_{\mathrm{match}}$ must be used in every channel; this identity does not
permit a separate sign choice for each decay.

\subsection{OPE quadrature and radial-endpoint convergence}
\label{app:ope_numerical_convergence}

Table~\ref{tab:appD_ope_convergence} gives the complex amplitudes underlying
the compact OPE convergence summary in Sec.~\ref{sec:validation}.  Both tests
use the reference $P_c(4312)\to\Lambda_c\bar D^*$ channel, the direct
baryon-line energy transfer, the pion-pole kernel, and
$\Lambda_\pi=0.50~\mathrm{GeV}$. The quadrature-order comparison keeps
$p_{\max}^{\pi}=2~\mathrm{GeV}$ fixed.  In the radial-endpoint comparison,
$N_\pi=256$ is used for both endpoints so that the high-momentum tail is
isolated from the quadrature-order difference.

\begin{table}[!h]
	\caption{Complex amplitudes in the detailed OPE convergence comparison.
		The first two rows vary $N_\pi$ at
		$p_{\max}^{\pi}=2~\mathrm{GeV}$; the last two vary
		$p_{\max}^{\pi}$ at $N_\pi=256$.  Amplitudes are in
		$\mathrm{GeV}^{-1/2}$.}
	\label{tab:appD_ope_convergence}
	\footnotesize
	\begin{ruledtabular}
		\begin{tabular}{lc}
			Setting & $T_{00,\pi,\mathrm{pole}}$ \\
			\colrule
			$N_\pi=64$
			& $0.463697754671+0.274212282530i$ \\
			$N_\pi=128$
			& $0.463697656416+0.274212282530i$ \\
			$p_{\max}^{\pi}=2~\mathrm{GeV}$
			& $0.463697650177+0.274212282530i$ \\
			$p_{\max}^{\pi}=3~\mathrm{GeV}$
			& $0.463697650361+0.274212282530i$ \\
		\end{tabular}
	\end{ruledtabular}
\end{table}

The reference result reported in Sec.~\ref{sec:results} uses $N_\pi=64$ and
$p_{\max}^{\pi}=2~\mathrm{GeV}$.  The absolute differences between the two
quadrature orders and between the two radial endpoints are, respectively,
$9.83\times10^{-8}$ and $1.84\times10^{-10}~\mathrm{GeV}^{-1/2}$.
They are many orders of magnitude smaller
than the changes generated by the cutoff, energy-transfer prescription, and
contact treatment.  The numerical stability demonstrated here therefore
does not reduce the matching ambiguities attached to the exploratory OPE
scenarios.

\bibliography{pc_open_charm_references_updated}

\end{document}